\documentclass[a4paper,11pt]{article}
\usepackage{rotating}
\usepackage{jheppub} 
\usepackage[utf8]{inputenc}
\usepackage[T1]{fontenc} 
\usepackage{placeins} 
\usepackage{chngpage}
\usepackage{amsmath,amsfonts}
\usepackage{graphicx}
\usepackage{color}
\usepackage{xcolor}
\usepackage{relsize}
\usepackage{epstopdf}
\usepackage{hyperref}
\usepackage{mathrsfs}
\usepackage{ragged2e}
\usepackage{amssymb}
\usepackage{placeins}
\usepackage[normalem]{ ulem }
\usepackage{amsthm}
\usepackage{comment}
\usepackage{graphics}
\usepackage{caption}
\usepackage{subcaption}
\usepackage{verbatim}
\usepackage{cprotect}
\usepackage{diagbox}

\usepackage{tabularx,booktabs}
\usepackage{hhline}
\usepackage{multicol}
\usepackage{array,multirow}
\usepackage{booktabs}
\usepackage{colortbl}
\usepackage{float}
\usepackage{verbatimbox}
\usepackage{adjustbox}
\newcommand{\crowcolor}{\rowcolor[rgb]{0.9,0.9,0.9}}
\newcommand{\ctoprule}{\toprule[0.5mm]}
\newcommand{\cbottomrule}{\bottomrule[0.5mm]}

\def\gsim{\raise0.3ex\hbox{$\;>$\kern-0.75em\raise-1.1ex\hbox{$\sim\;$}}}
\def\lsim{\raise0.3ex\hbox{$\;<$\kern-0.75em\raise-1.1ex\hbox{$\sim\;$}}}
\def \znbb {0\nu\beta\beta}
\def\qslash{q \hspace{-0.5em}/\;\:}

\makeatletter
\gdef\@fpheader{}
\vspace*{0cm}
\makeatother

\begin{document}

\title{Tree-level UV completions for $N_R$SMEFT $d=6$ and $d=7$ operators}

\author[a]{Rebeca Beltr\'an,}
\author[b]{Ricardo Cepedello,}
\author[a]{Martin Hirsch}
\affiliation[a]{Instituto de F\'isica Corpuscular (IFIC), 
Universidad de Valencia-CSIC, E-46980 Valencia, Spain}
\affiliation[b]{Institut f\"ur Theoretische Physik und Astrophysik, 
Universit\"{a}t W\"{u}rzburg, 97074 Würzburg, Germany}

\emailAdd{rebeca.beltran@ific.uv.es}
\emailAdd{ricardo.cepedello@physik.uni-wuerzburg.de}
\emailAdd{mahirsch@ific.uv.es}

\date{\today}% It is always \today, today,
             % but any date may be explicitly specified 

\abstract{ 

We study ultra-violet completions for operators in standard model
effective field theory extended with right-handed neutrinos
($N_R$SMEFT). Using a diagrammatic method, we generate systematically
lists of possible tree-level completions involving scalars, fermions
or vectors for all operators at $d=6$ and $d=7$, which contain at
least one right-handed neutrino. We compare our lists of possible UV
models to the ones found for pure SMEFT. We also discuss how the
observation of LNV processes via $N_R$SMEFT operators at the LHC can
be related to Majorana neutrino masses of the standard model neutrinos.

}

\keywords{SMEFT, UV completions, right-handed neutrinos}
             
\maketitle

% !TEX root = ../NR_decomp.tex

\section{Introduction\label{sec:intro}}

Experimental searches for heavy neutral leptons (HNLs) have gained a
lot of momentum in the past few years, for recent reviews see for
example \cite{Curtin:2018mvb,Alimena:2019zri}.  Minimal models of HNLs
assume only that some nearly singlet fermion exists with a small
coupling to gauge bosons. This setup is motivated experimentally as it
gives a simple two parameter extension (one mixing angle and one mass)
of the standard model (SM), describing the experimental sensitivity, which can then be
easily compared among different searches.

At the same time, there has also been a revival of interest in HNLs
from theory. Here, the main motivation is usually the connection HNLs
might have with the non-zero neutrino masses observed in oscillation
experiments. In the minimal type-I seesaw
~\cite{Minkowski:1977sc,Yanagida:1979as,GellMann:1980vs,Mohapatra:1979ia,
  Schechter:1980gr}, Majorana right-handed neutrinos ($N_R$'s) have
only one coupling to SM particles: The Yukawa coupling to leptons and
the Higgs field. After electro-weak symmetry breaking, this setup
leads to Majorana neutrino masses for the active neutrinos and HNL
states with a coupling suppressed by a small mixing angle, typically
$|V|^2 \propto m_{\nu}/M_M$, where $M_M$ is the mass of the right-handed 
neutrino. 

Compared to the expected experimental sensitivities this mixing is
quite small, but prospects are much better in many non-minimal models:
Inverse \cite{Mohapatra:1986bd} and linear seesaw
\cite{Akhmedov:1995ip,Akhmedov:1995vm}, also models with a new $Z'$
\cite{Deppisch:2018eth,Amrith:2018yfb} or leptoquarks
\cite{Dorsner:2016wpm}, to mention a few examples. However, given the
large number of possible BSM extensions involving $N_R$'s and the
absence (so far) of any BSM physics at the LHC, a more practical
ansatz to parametrise the phenomenolgy of $N_R$'s is to use effective
field theory, in particular $N_R$SMEFT, i.e.  standard model effective
theory extended with right-handed neutrinos.

The operator basis for $N_R$SMEFT is known now up to $d=9$
\cite{Li:2021tsq}. Lower dimensional operators have been studied in a
number of papers earlier, for example $d =5$
~\cite{delAguila:2008ir,Aparici:2009fh,Graesser:2007yj}, $d=6$
~\cite{delAguila:2008ir,Bell:2005kz,Graesser:2007pc} and $d=7$
~\cite{Bhattacharya:2015vja,Liao:2016qyd}. The phenomenology for
$N_R$SMEFT is also a very active area of research.
Ref.~\cite{Alcaide:2019pnf}, studied constraints on the four-fermion
operators involving $N_R$ assuming a stable $N_R$.  For promptly
decaying $N_R$'s see \cite{Butterworth:2019iff}, other decay modes
have been studied in \cite{Duarte:2015iba,Duarte:2016miz}.  $N_R$'s as
long-lived particles at the LHC in $N_R$SMEFT have been discussed in
\cite{Cottin:2021lzz,Beltran:2021hpq,Beltran:2022ast,Beltran:2023nli}.
The very recent paper \cite{Fernandez-Martinez:2023phj} gives a
reinterpretation of many previous HNL searches in terms of $N_R$SMEFT
operators. Further collider studies of $N_R$SMEFT include
Refs.~\cite{Duarte:2014zea,Duarte:2016caz,Han:2020pff,
  Barducci:2020ncz,Barducci:2020icf}.

While different UV models for $N_R$SMEFT operators at $d=5$ and $d=6$
have been mentioned in the literature, what is still lacking is a
systematic decomposition of all $d=6$ and $d=7$ operators, i.e. an
attempt to give the complete list of one and two particle extensions
of the SM, that can generate the complete set of operators in the
ultra-violet. To provide such a systematic particle ``dictionary''
constitutes the basic motivation for our current work. In this context we
need to mention \cite{Bischer:2019ttk}, which has some overlap with
our paper. In \cite{Bischer:2019ttk} a list of leptoquark states for
$N_R$SMEFT operators at $d=6$ has been derived and we agree with these
results.  Also important for us is \cite{deBlas:2017xtg}. Here, the
authors have presented the complete dictionary of one field extensions
of the SM particle content for pure SMEFT at $d=6$. We will compare
our results to \cite{deBlas:2017xtg} and comment on the differences
between $N_R$SMEFT and SMEFT dictionaries in section \ref{sect:d6}.

The list of $d=7$ operators and their tree-level completions is
presented in section \ref{sect:d7}.  At $d=7$ level, all operators
with $N_R$'s violate lepton number by two units.\footnote{Assuming
  that the lepton number of $N_R$ is $L(N_R)=+1$. There is a subtlety
  involved in this assumption, which is discussed in section
  \ref{sec:lnv}.}  Lepton number violation (LNV) in two units should
always generate Majorana neutrino masses for the active neutrinos of
the SM. The most famous example of this connection is the so-called
``black-box theorem'' \cite{Schechter:1981bd}, where it was shown that
the observation of neutrinoless double beta decay ($\znbb$ decay)
guarantees that Majorana neutrino masses are generated at some level
in perturbation theory. Similarly, if LNV is observed in a process
involving $N_R$'s at the LHC, the active neutrinos must have Majorana
masses. We will discuss this in detail in section \ref{sec:lnv}.

The rest of this paper is organised as follows. In the next section,
we first discuss the basics of the diagrammatic method.  We then give
the list of $d=6$ operators and their decomposition in section
\ref{sect:d6} and for $d=7$ operators in \ref{sect:d7}. Baryon number
violating operators are discussed separately in \ref{sect:bnv}.
Section \ref{sec:lnv} is then devoted to a discussion of LNV in
$N_R$SMEFT, before we close with a short summary. In the appendix we
provide Lagrangians for all the UV models we discussed in the main
text.

% !TEX root = ../NR_decomp.tex

\section{Decompositions\label{sec:decomp}}

\begin{table}[t]
\centering
\renewcommand*{\arraystretch}{1.1}
\begin{tabular}{@{\hspace{\tabcolsep}}l >{\centering\arraybackslash}p{1.6cm} >{\centering\arraybackslash}p{1.3cm}>{\centering\arraybackslash}p{1.3cm}>{\centering\arraybackslash}p{1.3cm}>{\centering\arraybackslash}p{1.3cm}@{\hspace{\tabcolsep}}}
\ctoprule
\crowcolor
Name  & $\mathcal{S}$      & $\mathcal{S}_1$    & $\varphi$             & $\Xi$              & $\Xi_1$            \\% \cmrule
Irrep & $(1,1,0)$          & $(1,1,1)$          & $\left(1,2,\frac{1}{2}\right)$      & $(1,3,0)$          & $(1,3,1)$          \\
$d=6$  & $\circ $  & $\circ $ & $\circ $  & $\circ $  & $\circ $  \\ \cbottomrule
\noalign{\vskip 1mm} 
\ctoprule
\crowcolor
Name  & $\mathcal{\omega}_1$      & $\mathcal{\omega}_2$    & $\Pi_1$             & $\Pi_7$              & $\zeta$            \\% \cmrule
Irrep & $\left(3,1,-\frac{1}{3}\right)$          & $\left(3,1,\frac{2}{3}\right)$          & $\left(3,2,\frac{1}{6}\right)$      & $\left(3,2,\frac{7}{6}\right)$          & $\left(3,3,-\frac{1}{3}\right)$          \\
$d=6$  & $\circ $  & $\circ $  & $\circ $  &  &  \\ \cbottomrule
\end{tabular}
\caption{New scalars contributing to $d=6$ and $d=7$ $N_R$SMEFT operators at tree-level. Only fields marked with a circle contribute to $d=6$, the remaining ones appear in models for $d=7$ operators. Field names follow the conventions of \cite{deBlas:2017xtg}.}
\label{tab:newS}
\end{table}

\begin{table}[t]
\centering
\renewcommand*{\arraystretch}{1.1}
\begin{tabular}{@{\hspace{\tabcolsep}}lccccccc@{\hspace{\tabcolsep}}}
\ctoprule
\crowcolor
Name & $\mathcal{N}$ & $E$ & $\Delta_1$ & $\Delta_3$ & $\Sigma$ &
$\Sigma_1$ & \\ Irrep & $(1,1,0)$ & $(1,1,-1)$ &
$\left(1,2,-\frac{1}{2} \right)$ & $\left(1,2,-\frac{3}{2} \right)$ &
$(1,3,0)$ & $(1,3,-1)$ & \\ $d=6$ & $\circ$ & & $\circ$ & & $\circ$ &
$\circ$ & \\ \cbottomrule
\noalign{\vskip 1mm} 
\ctoprule
\crowcolor
Name  & $U$              & $D$        & $Q_1$                       & $Q_5$                       & $Q_7$         & $T_1$       &  $T_2$ \\
Irrep & $\left(3,1,\frac{2}{3}\right)$        & $\left(3,1,-\frac{1}{3}\right)$ & $\left(3,2,\frac{1}{6}\right)$ & $\left(3,2,-\frac{5}{6}\right)$& $\left(3,2,\frac{7}{6}\right)$        & $\left(3,3,-\frac{1}{3}\right)$      & $\left(3,3,\frac{2}{3}\right)$  \\
$d=6$   &  &    &                   &                          &  &   &   \\ \cbottomrule
\end{tabular}
\caption{New vector-like fermions contributing to $d=6$, $d=7$ $N_R$SMEFT at tree-level.}
\label{tab:newF}
\end{table}

\begin{table}[t]
\centering
\renewcommand*{\arraystretch}{1.1}
\begin{tabular}{@{\hspace{\tabcolsep}}lcccccc@{\hspace{\tabcolsep}}}
\ctoprule
\crowcolor
Name  & $\mathcal{B}$              & $\mathcal{B}_1$        & $\mathcal{W}$                       & $\mathcal{W}_1$                       &  $\mathcal{L}_1$         &       $\mathcal{L}_3$  \\
Irrep & $(1,1,0)$        & $(1,1,1)$ & $\left(1,3,0 \right)$ & $\left(1,3,1 \right)$ & $\left(1,2,\frac{1}{2}\right)$        &     $\left(1,2,-\frac{3}{2}\right)$       \\
$d=6$   & $\circ$    & $\circ$                  &                          &   & $\circ$   &  \\ \cbottomrule
\noalign{\vskip 1mm} 
\ctoprule
\crowcolor
Name  &  $\mathcal{U}_1$              & $\mathcal{U}_2$        & $\mathcal{Q}_1$                       & $\mathcal{Q}_5$                       & $\mathcal{X}$         &      \\
Irrep &  $\left(3,1,-\frac{1}{3}\right)$        & $\left(3,1,\frac{2}{3}\right)$ & $\left(3,2,\frac{1}{6}\right)$ & $\left(3,2,-\frac{5}{6}\right)$ & $\left(3,3,\frac{2}{3}\right)$        &       \\
$d=6$   & $\circ$  & $\circ$    & $\circ$                    &                          &  &  \\ \cbottomrule
\end{tabular}
\caption{New vector bosons contributing to $d=6$, $d=7$ $N_R$SMEFT at tree-level. 
Two fields, $\mathcal{L}_1$ and $\mathcal{U}_1$, are special cases, see text.}
\label{tab:newV}
\end{table}
In this section, we introduce the diagrammatic method used to
systematically decompose $N_R$SMEFT operators at tree-level. The same
method has been used for studying the Weinberg operator at different
loop orders in \cite{Bonnet:2012kz, AristizabalSierra:2014wal,
  Cepedello:2018rfh,Cepedello:2019zqf} and for $1$-loop openings of
SMEFT four-fermion operators in
\cite{Cepedello:2022pyx,Cepedello:2023yao}. We will therefore be brief
in this description. Subsequently, we apply the diagrammatic method to
$N_R$SMEFT operators which include at least one $N_R$ at both $d=6$
(section \ref{sect:d6}) and $d=7$ (section \ref{sect:d7}). Operators
at $d=6$ and $d=7$ that violate baryon number are discussed separately
in section \ref{sect:bnv}.

\subsection{Diagrammatic method basics\label{sect:diag}}

The procedure of \textit{opening up} or \textit{exploding} 
\footnote{This term was introduced in \cite{Gargalionis:2020xvt} 
and refers to the process of expanding an operator into a series of UV
renormalisable models generating the initial operator at tree-level.}
EFT operators can be summarized in three main steps. 
First, given an effective operator with $n$ light fields, one
constructs all possible \textit{topologies} that can generate the
operator at some loop order. A topology is made up of $n$ external
legs connected through internal lines using only renormalisable
interactions, i.e. $3$- and $4$-point interaction vertices. The number
of internal lines depends on the topology and on loop order. In our case,
we consider only tree-level openings and all internal lines are
identified with BSM \textit{heavy} fields.

In the second step, one
assigns each light field in the operator to an external leg of the
topology, and Lorentz invariance fixes the Lorentz nature of the
internal lines to either scalar ($S$), fermion ($F$) or vector
($V$). All possible permutations of the external light fields give
rise to different combinations for the internal fields. At this stage,
the output consists of a set of \textit{diagrams} where every line has
definite Lorentz nature.

Finally, by imposing gauge invariance, in our
case the SM symmetry $SU(3)_C \times SU(2)_L \times U(1)_Y$, in all
the interaction vertices, one determines the heavy fields' quantum
numbers, which are uniquely fixed for tree-level openings. The
diagrams are then promoted to \textit{model diagrams} and the lists of
particles that can be accommodated as internal lines constitute
our \textit{models}.

In summary, for each effective operator of dimension $d$, the output
of the diagrammatic process is a set of models consisting of heavy BSM
particles, each of which gives a tree-level opening of the
operator. We collect in tables~\ref{tab:newS}~-~\ref{tab:newV} all the
BSM fields that appear in the model diagrams found for
$N_R$SMEFT $d=6$ and $d=7$ operators, classified according to their
Lorentz nature. Fields contributing to $d=6$ operators are marked, the
remaining fields appear only in models for $d=7$ operators.

Some additional comments are in order. We have implemented the
described method in a \texttt{Mathematica} code. The input for this
code is the list of fields and the number of derivatives contained in
the considered operator. Neither the Lorentz structure nor the field
contractions have to be specified. As a consequence, in cases where
the basis of operators allow more than one operator with the same
field content (and number of derivatives) the method yields a list of
model diagrams, each of which will contribute to at least one operator
in this set, but does not provide the information, to which specific
operator (or operators) the model will be matched.

We choose to do so, because the model lists found in this way are
complete, once one scans over all possible operators at a given level
of $d$. After this has been done, many models are found more than once,
reflecting the fact that in most cases BSM extensions will generate
not only one specific operator, but contribute to several operators,
when the correct matching for the model is calculated.  Matching of
tree-level models (as we consider here) could of course easily be
calculated ``by hand'', but recently the code \texttt{Matchete}
\cite{Fuentes-Martin:2022jrf} has been published, with which 
the matching can be done also automatically.  

Our code does not fix the operator list to any particular on-shell
basis. In consequence, we do not consider diagrams with light
bridges. Once the operator list, to which any given model is matched,
is reduced to the on-shell basis, operators containing diagrams with
light bridges will be automatically included. It is important to note,
however, that diagrams with light bridges do not present new models,
i.e. it is guaranteed that with the diagrammatic method, as discussed
here, no models are lost.

For tables~\ref{tab:newS}~-~\ref{tab:newV}, we follow the naming conventions of \cite{deBlas:2017xtg}. Most fields in our list of possible BSM particles have already appeared in that reference in a different context. Note that (i) $\Xi_1$ in neutrino physics is usually denoted with the symbol $\Delta$ and (ii) we use $\mathcal{N}$ for the heavy fermion $F(1,1,0)$ to distinguish it from the light $N_R$.  However, there are two special cases, the vectors $\mathcal{L}_1$ and $\mathcal{U}_1$. The latter does not appear in pure SMEFT at $d=6$ \cite{deBlas:2017xtg}, but this field is listed in \cite{Bischer:2019ttk}. 

On the other hand, $\mathcal{L}_1$ is more subtle. One can choose to treat vector fields as either gauge vectors or not. For general vector fields, one can write down a general list of interactions obeying simply Lorentz and gauge symmetries of the model under consideration. However, for heavy gauge vectors a certain set of these interactions is not allowed, see the discussion in \cite{Arzt:1994gp}.
In pure SMEFT at $d=6$, $\mathcal{L}_1$ can contribute to the matching only if the interaction term $\mathcal{L}_{1,\mu}^{\dagger} D^{\mu}H$ is included \cite{deBlas:2017xtg}. Such a term is not allowed for a gauge vector \cite{Arzt:1994gp}. However, in $N_R$SMEFT a $\mathcal{L}_1$ will contribute to the matching, even if it is forced to be a gauge vector. We note that while we do not give the list of possible gauge groups for the vectors in table~\ref{tab:newV}, $\mathcal{L}_1$ appears, for example, in $331$ models \cite{Singer:1980sw,Pisano:1992bxx,Frampton:1992wt,Fonseca:2016xsy}.

Finally, all heavy fermions appearing in the models are considered to be vector-like under the SM gauge group, even though just one of the chiralities might be needed to open up an operator. This assumption is motivated by experimental data. The observation of a SM-like Higgs 
boson at ATLAS \cite{ATLAS:2012yve} and CMS \cite{CMS:2012qbp} rules out the existence of a fourth chiral generation (in the minimal SM).

\subsection{Dictionary for $d=6$\label{sect:d6}}
%%%%%%%%%%%%%%%%%%%%%%%%%%
%  Table d=6                       
%%%%%%%%%%%%%%%%%%%%%%%%%%
\begin{table}[t]
\renewcommand*{\arraystretch}{1.5}
\centering
 \begin{adjustbox}{max width=\textwidth}
\begin{tabular}{|cc|cc|cc|}
\hline
\crowcolor
\multicolumn{2}{|c|}{$\psi^2 H^3$ (+h.c.)}                                                                                    & \multicolumn{2}{c|}{$(\overline{R}R)(\overline{R}R)$}                                                      & \multicolumn{2}{|c|}{$(\overline{L}L)(\overline{R}R)$}                                                  \\ \hline
\multicolumn{1}{|c|}{$\mathcal{O}_{LNH^3}$}  & $(\overline{L}N_R)\tilde{H}(H^\dagger H)$                                      & \multicolumn{1}{c|}{$\mathcal{O}_{NN}$}   & $(\overline{N_R}\gamma^\mu N_R)(\overline{N_R}\gamma_\mu N_R)$ & \multicolumn{1}{|c|}{$\mathcal{O}_{LN}$}   & $(\overline{L}\gamma^\mu L)(\overline{N_R}\gamma_\mu N_R)$ \\ \cline{1-2} \\[-7.8mm]
\multicolumn{2}{|c|}{\cellcolor[gray]{0.9} $\psi^2 H^2 D$ (+h.c.)}                                                                                  & \multicolumn{1}{c|}{$\mathcal{O}_{eN}$}   & $(\overline{e}_R\gamma^\mu e_R)(\overline{N_R}\gamma_\mu N_R)$ & \multicolumn{1}{c|}{$\mathcal{O}_{QN}$}   & $(\overline{Q}\gamma^\mu Q)(\overline{N_R}\gamma_\mu N_R)$ \\ \cline{1-2} \cline{5-6} \\[-7.7mm]
\multicolumn{1}{|c|}{$\mathcal{O}_{NH^2D}$}  & $(\overline{N_R}\gamma^\mu N_R)(H^\dagger i \overset{\longleftrightarrow}{D_\mu} H)$ & \multicolumn{1}{c|}{$\mathcal{O}_{uN}$}   & $(\overline{u}_R\gamma^\mu u_R)(\overline{N_R}\gamma_\mu N_R)$ & \multicolumn{2}{|c|}{\cellcolor[gray]{0.9} $(\overline{L}R)(\overline{L}R)$ (+h.c.)}                                          \\ \cline{5-6} 
\multicolumn{1}{|c|}{$\mathcal{O}_{NeH^2D}$} & $(\overline{N_R}\gamma^\mu e_R)(\tilde{H}^\dagger i D_\mu H)$                  & \multicolumn{1}{c|}{$\mathcal{O}_{dN}$}   & $(\overline{d}_R\gamma^\mu d_R)(\overline{N_R}\gamma_\mu N_R)$ & \multicolumn{1}{|c|}{$\mathcal{O}_{LNLe}$} & $(\overline{L}N_R)\epsilon(\overline{L}e_R)$               \\ \cline{1-2} \\[-7.8mm]
\multicolumn{2}{|c|}{\cellcolor[gray]{0.9} $(\overline{L}R)(\overline{R}L)$ (+h.c.)}                                                                & \multicolumn{1}{c|}{$\mathcal{O}_{duNe}$} & $(\overline{d}_R\gamma^\mu u_R)(\overline{N_R}\gamma_\mu e_R)$ & \multicolumn{1}{c|}{$\mathcal{O}_{LNQd}$} & $(\overline{L}N_R)\epsilon(\overline{Q}d_R)$               \\ \cline{1-4}
\multicolumn{1}{|c|}{$\mathcal{O}_{QuNL}$}   & $(\overline{Q}u_R)(\overline{N_R} L)$                                          & \multicolumn{1}{c|}{$\mathcal{O}_{NNNN}$} & $(\overline{N^c_R} N_R)(\overline{N^c_R} N_R)$                 & \multicolumn{1}{c|}{$\mathcal{O}_{LdQN}$} & $(\overline{L}d_R)\epsilon(\overline{Q}N_R)$               \\ \hline
\end{tabular}
\end{adjustbox}
\caption{$N_R$SMEFT operators at $d=6$ which can be generated 
at tree-level. Baryon number violating operators are discussed 
separately in section \ref{sect:bnv}.}
\label{tab:Opd6}
\end{table}
%%%%%%%%%%%%%%%%%%%%%%%%%%%%%%%%%%
%%%%%%%%%%%%%%%%
% Operator Class vs. Diagrams
%%%%%%%%%%%%%
\begin{figure}[t]
\centering
\renewcommand*{\arraystretch}{1.5}
\begin{adjustbox}{max width=\textwidth}
\begin{tabular}{@{\hspace{\tabcolsep}}rl@{\hspace{\tabcolsep}}}
$\psi^2 H^3$ \;:
& \parbox[c]{3.3cm}  {\includegraphics[width=0.9\linewidth]{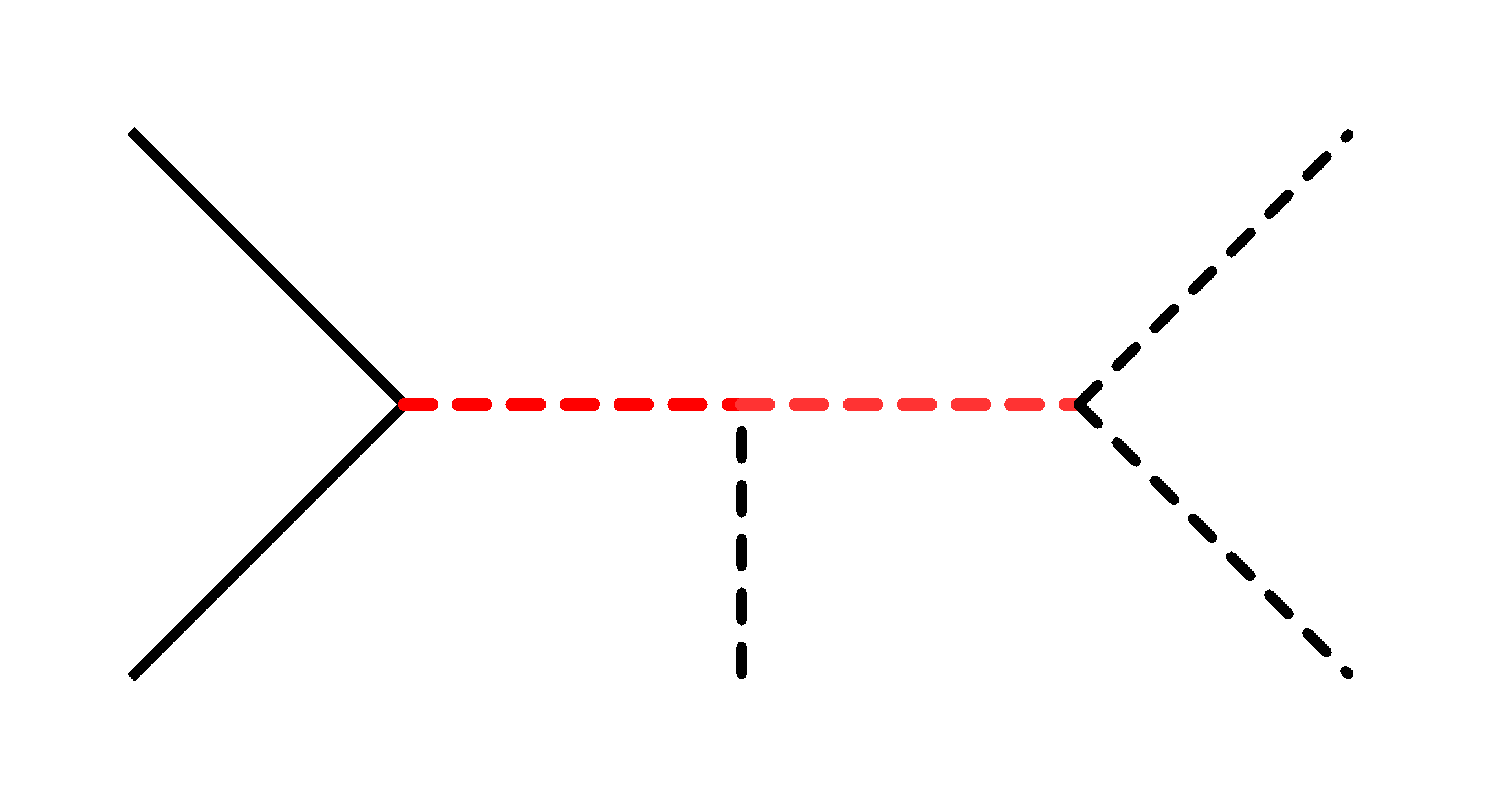}} \parbox[c]{3.3cm}  {\includegraphics[width=0.9\linewidth]{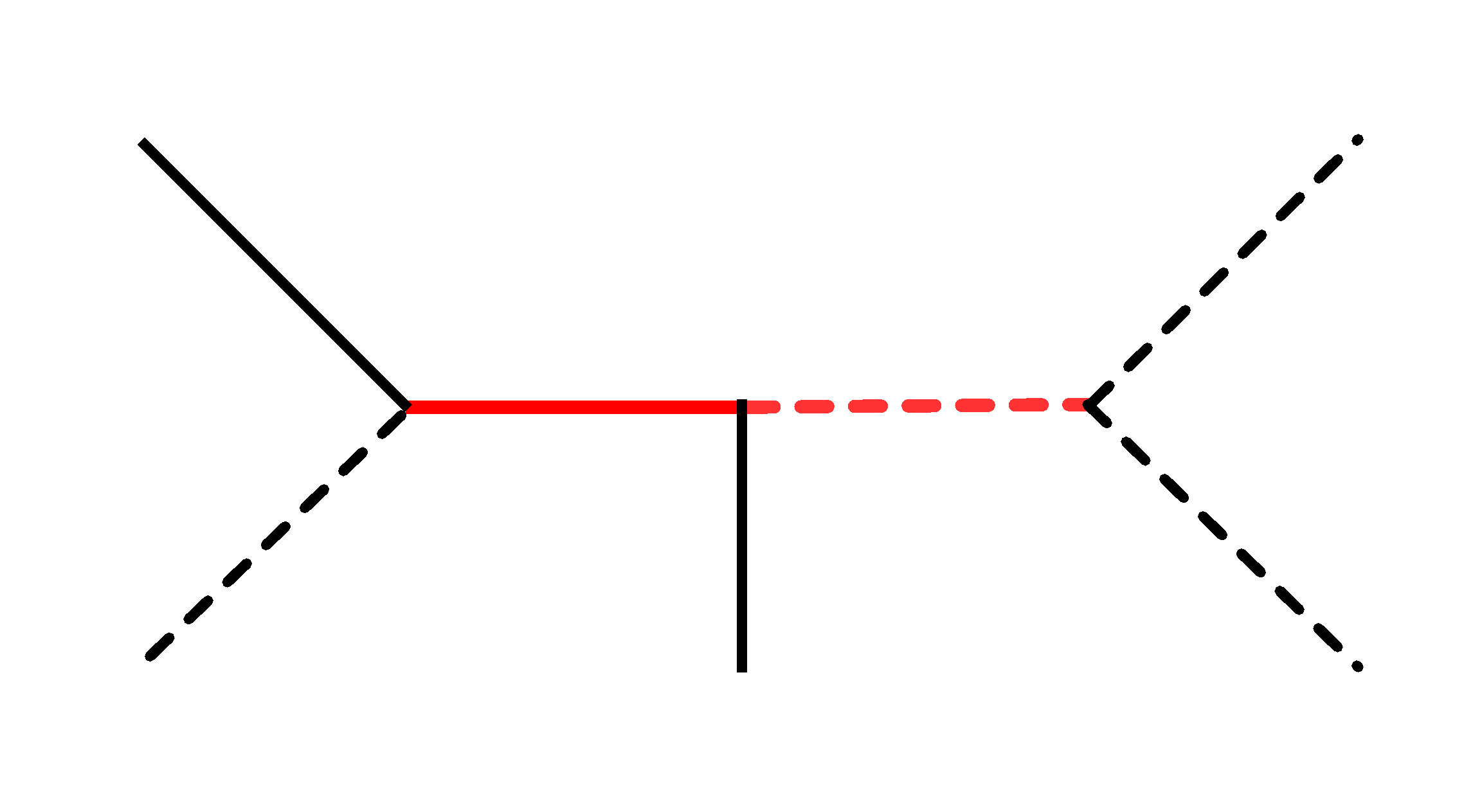}} \parbox[c]{3.3cm}  {\includegraphics[width=0.9\linewidth]{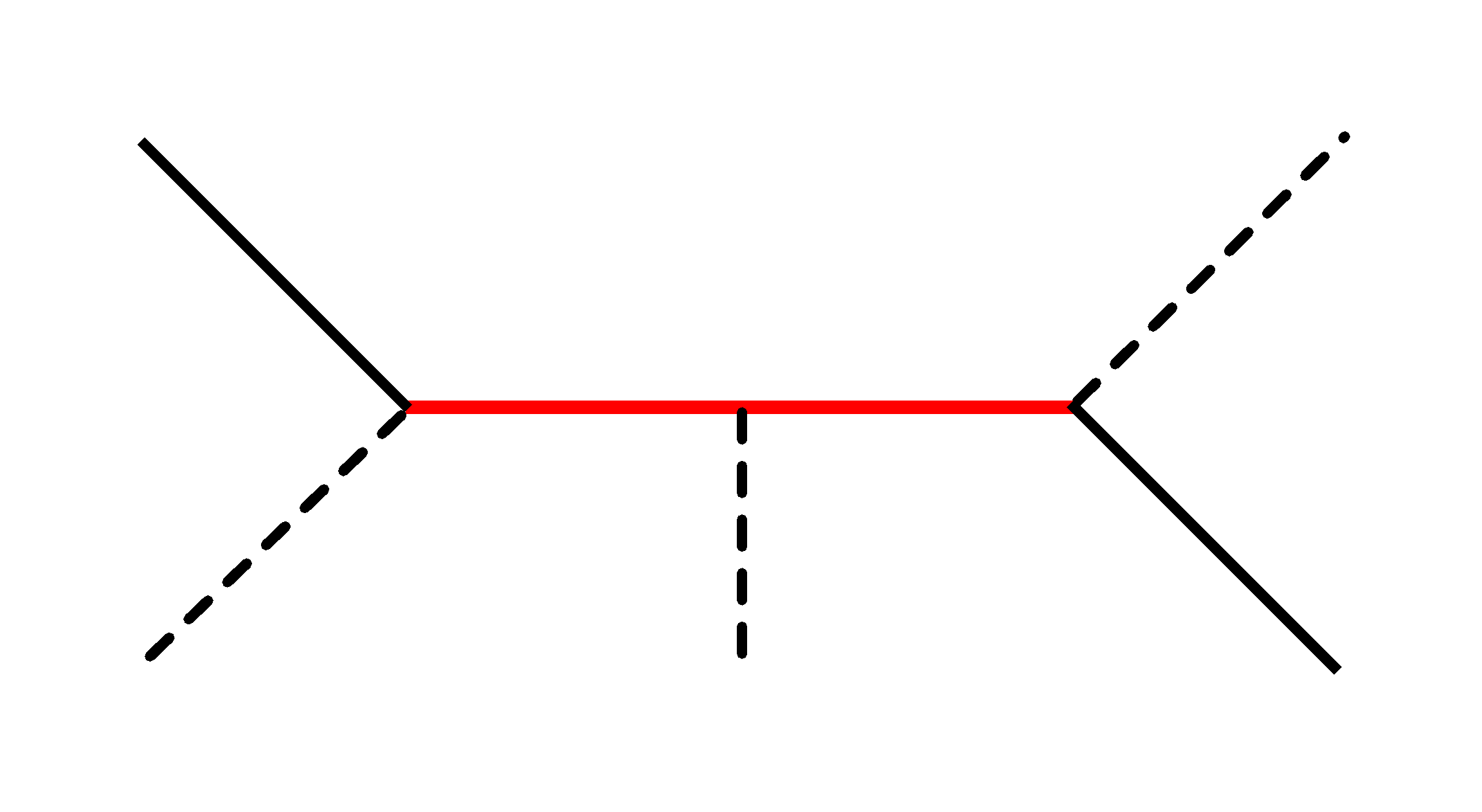}} \parbox[c]{3.3cm}  {\includegraphics[width=0.9\linewidth]{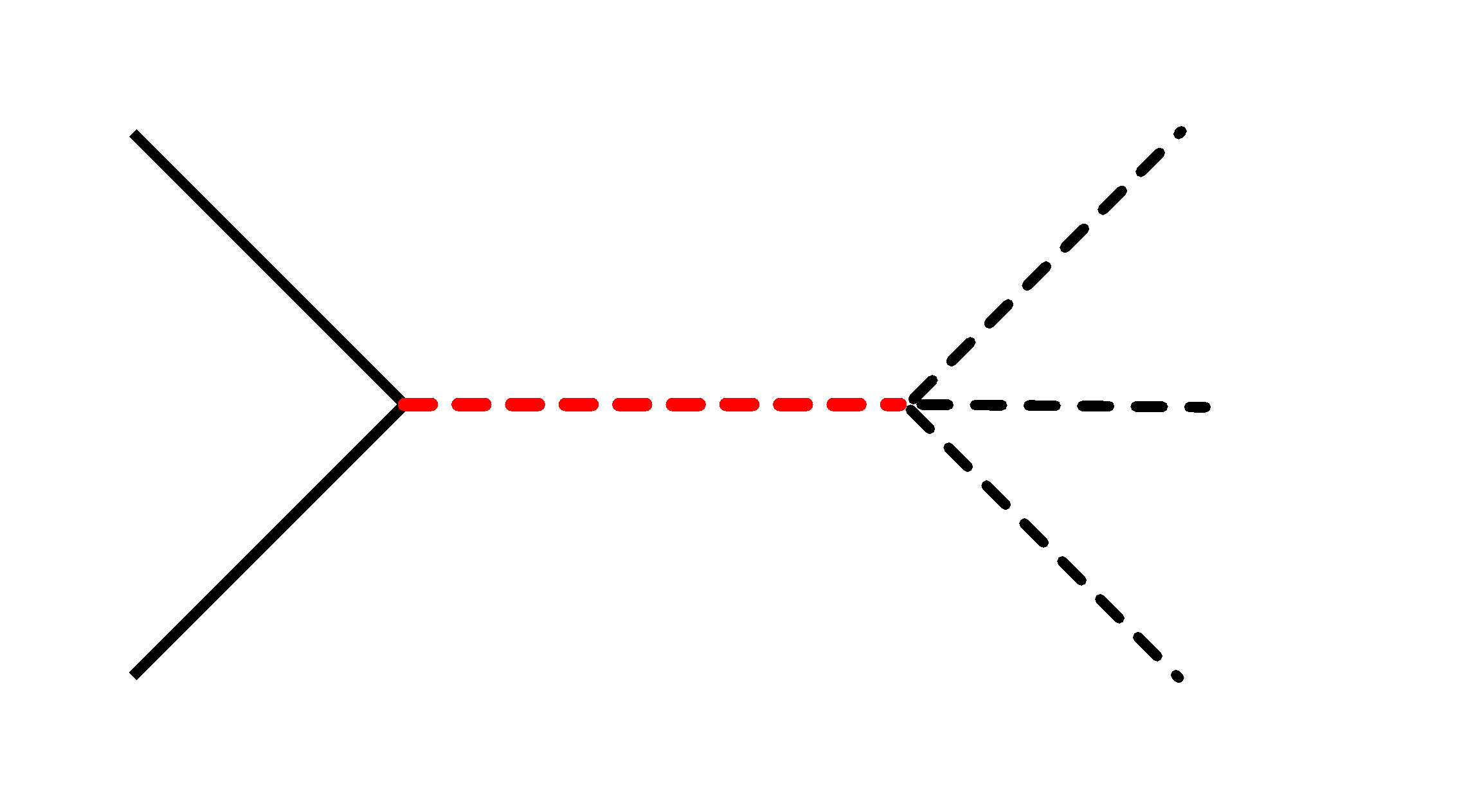}} \\ %\midrule
$\psi^4$ \;:
& \parbox[c]{3.4cm} {\includegraphics[width=0.9\linewidth]{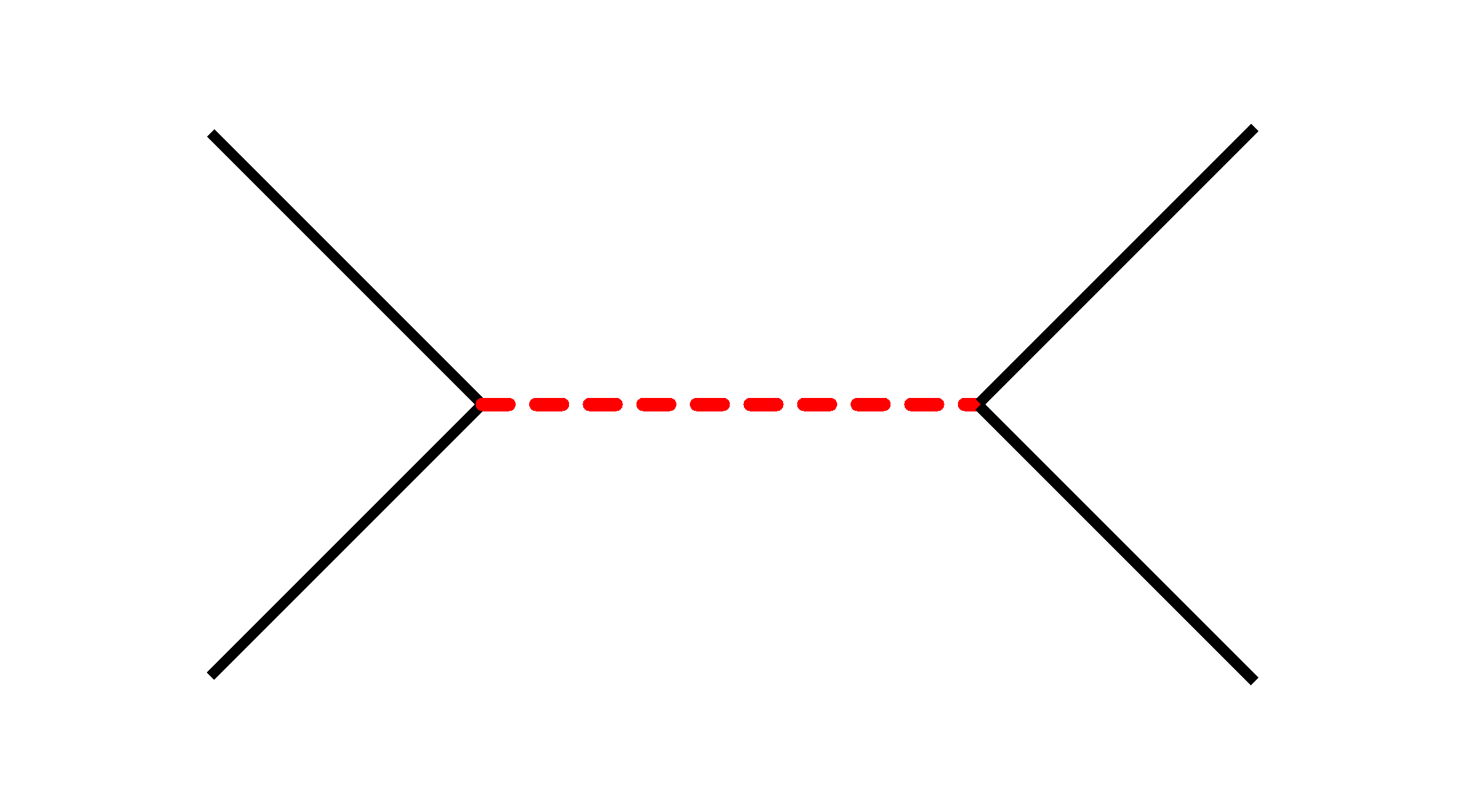}} \parbox[c]{3.4cm} {\includegraphics[width=0.9\linewidth]{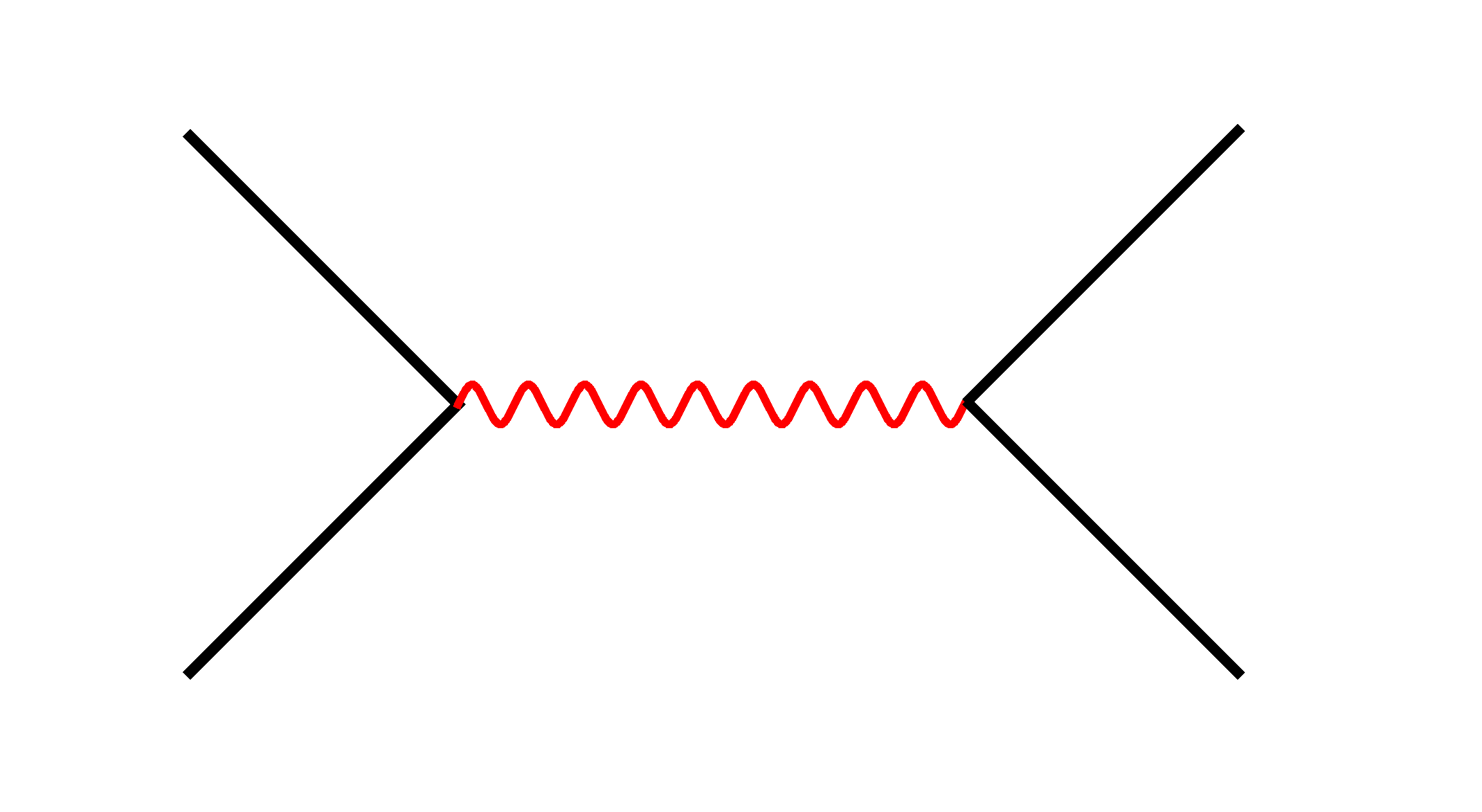}} \\ % \midrule
$\psi^2 H^2 D$ \;:
& \parbox[c]{3.4cm}  {\includegraphics[width=0.9\linewidth]{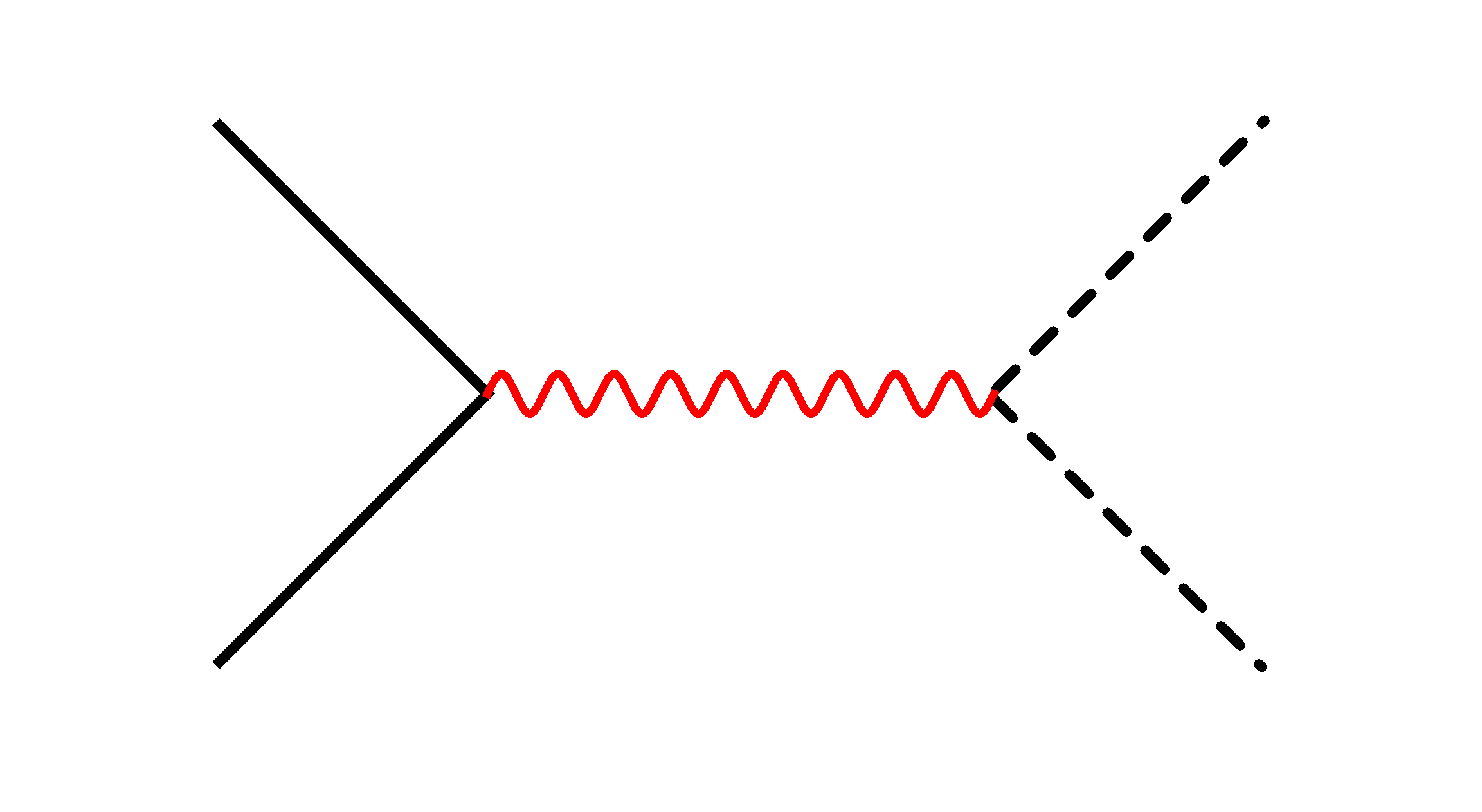}} \parbox[c]{3.4cm} {\includegraphics[width=0.9\linewidth]{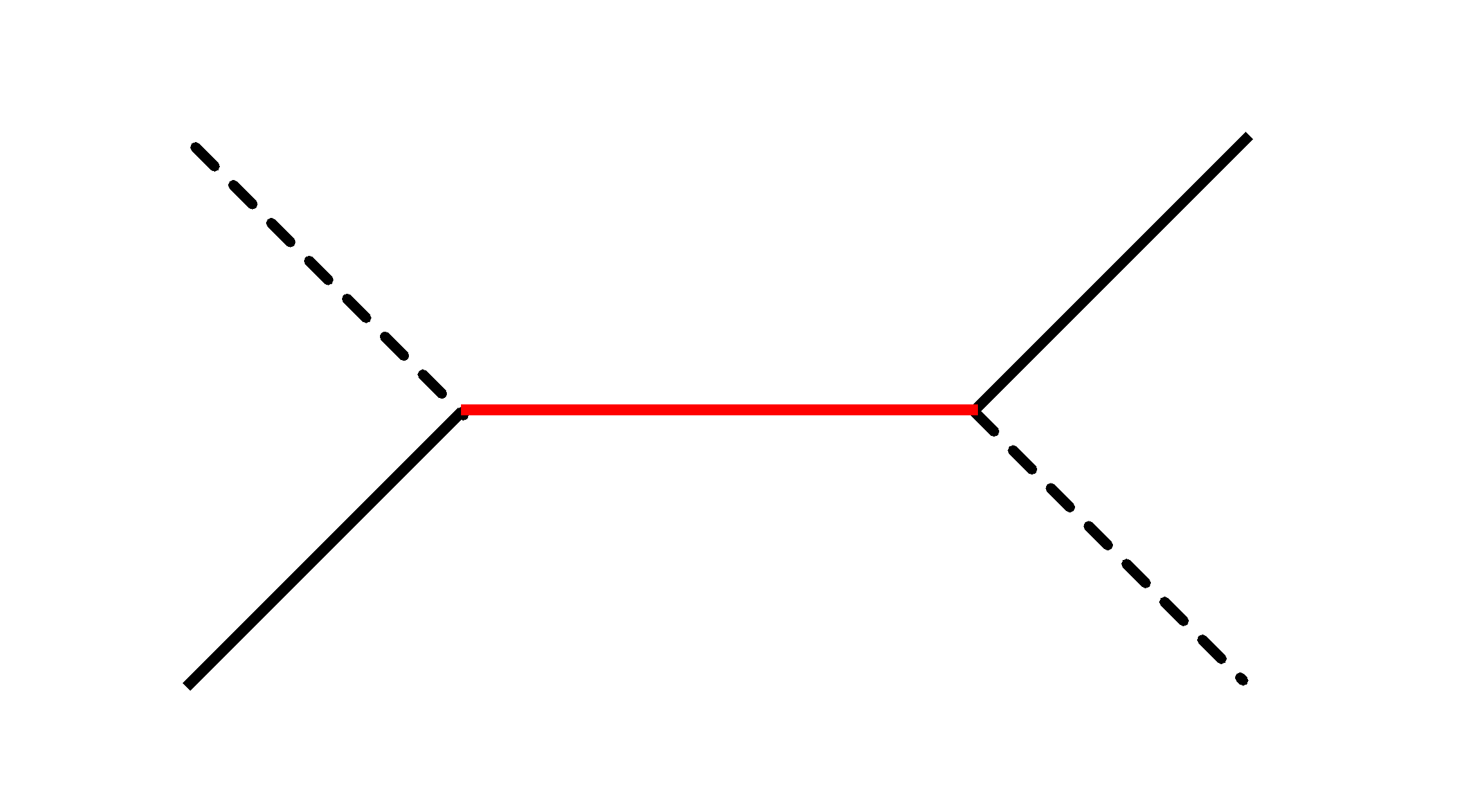}}  \\ %\cbottomrule
\end{tabular}%
\end{adjustbox}
\caption{Operator classes at $d=6$ and their respective tree openings at 
the diagram level. Solid lines correspond to fermions, dashed lines to
scalars and wavy lines to vectors. Red lines are for heavy fields.}
\label{fig:diagd6}
\end{figure}
%%%%%%%%%%%%%%%%%%%%%%%%%%%%%%%%%%%%%%
\begin{table}[t]
\centering
\renewcommand*{\arraystretch}{1.2}
\begin{tabular}{|@{\hspace{\tabcolsep}}cl@{\hspace{\tabcolsep}}|}
\hline
\crowcolor
Models         & Operators \\ \hline
$\mathcal{S}$ & $\mathcal{O}_{NN}$, $\mathcal{O}_{NNNN}$ \\
$\mathcal{S}_1$ & $\mathcal{O}_{LNLe}$, $\mathcal{O}_{eN}$ \\
$\varphi$ & $\mathcal{O}_{QuNL}$, $\mathcal{O}_{LNLe}$,
$\mathcal{O}_{LNQd}$, $\mathcal{O}_{LN}$, $\mathcal{O}_{LNH^3}$ \\
$\omega_1$ & $\mathcal{O}_{LNQd}$, $\mathcal{O}_{dN}$,
$\mathcal{O}_{duNe}$ \\ $\omega_2$ & $\mathcal{O}_{uN}$ \\ $\Pi_1$ &
$\mathcal{O}_{LNQd}$, $\mathcal{O}_{QN}$ \\ \hline $\Delta_1$ &
$\mathcal{O}_{NH^2D}$, $\mathcal{O}_{NeH^2 D}$ \\ \hline
$\mathcal{B}$ & $\mathcal{O}_{NH^2 D}$, $\mathcal{O}_{NN}$,
$\mathcal{O}_{eN}$, $\mathcal{O}_{uN}$, $\mathcal{O}_{dN}$,
$\mathcal{O}_{LN}$, $\mathcal{O}_{QN}$ \\ $\mathcal{B}_1$ &
$\mathcal{O}_{NeH^2 D}$, $\mathcal{O}_{eN}$, $\mathcal{O}_{duNe}$ \\
$\mathcal{L}_1$ & $\mathcal{O}_{LN}$ \\ $\mathcal{U}_1$ &
$\mathcal{O}_{dN}$ \\ $\mathcal{U}_2$ & $\mathcal{O}_{QuNL}$,
$\mathcal{O}_{uN}$, $\mathcal{O}_{duNe}$ \\ $\mathcal{Q}_1$ &
$\mathcal{O}_{QuNL}$, $\mathcal{O}_{QN}$ \\ \hline
\end{tabular}
\caption{One-particle decompositions for $d=6$ $N_R$SMEFT operators. We differentiate between scalar, fermion and vector models. The first column gives the particle, the second the operators 
that are generated at tree-level.}
\label{tab:modelsd6}
\vspace*{0.9em}
\begin{tabular}{|c|rl|}
\hline 
\crowcolor
$\psi^2 H^3$ & \multicolumn{2}{|l|}{Two-particle models} \\
\hline
\multicolumn{1}{|c|}{\multirow{3}{*}{$\mathcal{O}_{LNH^3}$}}  & $SS$ :       & $(\mathcal{S}, \varphi)$, $(\Xi_1, \varphi)$, $(\Xi, \varphi)$                                                            \\
& $FF$ : & $(\Delta_1, \mathcal{N})$, $(\Delta_1, \Sigma_1)$,
$(\Delta_1, \Sigma)$ \\ & $FS$ : & $(\mathcal{N},\mathcal{S})$,
$( \Delta_1 , \mathcal{S})$, $(\Delta_1 , \Xi_1)$, $(\Sigma_1
, \Xi_1)$, $(\Delta_1 , \Xi )$, $(\Sigma, \Xi)$ \\
\hline%\cbottomrule
\end{tabular}
\caption{Two-particle decompositions for the $d=6$ operator $\mathcal{O}_{LNH^3}$. 
There are three types of models according to the Lorentz nature of the heavy fields.}
\label{tab:modelsOLNH}
\end{table}

In table~\ref{tab:Opd6} we show the list of operators at $d=6$ that can be generated at tree-level. The operators are classified into three classes: $\psi^2 H^3$, $\psi^2 H^2 D$, and $\psi^4$.\footnote{The full list of four-fermion operators includes two additional operators that violate baryon number ($B$), these are discussed separately in section \ref{sect:bnv}.}  At $d=6$ there is a fourth operator class, $\psi^2 H X$, whose operators can only be
realised at loop level and hence are not listed here. For a complete basis of on-shell operators at $d=6$, see \cite{Liao:2016qyd}.

Figure~\ref{fig:diagd6} shows all possible diagrams that generate the three operator classes mentioned above. The first class, $\psi^2 H^3$, can be obtained through two different topologies, one of which contains two heavy fields (depending on their Lorentz nature we find
three diagrams corresponding to the possibilities: $SS$, $FF$, $SF$), leading in most cases to two-particle extensions of the SM. The other topology includes a 4-point interaction vertex and it contains just one heavy scalar field. The four-fermion class, $\psi^4$, can only be generated at tree-level with a scalar or a vector heavy propagator, depending on the chirality of the external fields. Finally, there are two openings for the operator class $\psi^2 H^2 D$ with just one internal field, one of which contains the derivative in the $VSS$ interaction vertex, and in the other, the derivative comes from the fermion propagator. Thus, all operators in table~\ref{tab:Opd6} can be generated at tree-level with one-particle SM extensions, except for the operator $\mathcal{O}_{LNH^3}$ belonging to $\psi^2 H^3$, for which some openings require two-particle models.

Here, our approach differs from the philosophy followed in \cite{deBlas:2017xtg}, since we insist on using only renormalisable vertices. The authors of \cite{deBlas:2017xtg}, on the other hand, consider only one particle extensions of the SM. However, already at $d=6$ in pure SMEFT there are some operators which require two BSM fields. The list of decompositions given in \cite{deBlas:2017xtg} for pure SMEFT is nevertheless complete, since \cite{deBlas:2017xtg} add also non-renormalisable operator (NRO) interactions at $d=5$ to their Lagrangian.\footnote{Our \texttt{Mathematica} code can handle also non-renormalisable interactions. We have checked for some concrete cases that we can reproduce the one-heavy-particle at a time results of \cite{deBlas:2017xtg}, once we include $d=5$ terms.}

We present in table~\ref{tab:modelsd6} the list of one-particle models found at $d=6$ and the corresponding operators they open at tree-level, while the two-particle models for $\mathcal{O}_{LNH^3}$ are shown separately in table~\ref{tab:modelsOLNH}, where we categorize the models based on the Lorentz nature of the involved fields. We note that the new field $\mathcal{U}_1$ contributes to only one  $d=6$ operator in $N_R$SMEFT, $\mathcal{O}_{dN}$. This is due to the existence of a single interaction vertex involving light fields and the heavy vector, given by $\mathcal{L} \propto \left( \overline{N_R} \gamma_\mu d_R \right) \mathcal{U}_1^{\mu \dagger}$.  

The Lagrangian terms of heavy fields including a $N_R$ are given in appendix~\ref{sec:app}.  Furthermore, we also collect there all gauge-invariant renormalisable terms that can be written down for $\mathcal{U}_1$ and the BSM fields appearing in tables~\ref{tab:newS}~-~\ref{tab:newV}, as well as for the vector $\mathcal{L}_1$. Recall, that the field $\mathcal{L}_1$ was also included in \cite{deBlas:2017xtg}, but as a non-gauge vector.  For $N_R$SMEFT, we find additional renormalisable interactions of this vector. We also present them in appendix \ref{sec:app}. The term containing solely light fields and $\mathcal{L}_1$, which contributes to the matching of the operator $\mathcal{O}_{LN}$, is given by $\mathcal{L} \propto \left( \overline{N_R^c} \gamma_\mu L \right) \mathcal{L}_1^{\mu}$.

Given the complete UV Lagrangian, one could perform the tree-level
matching onto the set of $N_R$SMEFT operators. While we do not perform
explicitly this matching, the process can be automated with computer
tools, such as \texttt{Matchete}~\cite{Fuentes-Martin:2022jrf}. We
added a notebook with some example models as an auxiliary file to
this paper.

\subsection{Dictionary for $d=7$\label{sect:d7}}
%%%%%%%%%%%%%%%%%%%%%%%%
%  Table d=7                      
%%%%%%%%%%%%%%%%%%%%%%%%%
\begin{table}[t]
\renewcommand*{\arraystretch}{1.8}
\centering
 \begin{adjustbox}{max width=\textwidth}
\begin{tabular}{|p{0.1\textwidth}>{\centering\arraybackslash}p{0.32\textwidth}|p{0.1\textwidth}>{\centering\arraybackslash}p{0.28\textwidth}|p{0.1\textwidth}>{\centering\arraybackslash}p{0.22\textwidth}|}
%|cc|cc|cc|
\hline
\crowcolor
\multicolumn{2}{|c|}{$\psi^2 H^3 D$}                                                                                                                                                      & \multicolumn{2}{|c|}{$\psi^4 H$}                                                                                                                           & \multicolumn{2}{|c|}{$\psi^4 H$}                                                                             \\ \hline
\multicolumn{1}{|c|}{\multirow{2}{*}{$\mathcal{O}_{NLH^3D}$}}   & $\epsilon_{ij}(\overline{N^c_R}\gamma_\mu L^i)(i D^\mu H^j)(H^\dagger H)$                                               & \multicolumn{1}{c|}{$\mathcal{O}_{LNLH}$}                   & $\epsilon_{ij} (\overline{L}\gamma_\mu L)(\overline{N^c_R} \gamma^\mu L^i) H^j$             & \multicolumn{1}{c|}{$\mathcal{O}_{LNeH}$} & $(\overline{L} N_R) (\overline{N^c_R} e_R) H$                   \\ \cline{3-4}
\multicolumn{1}{|c|}{}                                          & $\epsilon_{ij}(\overline{N^c_R}\gamma_\mu L^i) H^j (H^\dagger i \overset{\longleftrightarrow}{D^\mu} H)$                & \multicolumn{1}{c|}{\multirow{2}{*}{$\mathcal{O}_{QNLH}$}}  & $\epsilon_{ij} (\overline{Q}\gamma_\mu Q)(\overline{N^c_R} \gamma^\mu L^i) H^j$             & \multicolumn{1}{c|}{$\mathcal{O}_{eLNH}$} & $ H^\dagger (\overline{e_R} L) (\overline{N^c_R} N_R) $         \\ \cline{1-2} \\[-9.4mm]
\multicolumn{2}{|c|}{\cellcolor[gray]{0.9}$\psi^2 H^2 D^2$}                                                                                                                                                    & \multicolumn{1}{c|}{}                                       & $\epsilon_{ij} (\overline{Q}\gamma_\mu Q^i)(\overline{N^c_R} \gamma^\mu L^j) H$             & \multicolumn{1}{c|}{$\mathcal{O}_{QNdH}$} & $(\overline{Q} N_R) (\overline{N^c_R} d_R) H$                   \\ \cline{1-4}
\multicolumn{1}{|c|}{$\mathcal{O}_{NeH^2 D^2}$}                 & $\epsilon_{ij}(\overline{N^c_R} \overset{\longleftrightarrow}{D_\mu}  e_R)(H^i D^\mu H^j)$                              & \multicolumn{1}{c|}{$\mathcal{O}_{eNLH}$}                   & $\epsilon_{ij} (\overline{e_R}\gamma_\mu e_R)(\overline{N^c_R} \gamma^\mu L^i) H^j$         & \multicolumn{1}{c|}{$\mathcal{O}_{dQNH}$} & $ H^\dagger (\overline{d_R} Q) (\overline{N^c_R} N_R) $         \\ \cline{1-2}
\multicolumn{1}{|c|}{\multirow{2}{*}{$\mathcal{O}_{NH^2 D^2}$}} & $(\overline{N^c_R} \overset{\longleftrightarrow}{\partial_\mu}  N_R)(H^\dagger \overset{\longleftrightarrow}{D^\mu} H)$ & \multicolumn{1}{c|}{$\mathcal{O}_{dNLH}$}                   & $\epsilon_{ij} (\overline{d_R}\gamma_\mu d_R)(\overline{N^c_R} \gamma^\mu L^i) H^j$         & \multicolumn{1}{c|}{$\mathcal{O}_{QNuH}$} & $(\overline{Q} N_R) (\overline{N^c_R} u_R) \tilde{H}$           \\
\multicolumn{1}{|c|}{}                                          & $(\overline{N^c_R}  N_R)(D_\mu H)^\dagger D^\mu H$                                                                      & \multicolumn{1}{c|}{$\mathcal{O}_{uNLH}$}                   & $\epsilon_{ij} (\overline{u_R}\gamma_\mu u_R)(\overline{N^c_R} \gamma^\mu L^i) H^j$         & \multicolumn{1}{c|}{$\mathcal{O}_{uQNH}$} & $ \tilde{H}^\dagger (\overline{u_R} Q) (\overline{N^c_R} N_R) $ \\ \cline{1-2} \\[-9.4mm]
\multicolumn{2}{|c|}{ \cellcolor[gray]{0.9}$\psi^2 H^2 X$}                                                                                                                                                      & \multicolumn{1}{c|}{$\mathcal{O}_{duNLH}$}                  & $\epsilon_{ij} (\overline{d_R}\gamma_\mu u_R)(\overline{N^c_R} \gamma^\mu L^i) \tilde{H}^j$ & \multicolumn{1}{c|}{$\mathcal{O}_{LNNH}$} & $(\overline{L} N_R) (\overline{N^c_R} N_R) \tilde{H}$           \\ \cline{1-2}
\multicolumn{1}{|c|}{$\mathcal{O}_{NeH^2 W}$}                   & $(\epsilon \tau^I)_{ij} (\overline{N^c_R} \sigma^{\mu\nu} e_R)(H^i H^j) W_{\mu\nu}^I$                                   & \multicolumn{1}{c|}{$\mathcal{O}_{dQNeH}$}                  & $\epsilon_{ij} (\overline{d_R} Q^i)(\overline{N^c_R} e_R) H^j$                              & \multicolumn{1}{c|}{$\mathcal{O}_{NLNH}$} & $ \tilde{H}^\dagger (\overline{N_R} L) (\overline{N^c_R} N_R) $ \\ \cline{3-6}  \\[-9.3mm]
\multicolumn{1}{|c|}{$\mathcal{O}_{NH^2 B }$}                   & $(\overline{N^c_R} \sigma^{\mu\nu} N_R)(H^\dagger H) B_{\mu\nu}$                                                        & \multicolumn{1}{c|}{\multirow{2}{*}{$\mathcal{O}_{QuNeH}$}} & $(\overline{Q} u_R)(\overline{N^c_R} e_R) H$                                                & \multicolumn{2}{|c|}{\cellcolor[gray]{0.9} $\psi^2 H^4 $}                                                                          \\ \cline{5-6}
\multicolumn{1}{|c|}{$\mathcal{O}_{NH^2 W}$}                    & $ (\overline{N^c_R} \sigma^{\mu\nu} N_R)(H^\dagger \tau^I H) W_{\mu\nu}^I$                                              & \multicolumn{1}{c|}{}                                       & $(\overline{Q} \sigma_{\mu\nu} u_R)(\overline{N^c_R} \sigma^{\mu\nu} e_R) H$                & \multicolumn{1}{c|}{$\mathcal{O}_{NH^4}$} & $(\overline{N^c_R} N_R) (H^\dagger H )^2$                       \\ \hline
\end{tabular}
\end{adjustbox}
\caption{$N_R$SMEFT operators at $d=7$ that admit tree-level decompositions. All operators are non-hermitian, $(+\text{h.c.})$ is implicitly assumed.}
\label{tab:Opd7}
\end{table}
%%%%%%%%%%%%%%%%%%%%%%%%%%%%%%%%%%%

%%%%%%%%%%%%%%%%%%%%%%%%
% Op. Classes and DIAGRAMS
%%%%%%%%%%%%%%%%%%%%%%%%
\begin{figure}[t]
\centering
\renewcommand*{\arraystretch}{1.4}
 \begin{adjustbox}{max width=\textwidth}
\begin{tabular}{@{\hspace{\tabcolsep}}rl@{\hspace{\tabcolsep}}}
%\ctoprule
%\textbf{Operator Class}   & \multicolumn{1}{c}{\textbf{Diagrams}}                \\ \midrule
$\psi^2 H^3 D$  \;:  & \parbox[c]{3cm} {\includegraphics[width=0.9\linewidth]{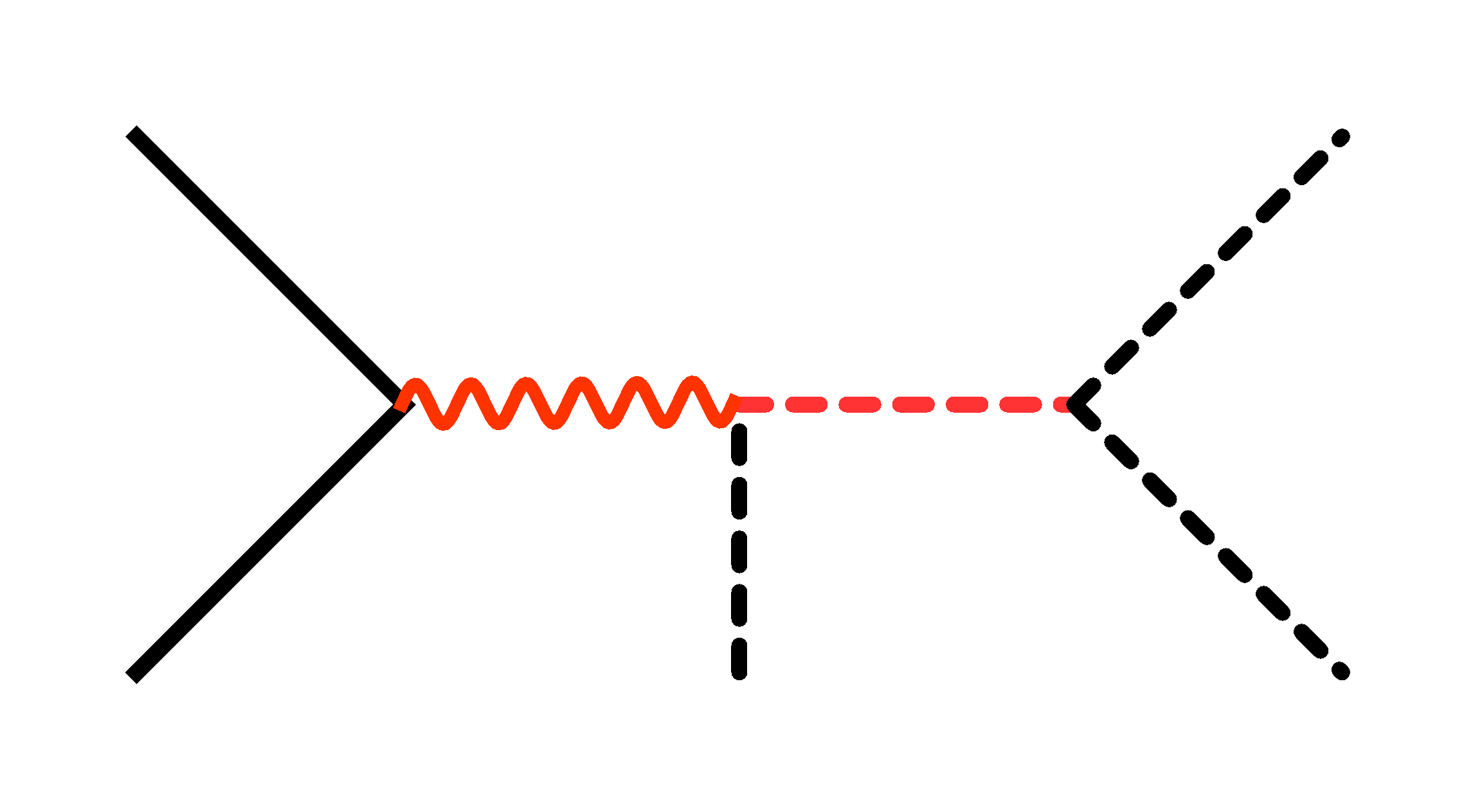}}  \parbox[c]{3cm} {\includegraphics[width=0.9\linewidth]{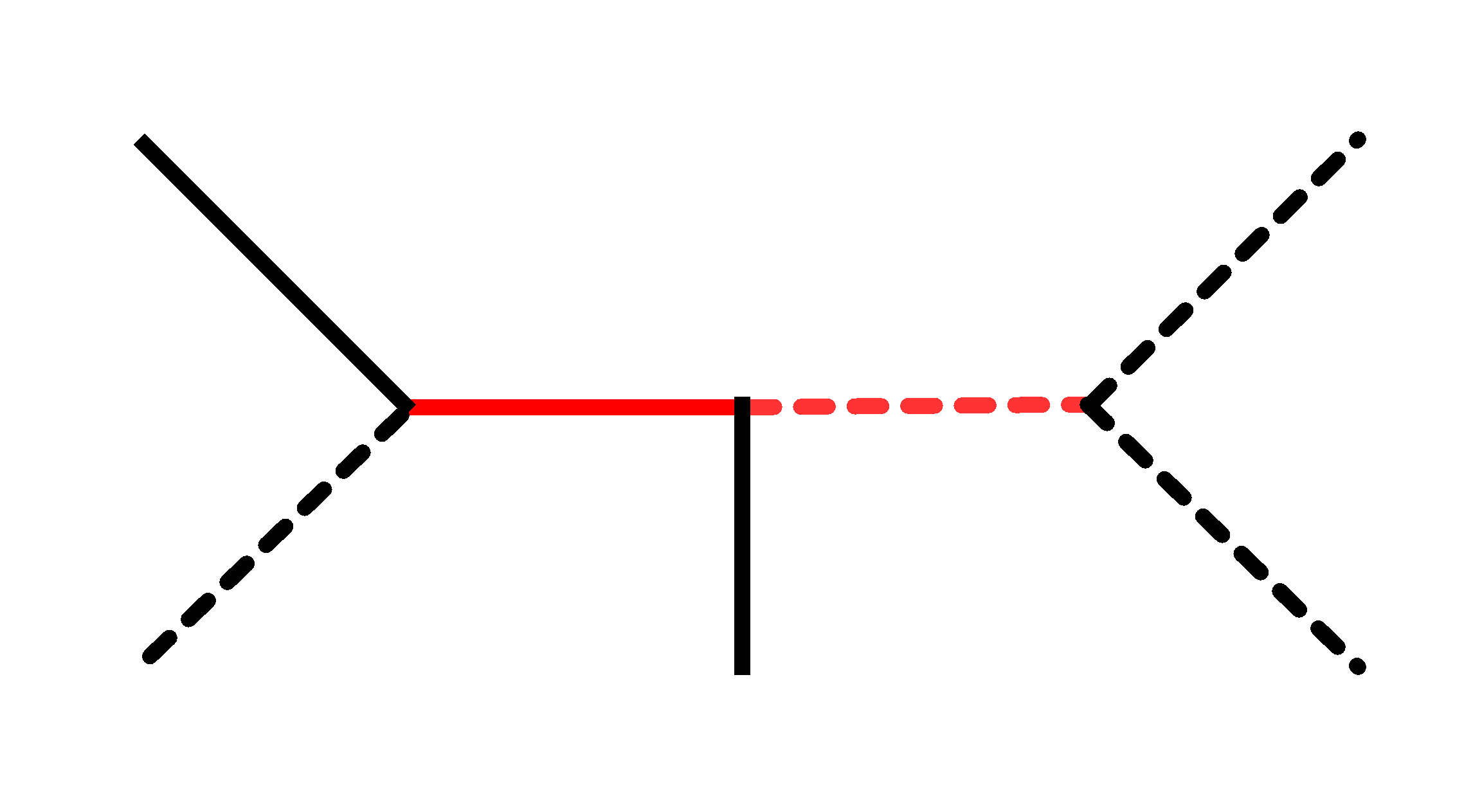}}  \parbox[c]{3cm}  {\includegraphics[width=0.9\linewidth]{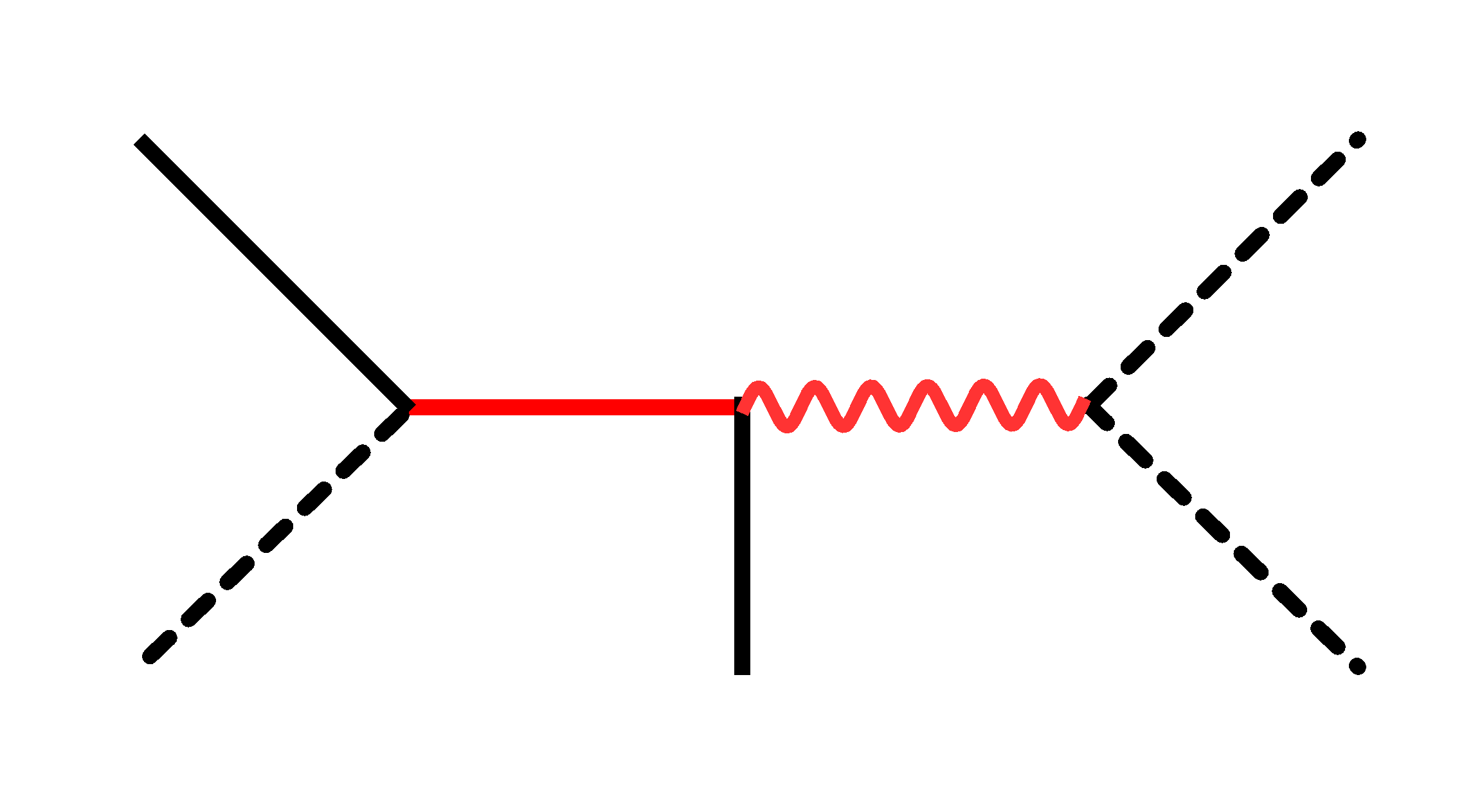}} \parbox[c]{3cm}  {\includegraphics[width=0.9\linewidth]{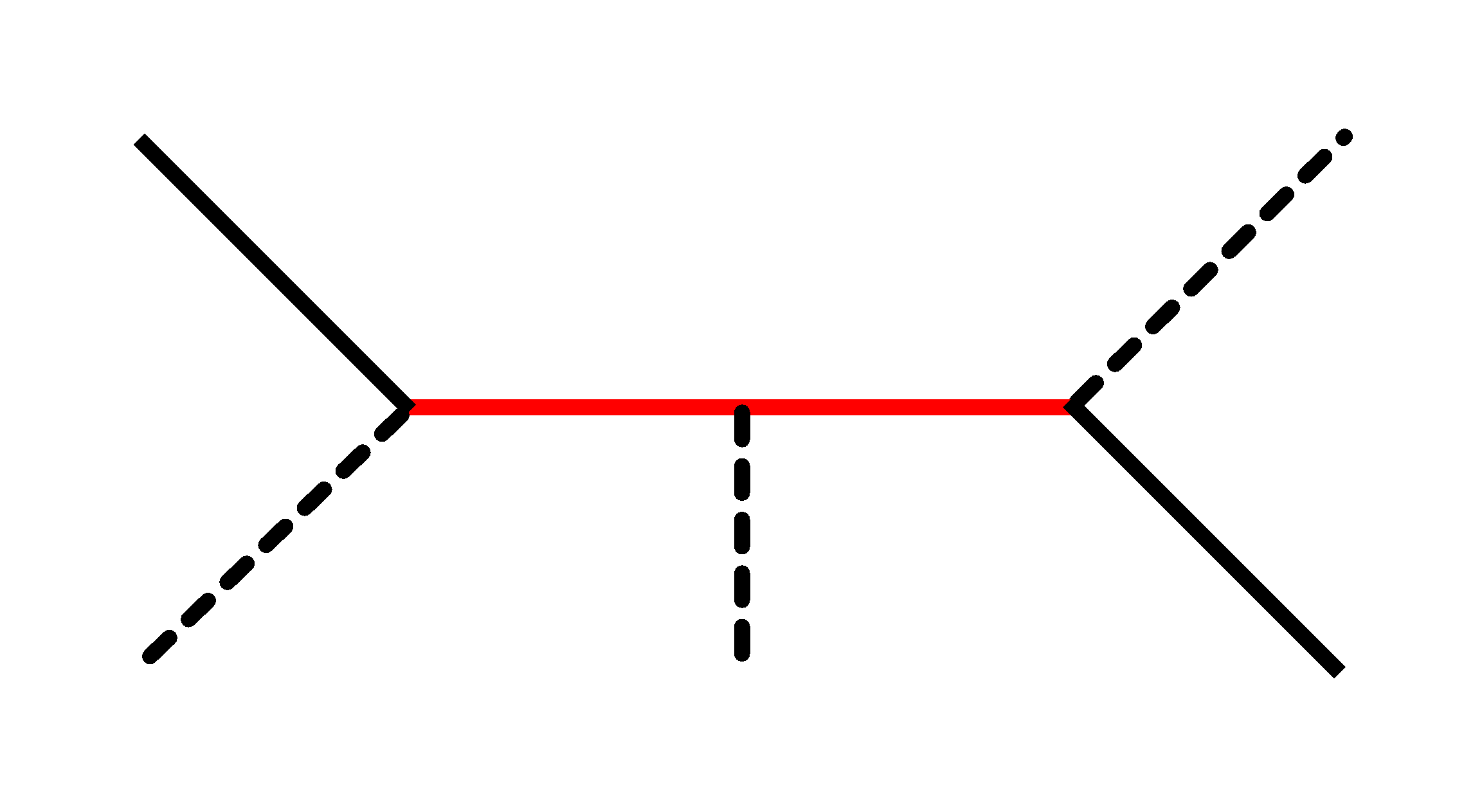}}    \\ %\midrule
%%%%%
$\psi^2 H^2 D^2$ \;:  &  \parbox[c]{3cm}  { \includegraphics[width=0.9\linewidth]{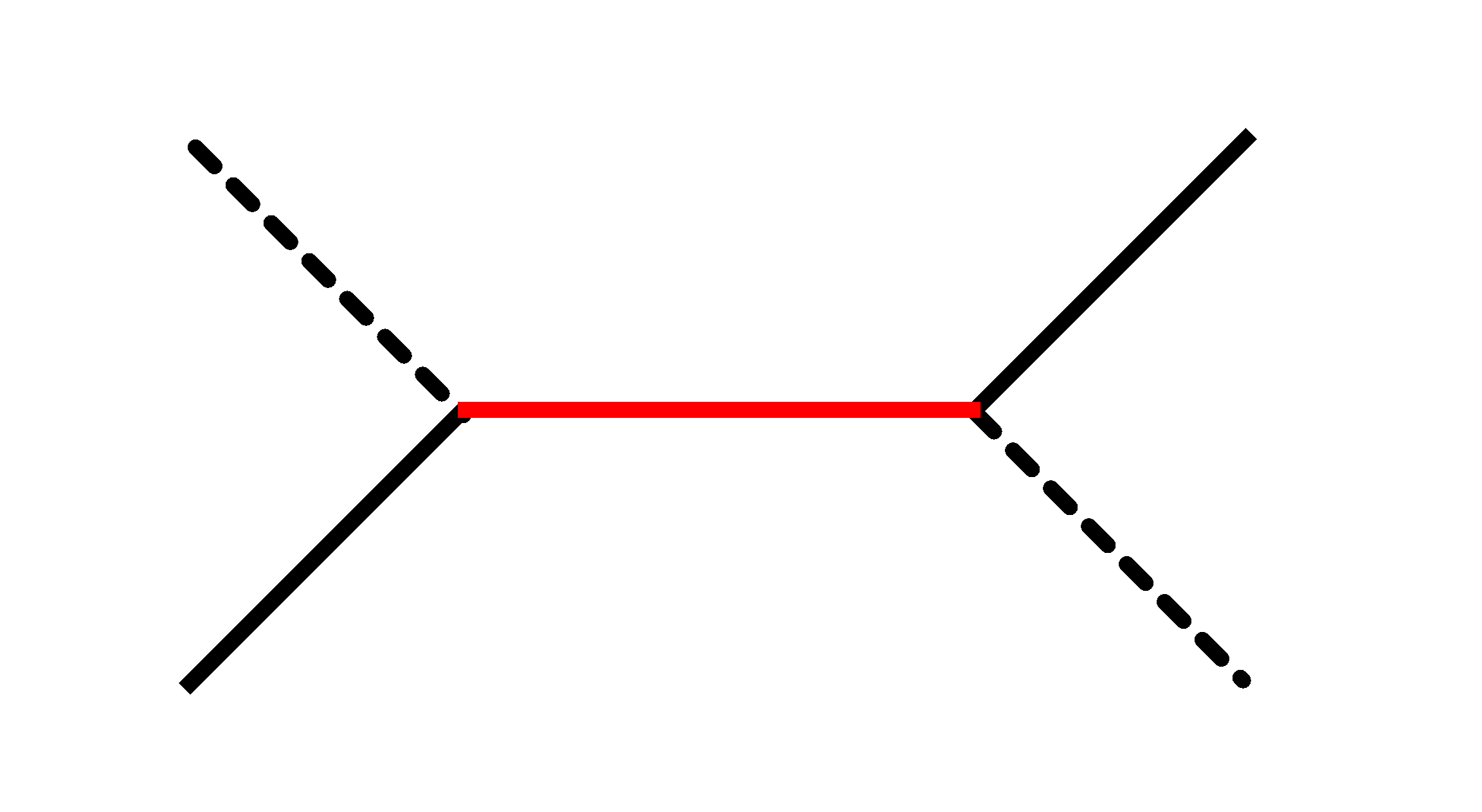}}    \; \parbox[c]{3cm} {\includegraphics[width=0.9\linewidth]{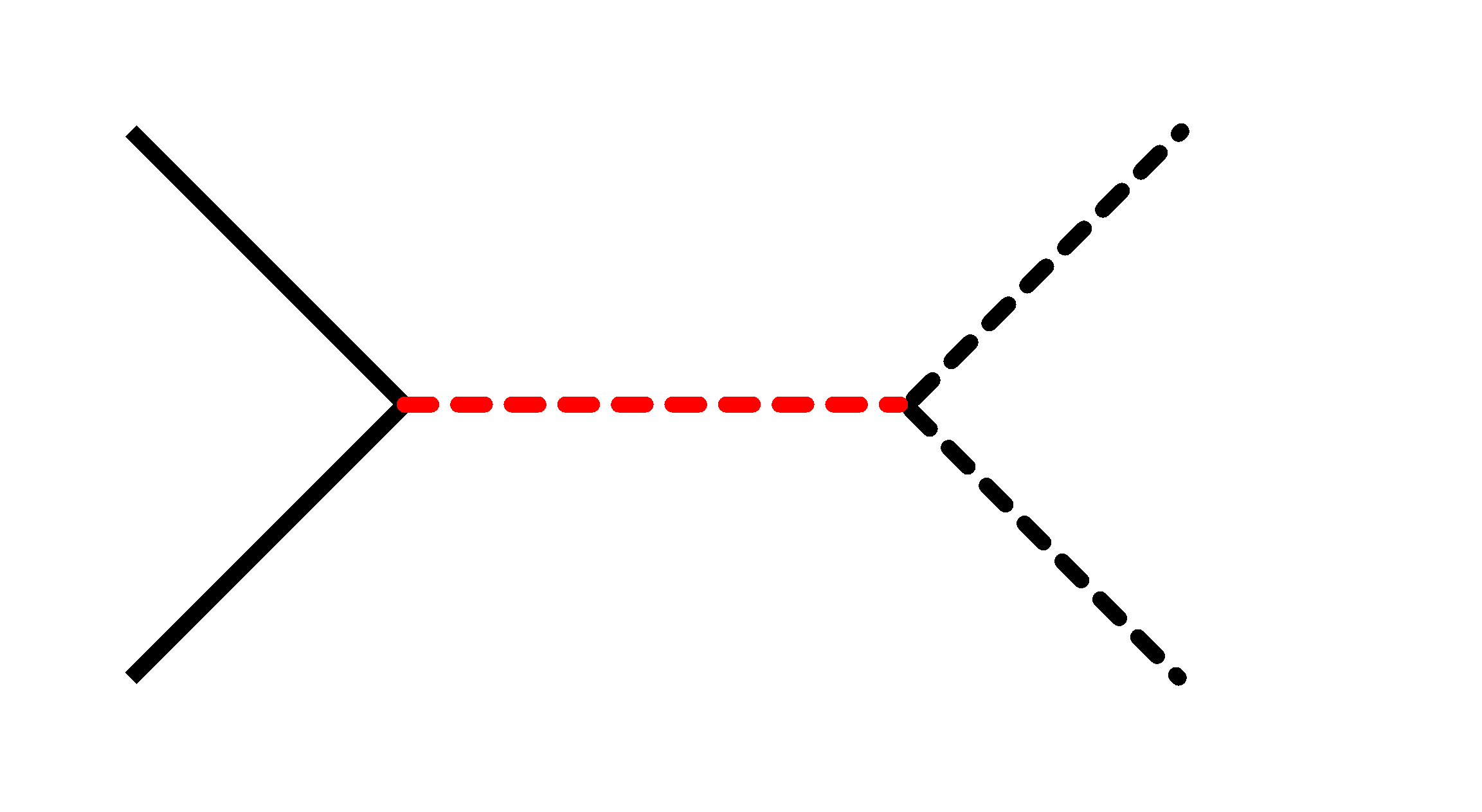}}         \\ %\midrule
%%%%%
$\psi^2 H^2 X$ \;:  & \parbox[c]{3cm}  {\includegraphics[width=0.9\linewidth]{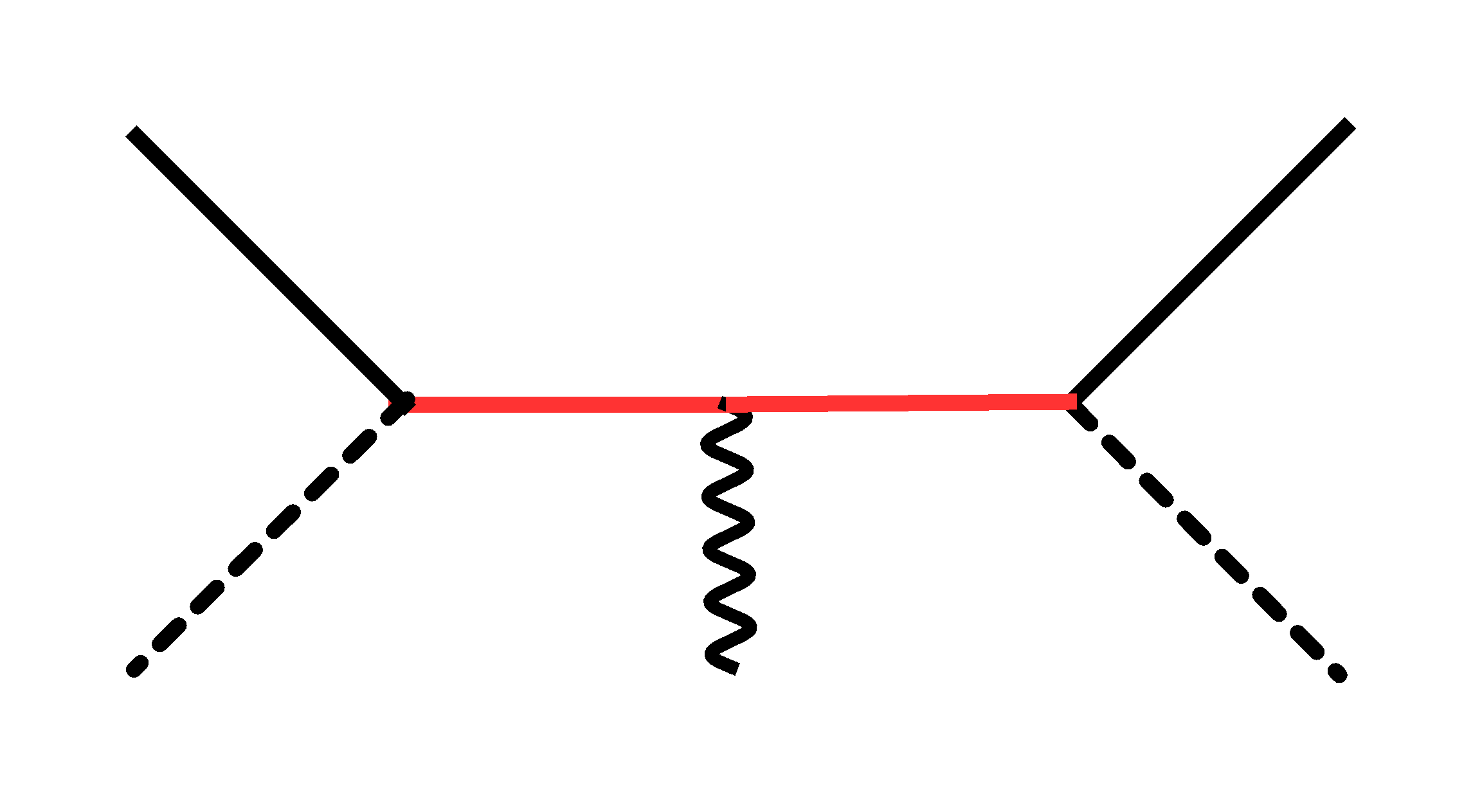}}        \\ %\midrule
%%%%%
$\psi^4 H$ \;:  & \parbox[c]{3cm} {\includegraphics[width=0.9\linewidth]{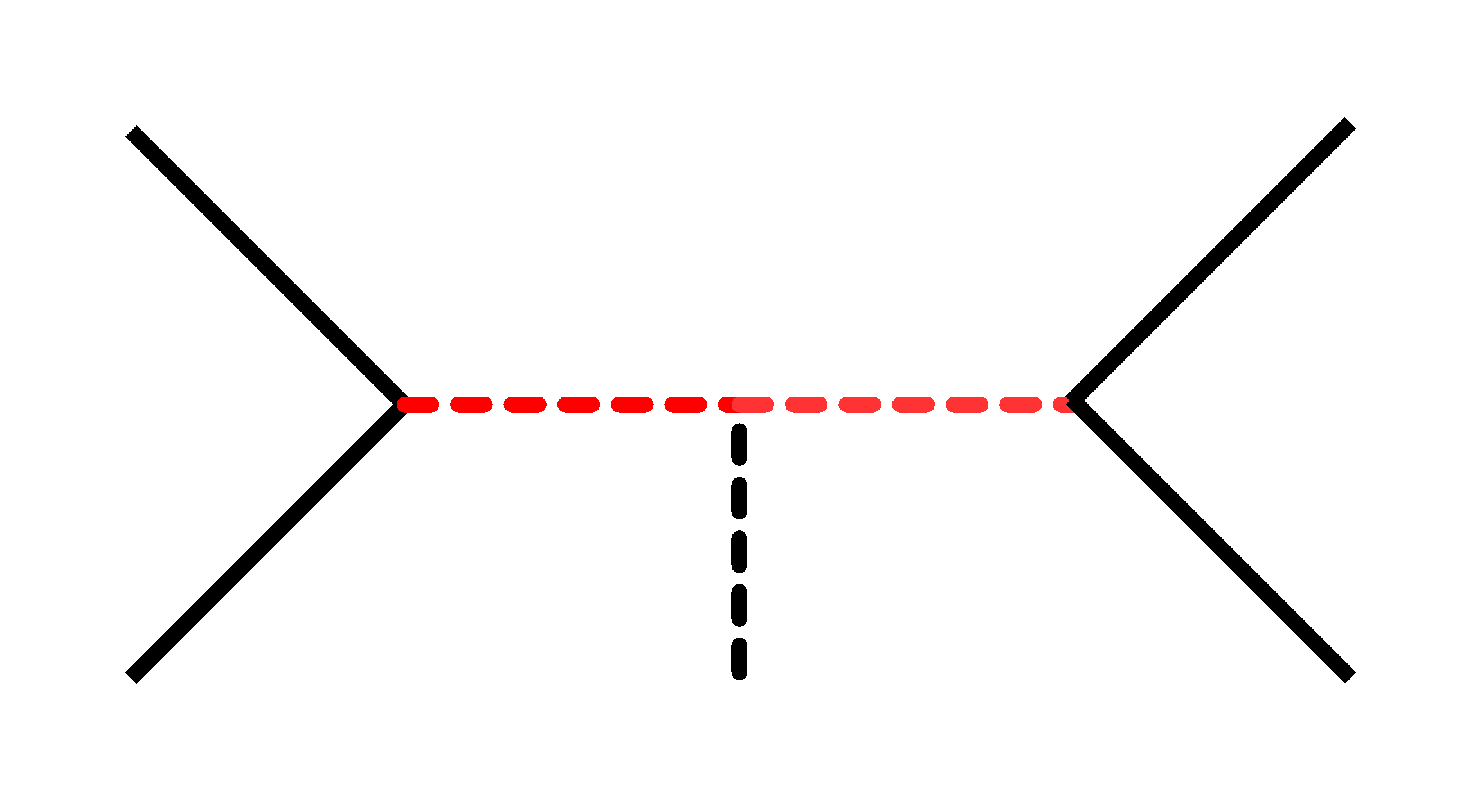}}     \parbox[c]{3cm}  {\includegraphics[width=0.9\linewidth]{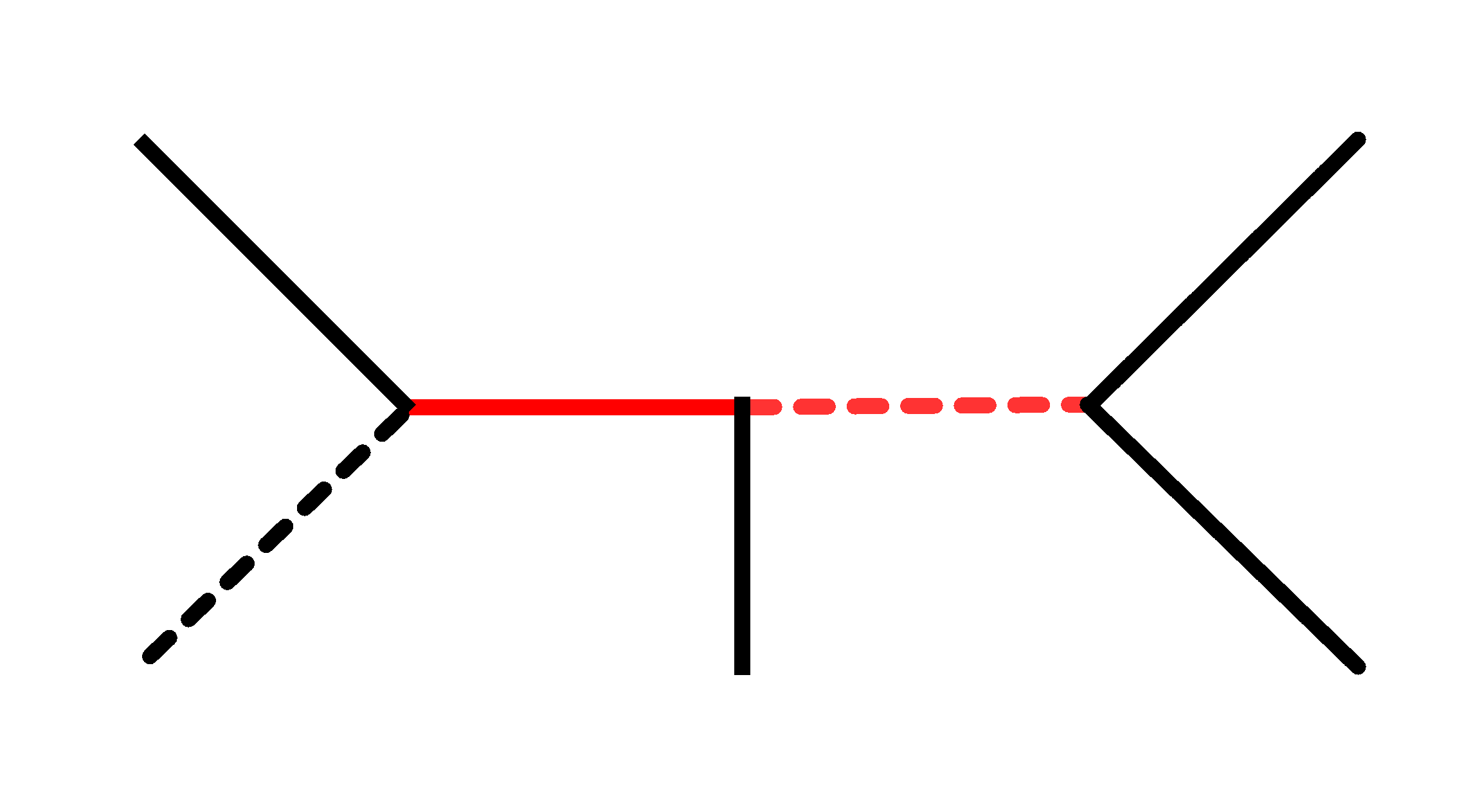}}  \parbox[c]{3cm} {\includegraphics[width=0.9\linewidth]{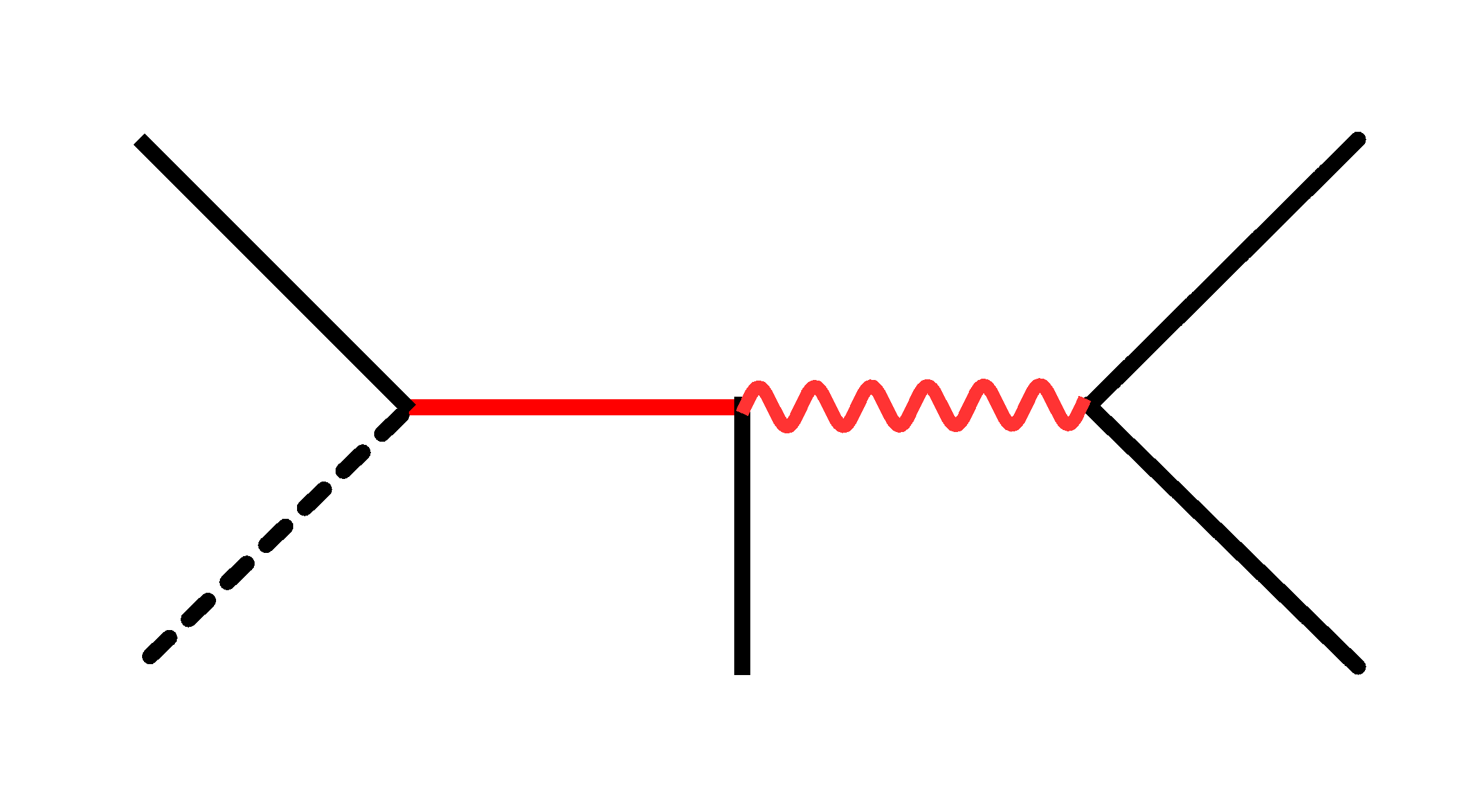}}       \\% \midrule
%%%%
$\psi^2 H^4$ \;: & \parbox[c]{3cm}  {\includegraphics[width=0.9\linewidth]{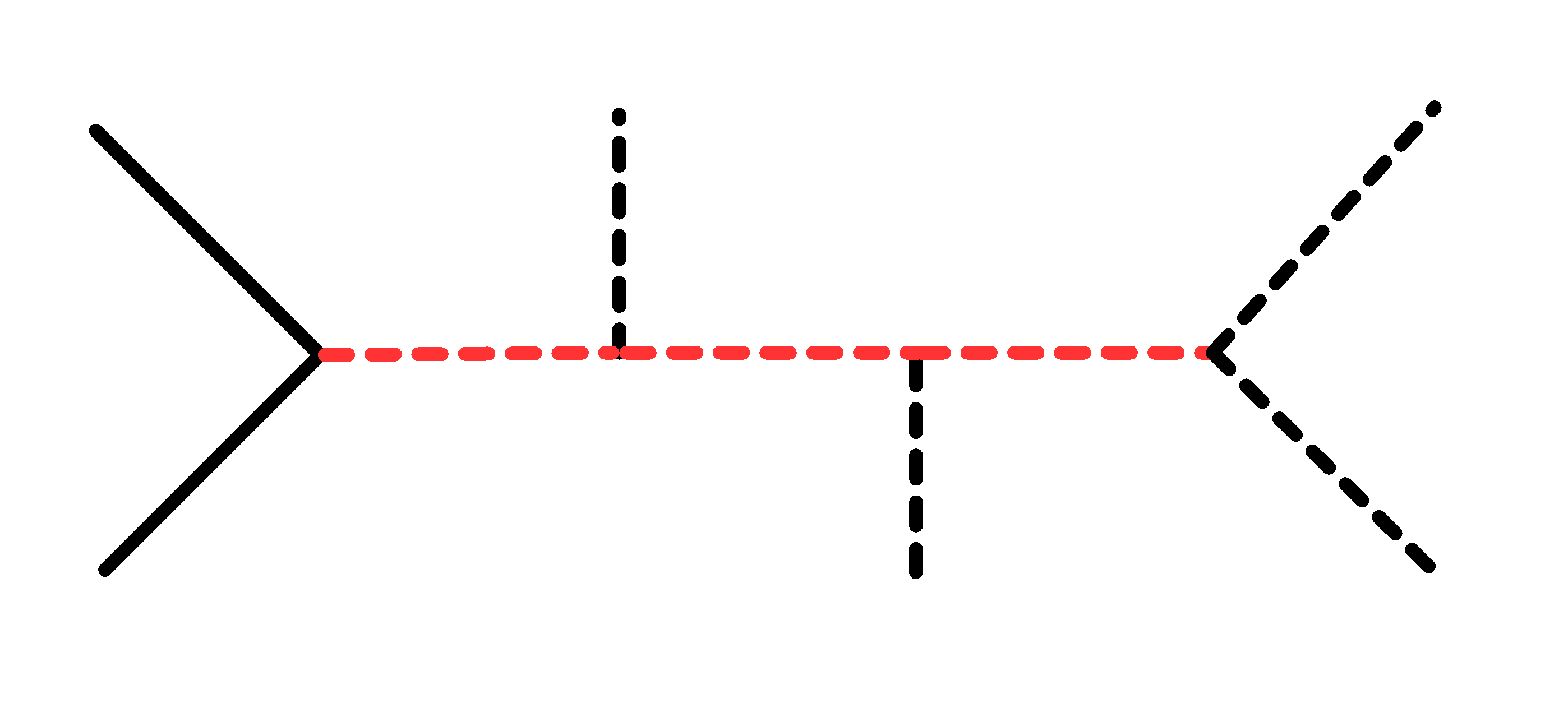}}   \parbox[c]{3cm}  {\includegraphics[width=0.9\linewidth]{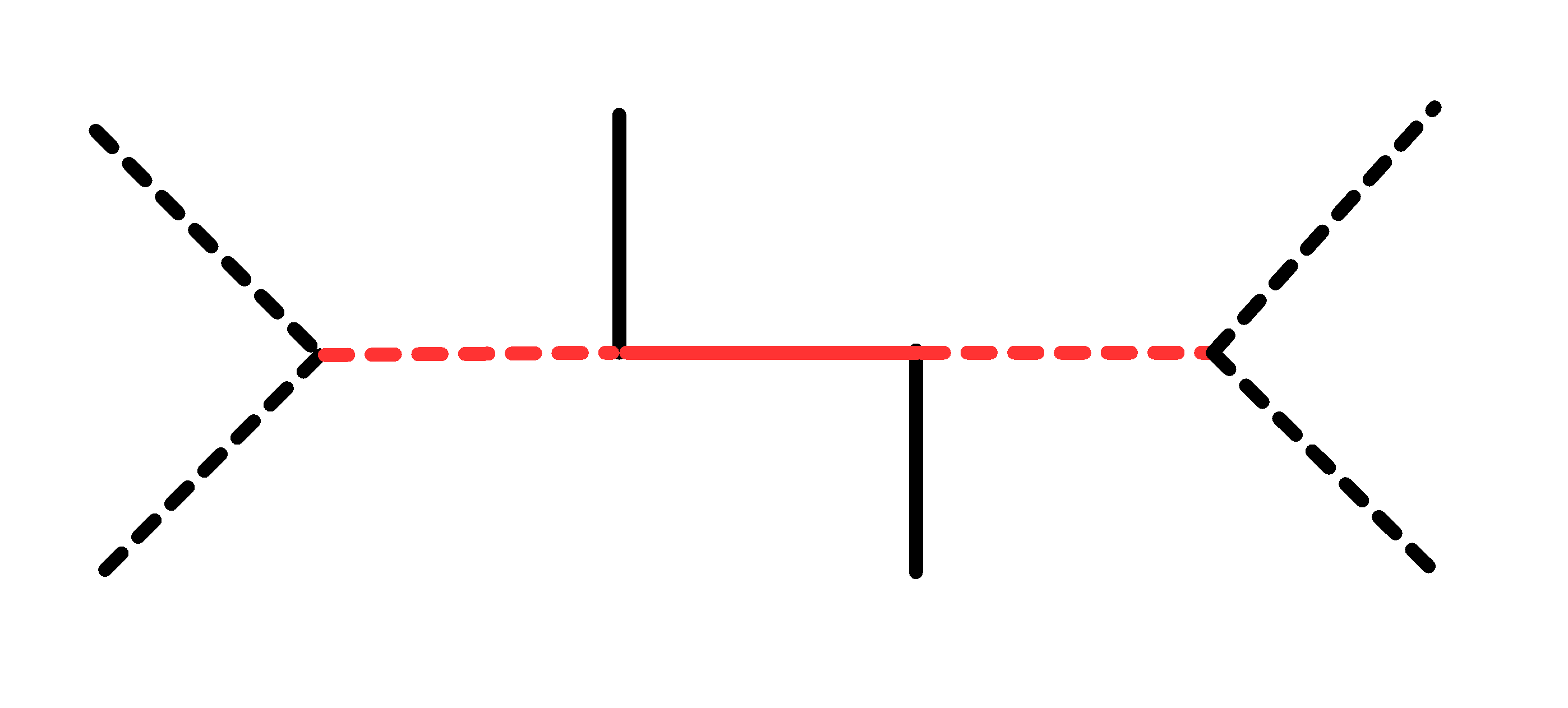}}   \parbox[c]{3cm}  {\includegraphics[width=0.9\linewidth]{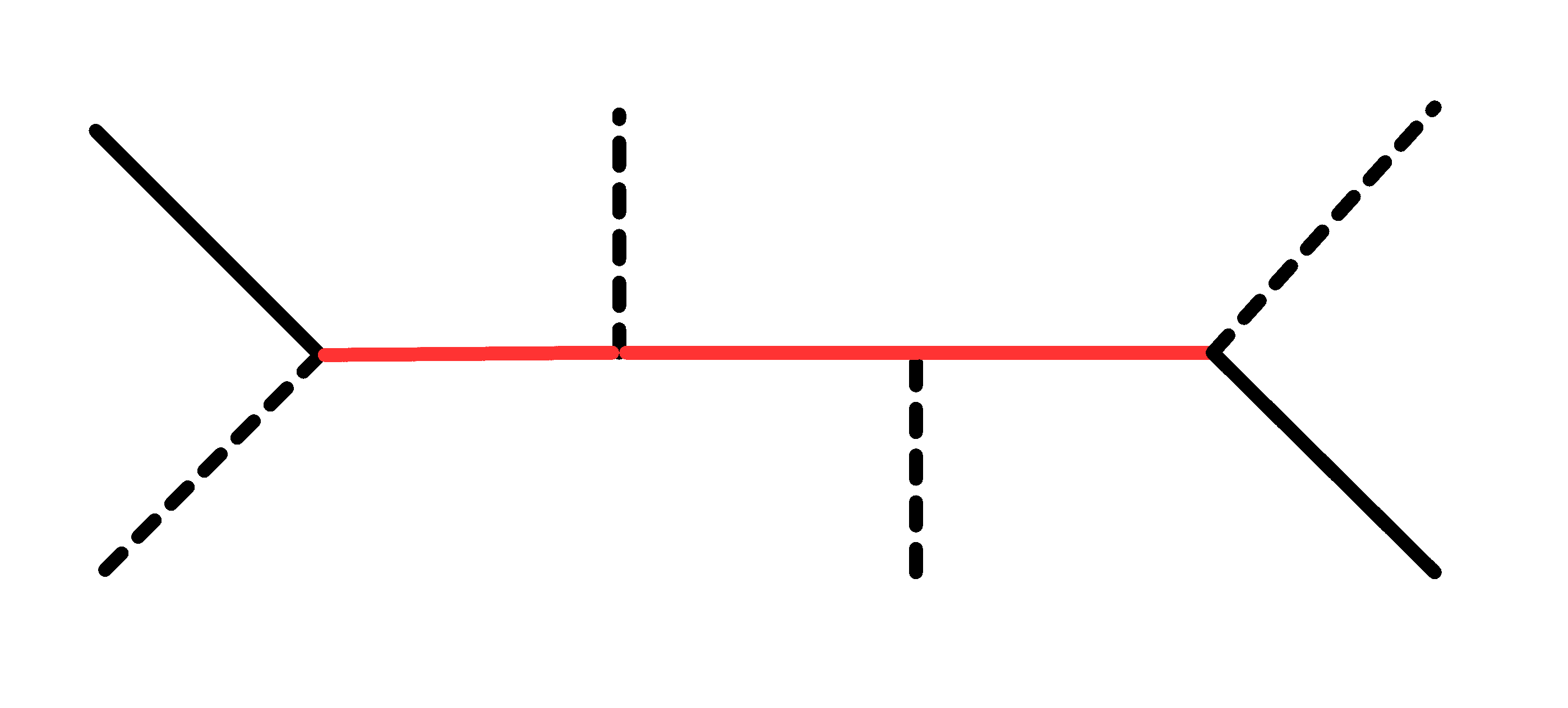}}  \parbox[c]{3cm}  {\includegraphics[width=0.9\linewidth]{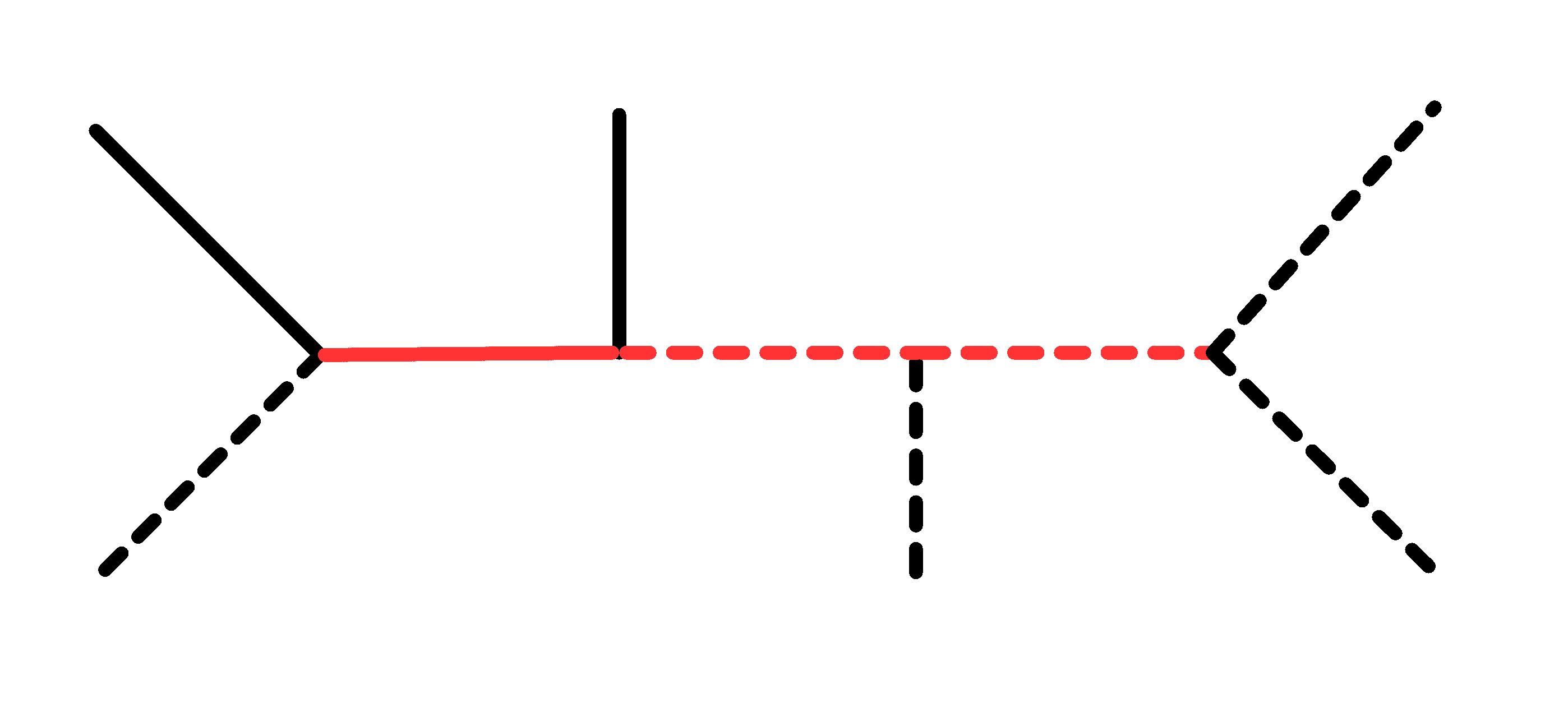}}   \\ 
& \parbox[c]{3cm}  {\includegraphics[width=0.9\linewidth]{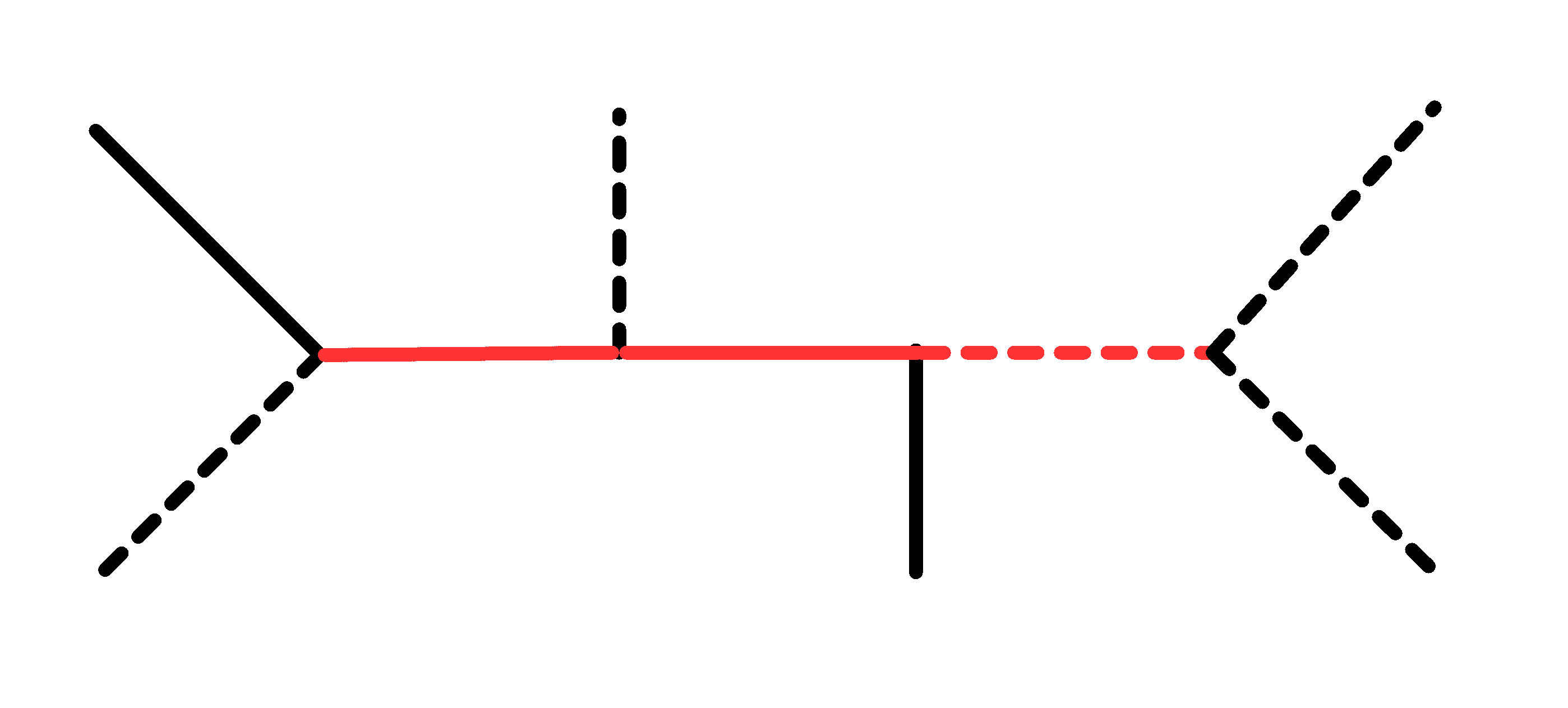}}  \parbox[c]{3cm}  {\includegraphics[width=0.9\linewidth]{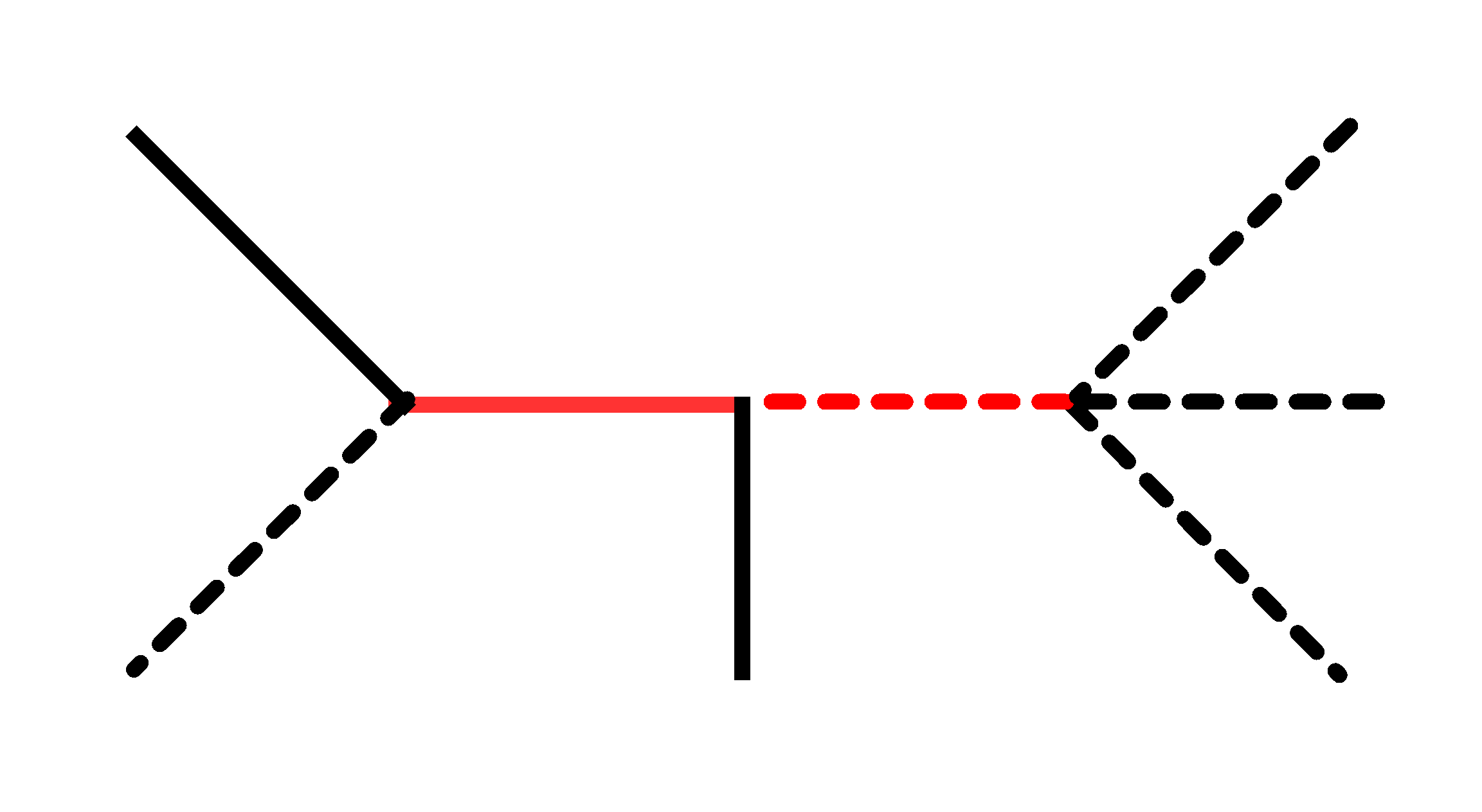}}    \parbox[c]{3cm}  {\includegraphics[width=0.9\linewidth]{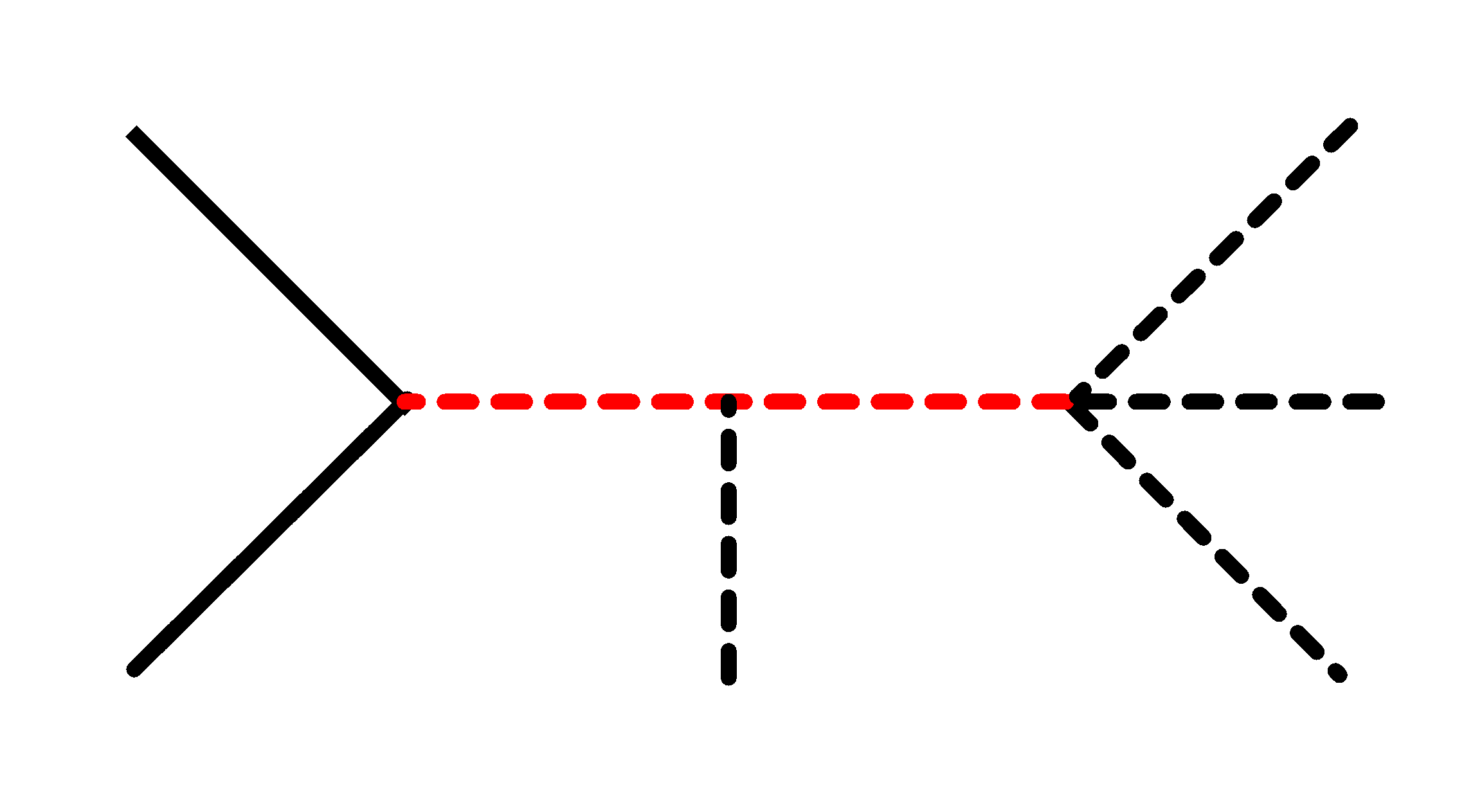}}   \parbox[c]{3cm}  {\includegraphics[width=0.9\linewidth]{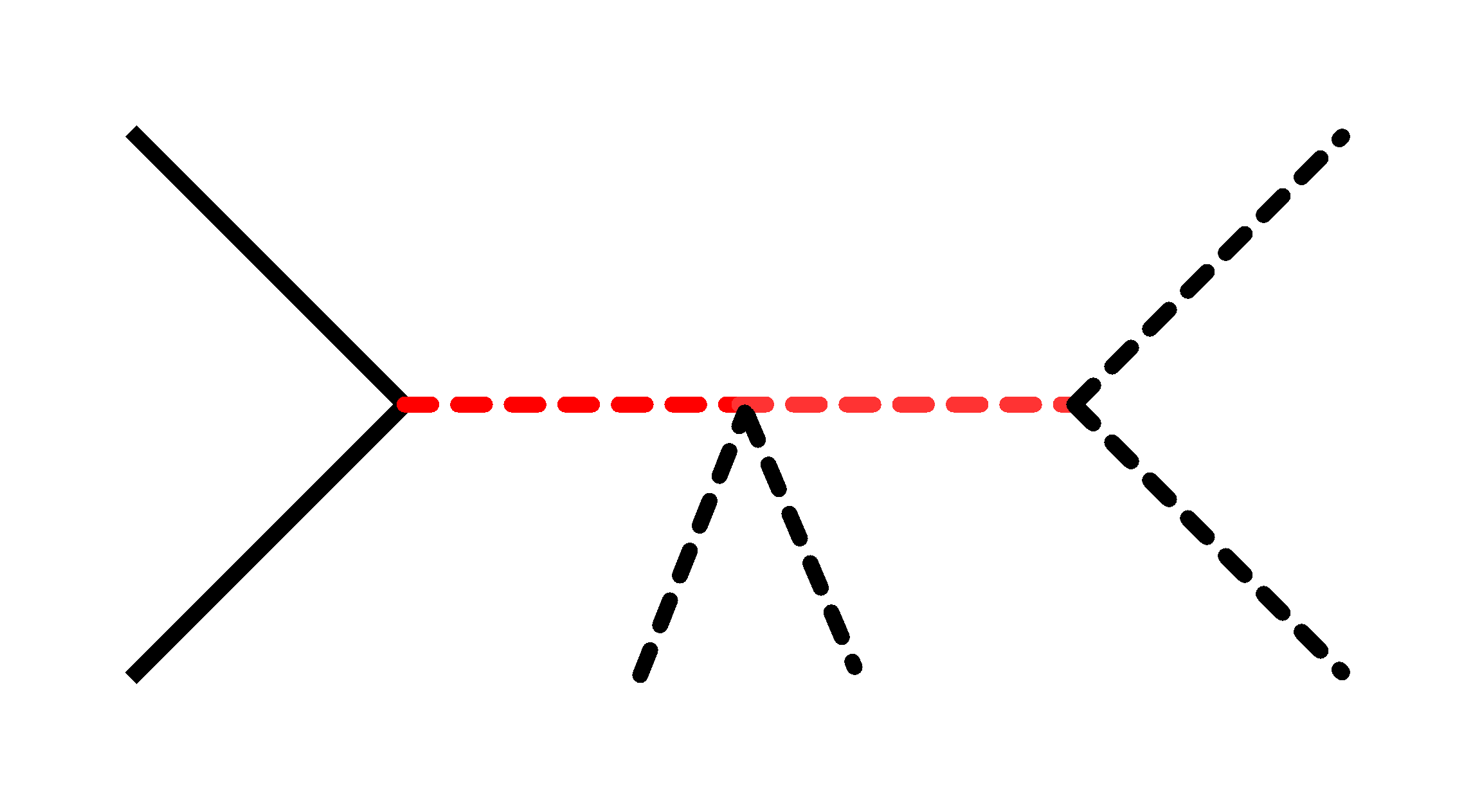}}     \\
   & \parbox[c]{3cm}  {\includegraphics[width=0.9\linewidth]{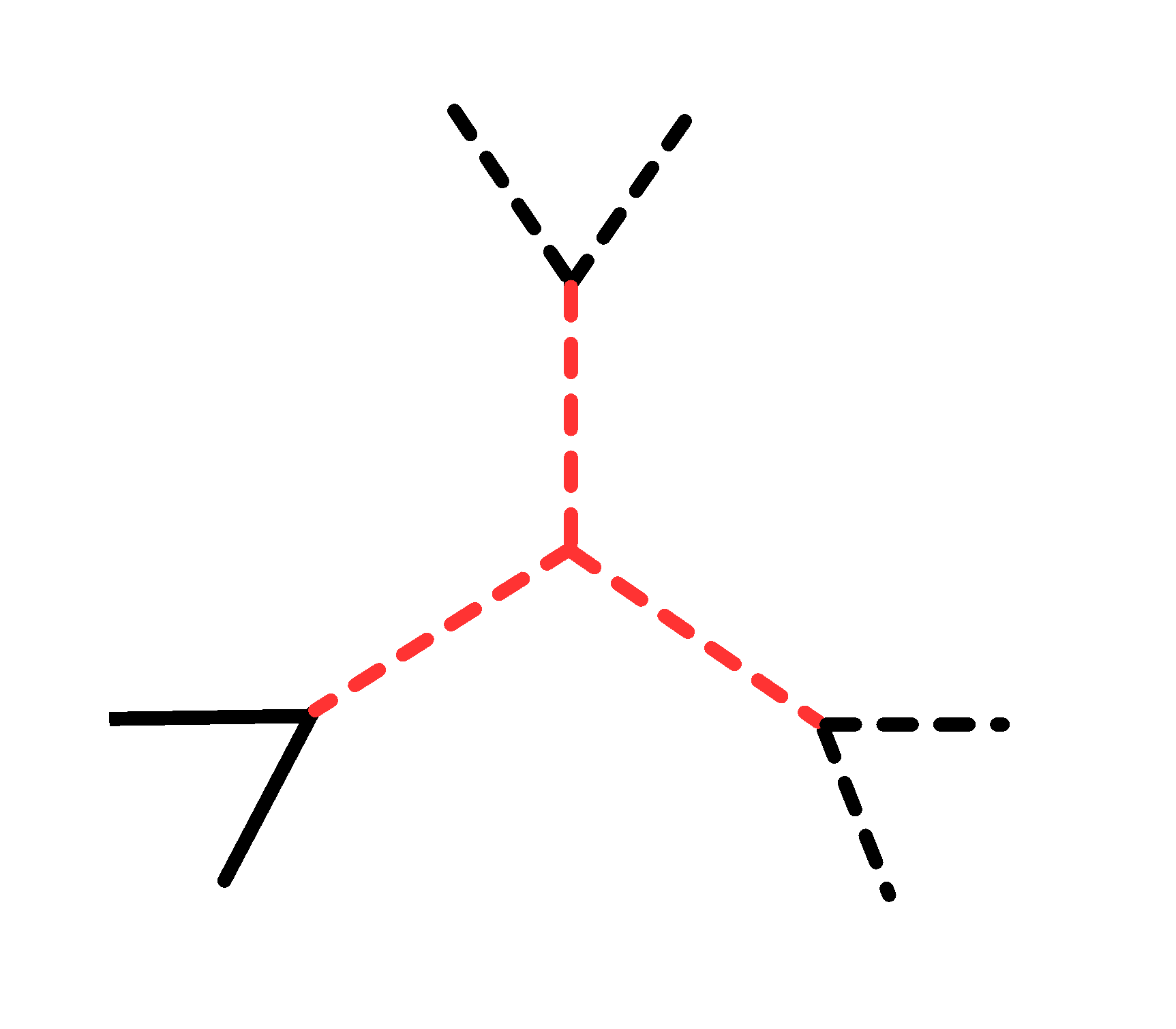}}   \parbox[c]{3cm}  {\includegraphics[width=0.9\linewidth]{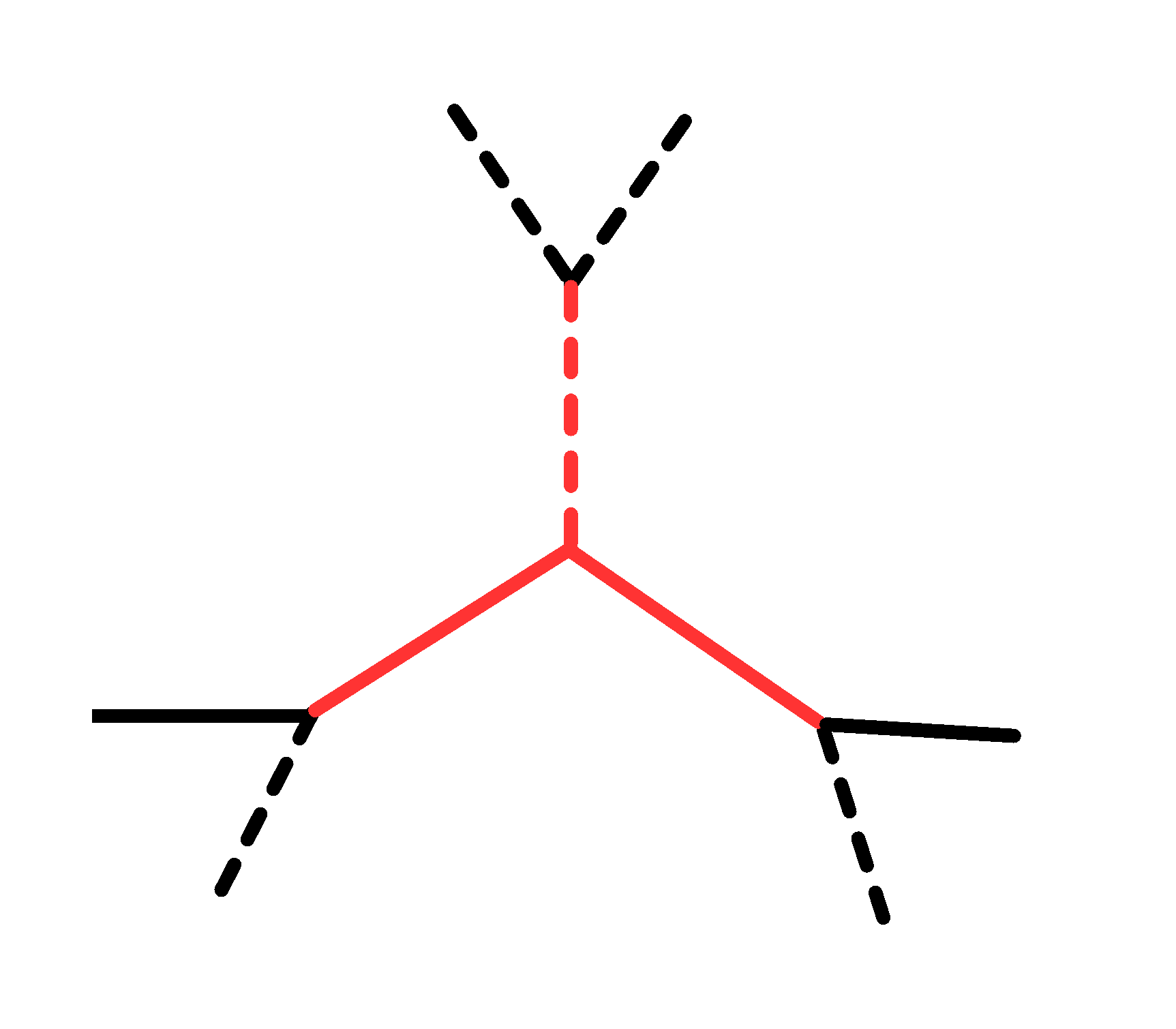}}       \\ %\cbottomrule
\end{tabular}
\end{adjustbox}
\caption{Diagrams generating different $d=7$ operator classes at tree-level. External light fields appear as black lines and heavy
propagators are marked in red.}
\label{fig:diagd7}
\end{figure}
%%%%%%%%%%%%%%%%%%%%%%%%

We next discuss the results for $d=7$ $N_R$SMEFT operators. There are eight different operator classes, but only five of them can be generated at tree-level.  Operators with two fermions containing derivatives or strength field tensors can only come at tree-level if they have at least two scalars \cite{Arzt:1994gp}. Operators with four fermions and a derivative cannot be generated either at tree-level, since the potential diagrams with four external legs (see diagrams for class $\psi^4$ in figure~\ref{fig:diagd6}) do not contain any derivative in the interaction vertices. Hence the classes $\psi^2 H^3 D$, $\psi^2 H^2 D^2$ and $\psi^2 H^2 X$ are tree-level generated, whereas $\psi^2 H D X$, $\psi^4 D$ and $\psi^2 X^2$ can only be opened via loops. The remaining classes generated at tree-level contain only fermions and scalars: $\psi^4 H$ and $\psi^2 H^4$. All the operators which can be decomposed at tree-level are shown in table~\ref{tab:Opd7}.\footnote{Again, baryon number violating operators, in class $\psi^4 H$, are treated separately in section \ref{sect:bnv}.} The complete list of on-shell operators at $d=7$ can be found, for example, in \cite{Liao:2016qyd}.

The diagrams generating the listed operators are represented in figure~\ref{fig:diagd7}, covering all possible configurations for the considered operators in $N_R$SMEFT. As before, we adopt the convention of using solid, dashed, and wavy lines to denote fermion, scalar, and vector fields, respectively, and black for light fields, while red lines denote heavy fields.

In most diagrams, there are now two or even three heavy propagators. However, the operator class $\psi^2 H^2 D^2$ can be generated with just one heavy propagator. Consequently, at $d=7$, we encounter models with up to three new particles. Regarding the Lorentz nature of the heavy fields appearing in the different models, we find that for the operators belonging to the $\psi^2 H^2 X$ class, only heavy fermions are allowed and for the ones in $\psi^2 H^2 D^2$ and $\psi^2 H^4$, only scalars and fermions can be drawn as internal lines. In the remaining classes, $\psi^2 H^3 D$ and $\psi^4 H$, the models consist of a pair of fields that are different combinations of scalars, fermions and vectors.

As the number of internal lines increases in the opening of diagrams of
higher dimensional operators, so does the possible number of
models. In total we count 224 UV-decompositions for the $d=7$
operators in table~\ref{tab:Opd7}, but the same particle content can
give rise to more than one $d=7$ operator in many cases. The total
number of different particle combinations for the listed operators is
then found to be 112. We present the results for each operator
individually in tables~\ref{tab:modelsd7_a}~-~\ref{tab:modelsd7_b},
classifying the models by the number of BSM particles they contain
(one, two or three). Additionally, we group the models based on the
Lorentz nature of the heavy propagators depicted in the diagrams.

Among these models, there are only two one-particle models: the scalar $\mathcal{S}$ and the fermion $\Delta_1$. The number of two-particle models increases to 102, consisting of 42 $FS$ models and 45 $FV$ models. The remaining models include three $FF$, nine $SS$, and three $SV$ configurations. As we indicated in figure~\ref{fig:diagd7}, we did not find any $VV$ model. Finally, there are eight three-particle models, which generate the operator $\mathcal{O}_{NH^4}$. It is worth noting that subsets of particles within the three-particle models can also generate other operators.

\begin{table}[t]
\centering
\renewcommand*{\arraystretch}{1.1}
 \begin{adjustbox}{max width=\textwidth}
 \begin{tabular}{|p{0.1\textwidth}|r p{0.26\textwidth}|c|r  p{0.3\textwidth}|}
\hline
\crowcolor
\multicolumn{1}{|c|}{$\psi^2 H^2 D^2$}  & \multicolumn{2}{l|}{Models} & $\psi^2 H^2 X$  & \multicolumn{2}{l|}{Models} \\
\hline
$\mathcal{O}_{NeH^2D^2}$                 & $F$ \;: & $\Delta_1$   & $\mathcal{O}_{NeH^2W}$ & $F$ \;: & $\Delta_1$ \\ \hline
\multirow{2}{*}{$\mathcal{O}_{NH^2D^2}$} & $S$ \;: & $\mathcal{S}$ & $\mathcal{O}_{NH^2B}$  & $F$ \;: & $\Delta_1$ \\  \cline{4-6} 
                                         & $F$ \;: & $\Delta_1$   & $\mathcal{O}_{NH^2W}$  & $F$ \;: & $\Delta_1$ \\ \hline
\end{tabular}
\end{adjustbox}
\bigskip
 \begin{adjustbox}{max width=\textwidth}
\begin{tabular}{|>{\centering\arraybackslash}p{0.1\textwidth}|rl|}
\hline

\crowcolor
 \multicolumn{1}{|c|}{$\psi^4 H$}  & \multicolumn{2}{|l|}{Models}  \\
\hline
\multirow{3}{*}{$\mathcal{O}_{LNLH}$}  & $SS$ \;: & $(\mathcal{S}_1,\varphi)$ $(\varphi,\Xi_1)$  \\
                                       & $FS$  \;: & $(E,\mathcal{S}_1)$ $(\Sigma_1,\Xi _1)$ $(\Delta_1, \mathcal{S}_1)$ $(\Delta_1,\Xi_1)$ $(\mathcal{N}, \varphi)$ $(\Sigma,\varphi)$ \\
                                       & $FV$  \;: & $(\mathcal{N},\mathcal{B})$ $(\Sigma,\mathcal{W})$ $(\mathcal{N},\mathcal{L}_1)$ $(\Sigma,\mathcal{L}_1)$ $(\Delta_1,\mathcal{B})$ $(\Delta_1,\mathcal{W})$ $(E,\mathcal{L}_1)$ $(\Sigma_1,\mathcal{L}_1)$  \\ \hline
\multirow{3}{*}{$\mathcal{O}_{QNLH}$}  & $SS$ \;: & $(\omega_1, \Pi_1)$ $(\Pi_1, \zeta)$  \\
                                       & $FS$ \;: & $(D,\omega_1)$ $(T_1,\zeta)$ $(\Delta_1,\omega_1)$ $(\Delta_1,\zeta)$ $(\mathcal{N},\Pi_1 )$ $(\Sigma,\Pi_1 )$ $(U,\Pi_1)$ $(T_2, \Pi_1 )$ \\
                                       & $FV$ \;: & $(U,\mathcal{U}_2)$ $(T_2,\chi)$ $(U,\mathcal{L}_1)$ $(T_2,\mathcal{L}_1)$ $(\mathcal{N},\mathcal{B})$ $(\Sigma, \mathcal{W})$ $(\mathcal{N},\mathcal{Q}_1)$ $(\Sigma,\mathcal{Q}_1)$ \\
  &  &    $(\Delta_1,\mathcal{U}_2)$ $(\Delta_1,\chi)$ $(D,\mathcal{L}_1)$ $( T_1,\mathcal{L}_1)$  $(\Delta_1,\mathcal{B})$ $( \Delta_1,\mathcal{W})$ $(D,\mathcal{Q}_1)$ $(T_1,\mathcal{Q}_1)$    \\ \hline
\multirow{3}{*}{$\mathcal{O}_{eNLH}$}  & $SS$ \;: & $(\mathcal{S}_1, \varphi)$   \\
                                       & $FS$ \;:  & $(\mathcal{N},\mathcal{S}_1)$ $(\Delta_1, \varphi)$  $(\Delta_3,\mathcal{S}_1)$ \\
                                       & $FV$ \;: & $(\mathcal{N}, \mathcal{B}) \ (\mathcal{N}, \mathcal{B}_1) \ (\Delta_3, \mathcal{L}_3) \
(\Delta_3, \mathcal{L}_1) \ (\Delta_1, \mathcal{B}) \ (\Delta_1, \mathcal{B}_1) \
(\Delta_1, \mathcal{L}_3) \ (\Delta_1, \mathcal{L}_1)$ \\\hline
\multirow{3}{*}{$\mathcal{O}_{dNLH}$}  & $SS$ \;: &  $(\omega_1, \Pi_1)$ \\
                                       & $FS$ \;: & $ (Q_1, \Pi_1) \ (\Delta_1, \Pi_1) \ (Q_5, \omega_1) \ (\mathcal{N}, \omega_1)$ \\
                                       & $FV$ \;: & $(\mathcal{N}, \mathcal{B}) \ (\mathcal{N}, \mathcal{U}_1) \ (Q_5, \mathcal{Q}_5) \ (Q_5, \mathcal{L}_1) \ (\Delta_1, \mathcal{B}) \ (Q_1, \mathcal{U}_1) \ (\Delta_1, \mathcal{Q}_5) \ (Q_1, \mathcal{L}_1)$ \\\hline
\multirow{3}{*}{$\mathcal{O}_{uNLH}$}  & $SS$ \;: &  $(\omega_2, \Pi_7)$\\
                                       & $FS$ \;: & $(Q_7, \Pi_7)\ (\Delta_1, \Pi_7) \ (\mathcal{N}, \omega_2) \ (Q_1, \omega_2) $ \\
                                       & $FV$ \;: & $(Q_1, \mathcal{Q}_1) \ (Q_1, \mathcal{L}_1) \
(\mathcal{N}, \mathcal{B}) \ (\mathcal{N}, \mathcal{U}_2) \ (\Delta_1, \mathcal{Q}_1) \
(Q_7, \mathcal{L}_1) \ (\Delta_1, \mathcal{B}) \
(Q_7, \mathcal{U}_2) $ \\\hline
\multirow{3}{*}{$\mathcal{O}_{duNLH}$} & $SS$ \;: & $(\omega_2, \Pi_1)$ \\
                                       & $FS$ \;: & $(Q_1, \Pi_1) \ (\Delta_1, \Pi_1) \ (Q_1, \omega_2) \ (E, \omega_2) $ \\
                                       & $FV$ \;: &  $ (E, \mathcal{B}_1) \ (E, \mathcal{U}_1) \
(Q_1, \mathcal{Q}_1) \ (Q_1, \mathcal{L}_1) \ (\Delta_1, \mathcal{B}_1) \ (Q_1, \mathcal{U}_1) \ (\Delta_1, \mathcal{Q}_1)$ \\ \hline
\multirow{3}{*}{$\mathcal{O}_{dQNeH}$} & $SS$ \;: &  $(\mathcal{S}_1,\varphi)$ \\
                                       & $FS$ \;: & $(\Delta_1, \varphi) \ (Q_5, \mathcal{S}_1) \ (U, \mathcal{S}_1) $ \\
                                       & $FV$ \;: & $(U, \mathcal{U}_2) \ (U, \mathcal{U}_1) \ (Q_5, \mathcal{Q}_5) \ (Q_5, \mathcal{Q}_1) \
(\Delta_1, \mathcal{U}_2) \ (\Delta_1, \mathcal{U}_1) \ (\Delta_1, \mathcal{Q}_5) \
(\Delta_1, \mathcal{Q}_1) $ \\ \hline
\multirow{2}{*}{$\mathcal{O}_{QuNeH}$} & $SS$ \;: & $ (\mathcal{S}_1,\varphi) \ (\omega_1,\Pi_1) \  (\omega_2, \Pi_7) $ \\
                                       & $FS$ \;: & $ (Q_7, \Pi_7) \ (\Delta_1, \Pi_7) \ (D, \omega_1) \ (\Delta_1, \omega_1) \ (D, \mathcal{S}_1) \ (Q_7, \mathcal{S}_1) \ (\Delta_1, \varphi) \ (\Delta_1, \Pi_1) \ $  \\
& &  $(Q_7, \Pi_1) \ (\Delta_1, \omega_2) \ (D, \omega_2)$    \\                                   
\hline
\end{tabular}
\end{adjustbox}
\caption{$N_R$SMEFT $d=7$ operators in classes $\psi^2 H^2 D^2$, $\psi^2 H^2 X$ and $\psi^4H$ and the corresponding  models generating them at tree-level. }
\label{tab:modelsd7_a}
\end{table}

\begin{table}[t]
\centering
\renewcommand*{\arraystretch}{1.1}
 \begin{adjustbox}{max width=\textwidth}
\begin{tabular}{|>{\centering\arraybackslash}p{0.1\textwidth}|rl|}
\hline
 \rowcolor[gray]{.9}
$\psi^2 H^3 D$ & \multicolumn{2}{|l|}{Models}  \\
\hline
\multicolumn{1}{|c|}{\multirow{4}{*}{$\mathcal{O}_{NLH^3D}$}}   & $SV$ \;: & \multicolumn{1}{l|}{$(\mathcal{S},\mathcal{L}_1)$ $(\Xi,\mathcal{L}_1)$ $(\Xi_1,\mathcal{L}_1)$ }  \\
 & $FS$ \;: & \multicolumn{1}{l|}{$(\mathcal{N},\mathcal{S})$ $(\Sigma, \Xi)$ $(\Sigma_1, \Xi_1)$ $(\Delta_1,\mathcal{S})$ $(\Delta_1,\Xi)$ $(\Delta_1, \Xi_1)$} \\
	 & $FV$ \;: & \multicolumn{1}{l|}{$(\mathcal{N}, \mathcal{B})$ $(\Sigma, \mathcal{W})$ $(\Sigma_1,\mathcal{W}_1)$ $(\Delta_1, \mathcal{B})$ $(\Delta_1,\mathcal{W})$ $(\Delta_1, \mathcal{B}_1)$ $(\Delta_1,\mathcal{W}_1)$} \\
	 & $FF$ \;:  &  \multicolumn{1}{l|}{$(\mathcal{N},\Delta_1)$ $(\Delta_1,\Sigma)$ $(\Delta_1,\Sigma_1)$} \\ \hline
\hline
\rowcolor[gray]{.9}
 \multicolumn{1}{|c|}{$\psi^2 H^4$}  & \multicolumn{2}{|l|}{Models}  \\
\hline
\multicolumn{1}{|c|}{\multirow{7}{*}{$\mathcal{O}_{NH^4}$}}        & $S$ \;:   & \multicolumn{1}{l|}{$\mathcal{S}$ }\\
                                                             & $SS$ \;:  & \multicolumn{1}{l|}{ $( \mathcal{S},\varphi)$ $(\mathcal{S},\Xi_1)$ $(\mathcal{S},\Xi)$ } \\
                                                             & $FS$ \;: & \multicolumn{1}{l|}{$(\mathcal{N},\mathcal{S})$ $(\Sigma, \Xi)$ $(\Sigma_1, \Xi_1)$ $(\Delta_1,\mathcal{S})$ $(\Delta_1,\Xi)$ $(\Delta_1, \Xi_1)$ $(\Delta_1,\varphi)$} \\
                                                             & $FF$ \;: & \multicolumn{1}{l|}{ $(\mathcal{N},\Delta_1)$ $(\Delta_1,\Sigma)$ $(\Delta_1,\Sigma_1)$ } \\
                                                             & $SSS$ \;:  & \multicolumn{1}{l|}{$(\mathcal{S},\varphi ,\Xi_1)$  $(\mathcal{S},\varphi ,\Xi)$} \\
                                                             & $FFS$ \;: & \multicolumn{1}{l|}{$(\Delta_1,\Sigma_1,\Xi_1)$ $(\mathcal{N},\Delta_1,\mathcal{S})$ $(\Delta_1, \Sigma,\Xi)$} \\
                                                              & $FSS$ \;: & \multicolumn{1}{l|}{ $(\Delta_1,\mathcal{S},\varphi)$ $(\Delta_1,\varphi,\Xi )$ $(\Delta_1,\varphi,\Xi_1)$} \\
\hline
\hline
\rowcolor[gray]{.9}
 \multicolumn{1}{|c|}{$\psi^4 H$}  & \multicolumn{2}{|l|}{Models}  \\
\hline
\multirow{2}{*}{$\mathcal{O}_{LNeH}$}  & $SS$ \;: & $(\mathcal{S},\varphi)$ $(\mathcal{S}_1,\varphi)$  \\
                                       & $FS$  \;: & $(E,\mathcal{S})$ $(E,\mathcal{S}_1)$ $(\Delta_1,\varphi)$ $(\Delta_1,\mathcal{S})$ $(\Delta_1, \mathcal{S}_1)$ \\  \hline
\multirow{3}{*}{$\mathcal{O}_{eLNH}$}  & $SS$ \;: & $(\mathcal{S},\varphi)$\\
                                       & $FS$ \;: & $(\Delta_1, \mathcal{S}) \ (\Delta_1, \varphi) \ (E, \mathcal{S}_1) $\\
                                       & $FV$ \;: &$ (E, \mathcal{B}_1) \ (\Delta_1, \mathcal{L}_1) \ (\Delta_1, \mathcal{B}_1) $ \\ \hline
\multirow{2}{*}{$\mathcal{O}_{QNdH}$}  & $SS$ \;: &  $(\mathcal{S},\varphi)$ $(\omega_1, \Pi_1)$\\
                                       & $FS$ \;:  & $ (\Delta_1 , \varphi ) \ (D , \omega_1 )
(\Delta_1 , \omega_1 ) \ (Q_1 , \Pi_1 ) \ (\Delta _1 , \Pi_1 ) \ (Q_1 , \mathcal{S} ) \ (D, \mathcal{S} ) $ \\ \hline
\multirow{3}{*}{$\mathcal{O}_{dQNH}$}  & $SS$ \;: & $(\mathcal{S},\varphi)$ \\
                                       & $FS$ \;: & $(\Delta_1, \varphi) \ (Q_1, \mathcal{S}) \ (D, \mathcal{S}) $ \\
                                       & $FV$ \;: & $(\Delta_1, \mathcal{U}_1) \ (\Delta_1, \mathcal{Q}_1) \ (D, \mathcal{U}_1) \ (Q_1, \mathcal{Q}_1) $ \\\hline
\multirow{2}{*}{$\mathcal{O}_{QNuH}$}  & $SS$ \;: &  $(\mathcal{S},\varphi)$ $(\omega_2, \Pi_1)$\\
                                       & $FS$ \;: & $(Q_1, \Pi _1),$ $(\Delta _1, \Pi _1),$ $(U, \omega _2),$ $(\Delta _1, \omega _2),$ $(U, \mathcal{S}),$ $(Q_1, \mathcal{S})$ \\ \hline
\multirow{3}{*}{$\mathcal{O}_{uQNH}$} & $SS$ \;: &  $(\mathcal{S},\varphi)$ \\
                                       & $FS$ \;: &  $(\Delta_1, \varphi) \ (U, \mathcal{S}) \ (Q_1, \mathcal{S})  $\\
                                       & $FV$ \;: & $(Q_1, \mathcal{Q}_1) \ (U, \mathcal{U}_2) \ (\Delta_1, \mathcal{Q}_1) \ (\Delta_1, \mathcal{U}_2) $ \\\hline
\multirow{2}{*}{$\mathcal{O}_{LNNH}$} & $SS$ \;: &  $(\mathcal{S},\varphi)$ \\
                                       & $FS$ \;: & $(\mathcal{N},\mathcal{S}) \ (\Delta_1, \mathcal{S}) \ (\Delta_1, \varphi)  $ \\ \hline
\multirow{3}{*}{$\mathcal{O}_{NLNH}$} & $SS$ \;: & $(\mathcal{S},\varphi)$ \\
                                       & $FS$ \;: & $(\mathcal{N},\mathcal{S}) \ (\Delta_1, \mathcal{S}) \ (\Delta_1, \varphi)  $ \\ 
                                       & $FV$ \;: & $(\mathcal{N},\mathcal{B}) \ (\Delta_1, \mathcal{B}) \ (\Delta_1, \mathcal{L}_1)  $\\ \hline
\end{tabular}
\end{adjustbox}
\caption{Continuation of table~\ref{tab:modelsd7_a}. $N_R$SMEFT $d=7$ operators in $\psi^2 H^3D$, $\psi^2 H^4$, $\psi^4H$ and their tree-level decompositions.}
\label{tab:modelsd7_b}
\end{table}

\begin{table}[t]
\renewcommand*{\arraystretch}{1.2}
\centering
 \begin{adjustbox}{max width= \textwidth}
\begin{tabular}{|p{0.05\textwidth}>{\centering\arraybackslash}p{0.24\textwidth}|>{\raggedleft\arraybackslash}p{0.07\textwidth}>{\raggedright\arraybackslash}p{0.4\textwidth}|}%|cc|rl|
\hline
\rowcolor[gray]{.9}
\multicolumn{2}{|c|}{ $\psi^4$ \; $(d=6)$}                                             & \multicolumn{2}{|c|}{Models} \\ \hline
\multicolumn{1}{|c|}{\multirow{2}{*}{$\mathcal{O}_{QQdN}$}}  & \multirow{2}{*}{$\varepsilon_{ij} \left(\overline{Q_i^{c}} Q_j \right)\left(\overline{d^{c}_R} N_R\right)$} & $S$ \;:          &     $\omega_1$ \\
\multicolumn{1}{|c|}{}                                       &                   & $V$ \;:       &  $\mathcal{Q}_1$          \\ \hline
\multicolumn{1}{|c|}{$\mathcal{O}_{uddN}$}                   &   $ \left(\overline{u_R^{c}}d_R \right)\left(\overline{d_R^{c}} N_R\right)$                & $S$ \;:            &   $\omega_1$, $\omega_2$         \\ \hline
\hline
\rowcolor[gray]{.9}
\multicolumn{2}{|c|}{ $\psi^4 H$ \; $(d=7)$}                                           & \multicolumn{2}{|c|}{Models} \\ \hline
\multicolumn{1}{|c|}{\multirow{3}{*}{$\mathcal{O}_{QNddH}$}} & \multirow{3}{*}{$\varepsilon_{ij} \left(\overline{Q_i} N_R \right)\left(\overline{d_R} d_R^{c} \right) \tilde{H}_j$} & $SS$ \;:         &    $(\omega_2, \Pi_1)$        \\
\multicolumn{1}{|c|}{}                                       &                   & $FS$ \;:          & $(U,\omega_2) \ (\Delta_1,\omega_2) \ (Q_1, \Pi_1)$           \\
\multicolumn{1}{|c|}{}                                       &                   & $FV$ \;:          &  $(Q_1, \mathcal{Q}_1) \ (Q_1, \mathcal{U}_1) \ (\Delta_1, \mathcal{Q}_1) \ (U, \mathcal{U}_1) $          \\ \hline
\multicolumn{1}{|c|}{\multirow{3}{*}{$\mathcal{O}_{QNQH}$}}  & \multirow{3}{*}{$ \varepsilon_{ij} \left(\overline{Q_i} N_R \right)\left(\overline{Q_j} Q^{c} \right) H$} & $SS$ \;:          & $(\omega_1, \Pi_1) \ (\Pi_1, \zeta) $           \\
\multicolumn{1}{|c|}{}                                       &                   & $FS$  \;:         &  $(D,\omega_1) \  (\Delta_1, \omega_1) \ (T_1, \zeta) \ (\Delta_1, \zeta)$ \\
\multicolumn{1}{|c|}{}                                     
& & & $(D, \Pi_1) \ (T_1, \Pi_1) $          \\ \hline
\multicolumn{1}{|c|}{\multirow{4}{*}{$\mathcal{O}_{QNudH}$}} & \multirow{4}{*}{$ \left(\overline{Q} N_R \right)\left(\overline{u_R} d^{c}_R \right) H$ } & $SS$   \;:        &  $(\omega_1, \Pi_1)$          \\
\multicolumn{1}{|c|}{}                                       &                   & $FS$   \;:        &   $(D,\omega_1)  \ (\Delta_1, \omega_1 ) \ (Q_5, \Pi_1) \ (Q_1, \Pi_1) $            \\
\multicolumn{1}{|c|}{}                                       &                   & $FV$ \;:          &  $(Q_1, \mathcal{Q}_1) \ (Q_1, \mathcal{U}_1) \ (Q_5, \mathcal{Q}_5) \ (Q_5, \mathcal{U}_2) $     \\
\multicolumn{1}{|c|}{}                                      & & & $ (\Delta_1, \mathcal{Q}_1) \ (D,\mathcal{U}_1) \ (\Delta_1, \mathcal{Q}_5) \ (D, \mathcal{U}_2)   $       \\ \hline
\end{tabular}
\end{adjustbox}
\caption{Baryon number violating operators of $d=6$ and $d=7$ in $N_R$SMEFT and their tree-level decompositions. Models are classified in terms of the Lorentz nature of the fields.} 
\label{tab:BNV}
\end{table}

\subsection{Baryon number violating operators \label{sect:bnv}}

In this subsection we comment on baryon number violation (BNV) in
$N_R$SMEFT. We give the decompositions for the BNV operators here only
for completeness, since proton decay constraints (see below) usually
render these operators uninteresting for collider phenomenology.  The
BNV operators at $d=6$ and $d=7$ that can be opened at tree-level are
presented in table~\ref{tab:BNV}.\footnote{When defining the
  operators, $SU(3)$ contractions have been omitted. In all cases
  there should be the total antisymmetric tensor in colour indices,
  $\varepsilon_{\alpha \beta \sigma}$, contracted with the three quark
  fields' colour indices.} Next to each operator, we provide the list
of decompositions, classified according to the Lorentz nature of the
fields.  At $d=6$, there are two four-fermion operators that can be
opened at tree-level. At $d=7$ there are five BNV operators, two of
them belong to the $\psi^4 D$ class and can only be generated at loop
level, hence we don't list them. The three remaining operators are in
the $\psi^4 H$ category and can be opened through three different
diagrams, as was shown in figure~\ref{fig:diagd7}.

For the $d=6$ operators we find three one-particle decompositions: the two
scalars $\omega_1$, $\omega_2$, and the vector $\mathcal{Q}_1$.
Regarding the models for $d=7$ operators, we found 22 two-particle
models (3 $SS$, 10 $FS$ and 9 $FV$ models). Out of these 22, 17 models
appear also in the opening of other $d=7$ $N_R$SMEFT operators,
while 3 $FS$ and 2 $FV$ models are new. These are $(D, \Pi_1)$, 
$(T_1, \Pi_1)$, $(Q_5, \Pi_1)$ and $(Q_5, \mathcal{U}_2)$, 
$(D, \mathcal{U}_2)$, respectively. 

Note that the non-observation of proton decay sets stringent limits on
parameters of the models that allow for baryon number violating
processes. Let us discuss this first for a specific example. For
simplicity, we consider the scalar decomposition $\omega_1$ of the
$d=6$ operator $\mathcal{O}_{uddN}$. The relevant part of the UV
Lagrangian for our discussion reads
\begin{align}
 \mathcal{L} \propto & \; y_{ue}^{\omega_1} \left( \overline{u_R^c}e_R \right) \omega_1^{\dagger} + y_{Nd}^{\omega_1} \left( \overline{d^c_R} N_R \right) \omega_1^{\dagger} + y_{QL}^{\omega_1} \left( \overline{Q^c} L \right) \omega_1^\dagger  \nonumber \\
 & +  y_{ud}^{\omega_1} \left( \overline{u_R^c} d_R \right) \omega_1 + y_{Q}^{\omega_1} \left( \overline{Q^c} Q \right) \omega_1 +  \text{h.c.} + \ldots
 \label{eq:bnv}
\end{align}
The contribution to the operator $\mathcal{O}_{uddN}$ is given by the
matching relation
\begin{equation}
c_{uddN} = \frac{y_{Nd}^{\omega_1} \, y_{ud}^{\omega_1}}{\Lambda^2} \,,
\end{equation}
where we assumed $m_{\omega_1} = \Lambda$. Note that again we have
suppressed flavour indices; in full generality the Wilson coefficients
are matrices in flavour space.  This operator triggers the decay
process $p \rightarrow \pi^+ N$.\footnote{While $\omega_1$ can trigger
  decays such as $p \rightarrow \pi^+ N$ and $p \rightarrow K^+ N$,
  $\omega_2$ can only cause the decay $p \rightarrow K^+ N$, because
  the coupling $y_{dd}^{\omega_2}$ is anti-symmetric in flavour space.}  Experimental limits on the proton lifetime impose stringent
constraints on the combination $y_{Nd}^{\omega_1} \,
y_{ud}^{\omega_1}/ \Lambda^2$. For the current experimental bound on
this proton decay mode \cite{ParticleDataGroup:2022pth} the upper limit
is roughly
\begin{equation}\label{eq:limP}
y_{Nd}^{\omega_1} \, y_{ud}^{\omega_1} \lsim 10^{-26} \left(\frac{\Lambda}{\rm TeV}\right)^2 \,.
\end{equation}
Clearly, this constraint renders ${\cal O}_{uddN}$ unobservable for 
any foreseeable collider experiment. This does not, however, imply 
that the particle $\omega_1$ can not appear in other operators. 
Consider the following. In this simple model, BNV arises from the simultaneous presence of the
two couplings, $y_{Nd}^{\omega_1}$ and $y_{ud}^{\omega_1}$, since
there is no baryon number (B) that can be assigned to $\omega_1$ in a
way that preserves B in both interaction terms. For example, if we
assign $B(\omega_1)=1/3$, so that B is respected in the terms of the
first row of eq.~\eqref{eq:bnv}, the yukawa interactions of the second
row violate B. However, if we impose B conservation, either
$y_{Nd}^{\omega_1}$ or $y_{ud}^{\omega_1}$ could be non-zero,
depending on which $B(\omega_1)$ is assigned. In this case, $\omega_1$
can appear in the decomposition of other $N_R$SMEFT operators,
relevant for collider experiments.

The previous discussion can also be extended to decompositions for
$d=7$ operators, although additional subtleties arise as a larger
number of couplings enter in the matching relations. In summary, in
two particle models for $d=7$ operators there are three couplings
entering the matching conditions and usually it is sufficient to
forbid one of the three in order to ensure baryon number conservation.

Finally, we briefly comment on lepton number violation in BNV operators. 
The above discussion has not assumed any value for the lepton number 
of $N_R$, see also the discussion in the next section. If we use the 
standard assignment of $L(N_R)=1$, one can easily see that $d=6$ BNV 
operators violate $B+L$, while conserving $B-L$. Operators at $d=7$, 
on the other hand, violate $B-L$. This follows exactly the same 
pattern as found in SMEFT.

% !TEX root = ../NR_decomp.tex

\section{Lepton number violation (LNV)\label{sec:lnv}}

In this section we will discuss lepton number violation in $N_R$SMEFT
and how it is connected to {\em observable LNV processes} with charged
leptons at the LHC and to the Majorana masses of the SM neutrinos.
Before we turn to $N_R$SMEFT operators, however, it is instructive to
discuss the ``black box theorem'' of neutrinoless double beta decay
($\znbb$ decay) \cite{Schechter:1981bd}. We will recapitulate 
the basics in section \ref{sect:bb}.

In subsection \ref{sect:bbt}, using very similar arguments, we will
discuss how the observation of LNV $@$ LHC in processes involving a
$N_R$ and Majorana neutrino masses are related at the operator level.
For $N_R$SMEFT at $d=7$ we have given the possible decompositions at
tree-level in section \ref{sect:d7}. In subsection \ref{sect:lnvmdls}
we examine Majorana neutrino masses in the renormalisable UV models,
found in section \ref{sect:d7}.

\subsection{The black box theorem \label{sect:bb}}

\begin{figure}[t]
\begin{center}
\includegraphics[width=0.49\linewidth]{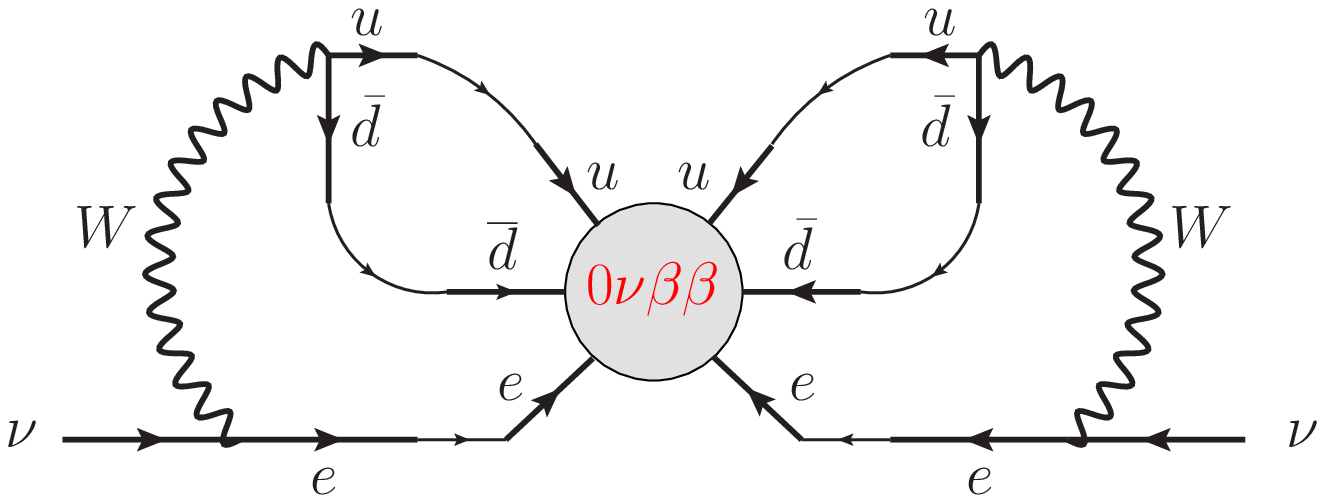}
\includegraphics[width=0.49\linewidth]{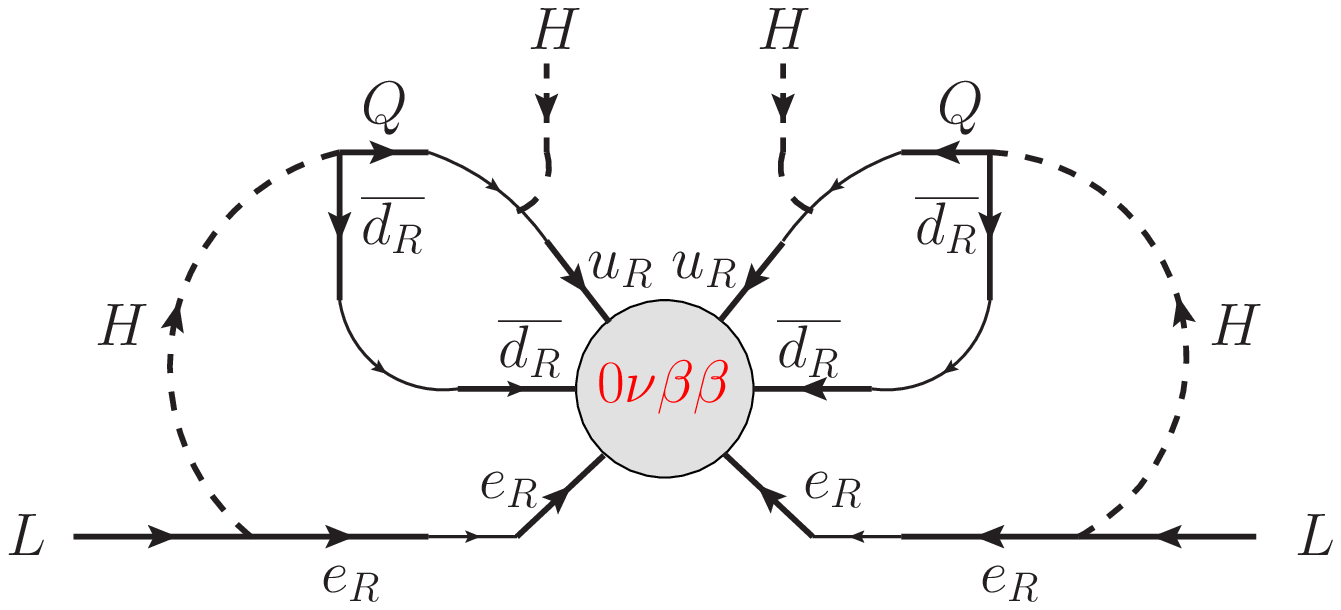}
\caption{Black box theorem of $\znbb$ decay graphically
  \cite{Schechter:1981bd}: Whatever is the underlying mechanism
  generating a non-zero neutrinoless double beta decay amplitude, will
  also generate a Majorana mass term for at least one of the SM
  neutrinos. Cutting the diagrams at the thinner lines leaves an
  operator contributing to $\znbb$ decay. To the left, diagram in the
  mass eigenstate basis. To the right, example diagram for the
  operator ${\cal O}^9_{ude}=u_R^2\overline{d_R}^2 e_R^2$ in the gauge
  basis. }
  \label{fig:bbt}
\end{center}
\end{figure}

It is known for a long time that the observation of $\znbb$ decay will
guarantee that at least one neutrino has a Majorana mass term
\cite{Schechter:1981bd}.  This is usually called the ``black box
theorem'', since the mechanism (or ``model'') underlying the $\znbb$
decay amplitude needs not to be known for this conclusion to hold, see
figure~\ref{fig:bbt}, left: At the quark level $\znbb$ decay is a
$d=9$ operator of the form $(\bar{d}\bar{d}uuee)$.  As the figure
shows, this operator can always be ``dressed'' with $W$-bosons to draw
a diagram that generates a radiative contribution to the Majorana
neutrino mass of a SM neutrino.

$\znbb$ decay is a low-energy process, thus the black box diagram is
usually drawn in the mass eigenstate basis, as done in
figure~\ref{fig:bbt}, left. It is instructive, however, to discuss the
black box theorem in terms of SMEFT operators. The neutrino Majorana
mass matrix is generated from the famous Weinberg operator, ${\cal
  O}_W = LLHH$, while contributions to the $\znbb$ decay amplitude
start at the $d=9$ level. $\Delta L=2$ operators (disregarding
operators with derivatives or field strength tensors) have been listed
up to $d=11$ in \cite{Babu:2001ex}. At $d=9$ there are six operators
relevant for $\znbb$ decay.\footnote{The mass mechanism of $\znbb$
  decay corresponds actually to a $d=11$ operator: 
${\cal O}^{11}_{{\bar Q}^2Q^2L^2H^2}$.}  The simplest six fermion
SMEFT operator for $\znbb$ decay is ${\cal O}^9_{u{\bar d}e}\propto
u_R^2\overline{d_R}^2 e_R^2$. The operator and its black box
connection is shown in figure~\ref{fig:bbt}, right. The SM Yukawa
couplings ${\bar L}e_RH$, ${\bar Q}d_RH$ and $\overline{u_R}QH$ are
used to close the loops. This example results in a 4-loop diagram, the
same as the black box diagram in the mass basis. The other five $d=9$
operators will result either in 2-loop or in 3-loop black box
diagrams.

Note that black box diagrams are in general divergent. Thus, one
expects that neutrino masses are generated at a lower loop level than
indicated by the respective black box diagram. However, at the level
of NROs it is not possible to decide at which level neutrino masses
are indeed generated, for this one needs to know the underlying UV
model. We will come back to this question in subsection
\ref{sect:lnvmdls}.

A 4-loop diagram will give only a tiny contribution to the neutrino
mass \cite{Duerr:2011zd} and, thus, the guaranteed, but minimal
contribution from the $\znbb$ decay black box diagram to the neutrino
masses is numerically much smaller than what is required to explain
neutrino oscillation data. However, given the finite number of
operators at $d=9$, one can open up the $\znbb$ decay operator(s) in
all possible ways, say, at tree-level \cite{Bonnet:2012kh}. The list
of possible UV ``models'' found in this exercise can then be examined
one-by-one and neutrino mass models from the tree- to the 4-loop level
emerge \cite{Helo:2015fba}. Whether any particular of these models can
or can not be the {\em main} contribution to the neutrino masses -- as
required for an explanation of oscillation data -- is then mainly a
question of the loop level at which the neutrino masses are generated
in the respective model, but in all cases a non-zero Majorana mass for
the SM neutrinos is guaranteed.

\subsection{A black box for $N_R$SMEFT and Majorana neutrino
  masses\label{sect:bbt}}

Before turning to $N_R$SMEFT operators, let us briefly discuss the
renormalisable Lagrangian for a simple model that adds a $N_R$ to the
SM. In the minimal type-I seesaw, there is a Majorana mass term and
the Lagrangian involving $N_R$ contains also a Yukawa coupling:
\begin{equation}\label{eq:lagSS1}
{\cal L}^{\rm Type-I} = -y_{\nu} \overline{N_R}L H 
- \frac{1}{2}M_M \overline{N_R^c}N_R + {\rm h.c.} 
\end{equation}
Note that the Majorana nature of the mass term is specific for the
type-I seesaw. The mass term for $N_R$ could also be of Dirac type,
the best-known example is the inverse seesaw \cite{Mohapatra:1986bd}.
Adding a second Weyl fermion to the model, call it $N_L$, one can
write a mass term $M_R \overline{N_L}N_R$ instead of $M_M$ in
eq.~\eqref{eq:lagSS1}. It is obvious that, assigning lepton number
$L(N_R)=1$ (and $L(N_L)=1$), the Yukawa term and $M_R$ conserve lepton
number, while $M_M$ violates $L$ by $\Delta(L)=2$.  However, in the
absence of $y_{\nu}$ one could assign $L(N_R)=0$ and this would
eliminate LNV from $M_M$. This may seem trivial, but we would like to
stress that LNV is always related to a mismatch in the lepton number
assignment of two or more terms in the Lagrangian in any model,
$N_R$SMEFT is no exception to this statement.

\begin{figure}[t]
\begin{center}
  \includegraphics[width=0.45\linewidth]{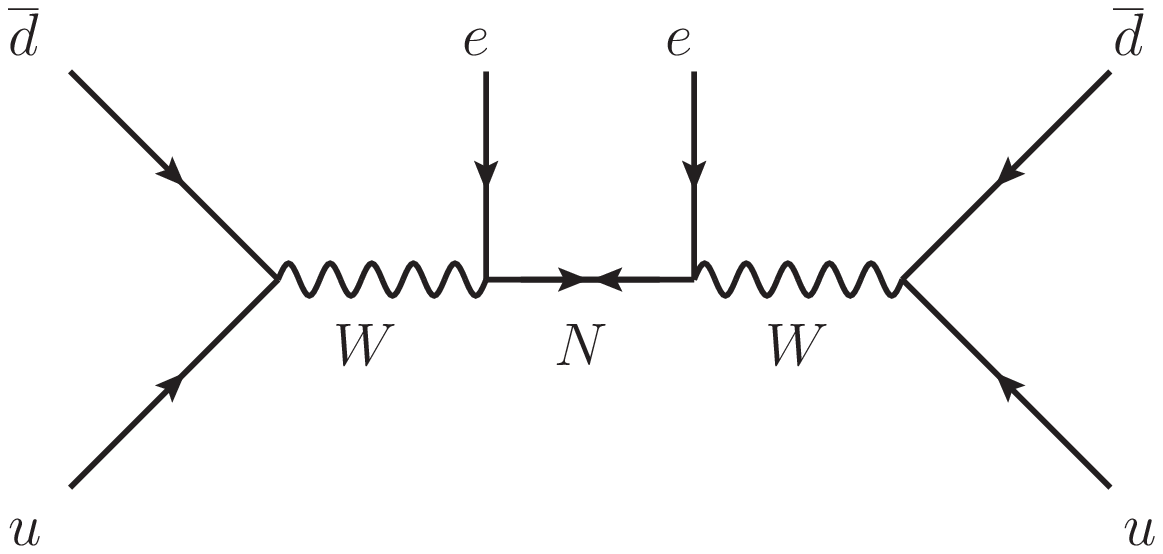}
  \includegraphics[width=0.45\linewidth]{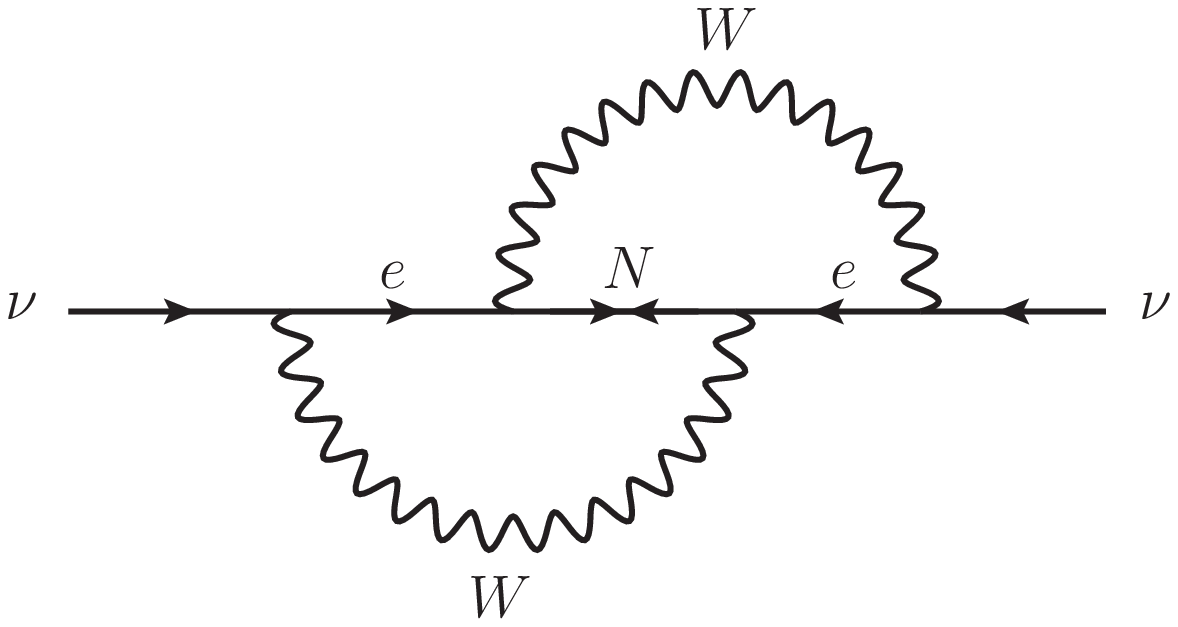}
  \caption{Lepton number violation at the LHC in a minimally extended
    variant of the SM (left) and a 2-loop neutrino mass diagram that will
  necessarily be generated at the same time.}
  \label{fig:bblhc}
\end{center}
\end{figure}

After electro-weak symmetry breaking, the simple model, defined by
eq.~\eqref{eq:lagSS1}, introduces a non-zero coupling of the (mostly)
singlet state $N$ \footnote{$N_R$SMEFT operators put the chirality of
  $N_R$ as right-handed. In the mass eigenstate basis $N$ has both,
  left- and right-handed components, so we do not add the subscript
  $R$ for mass eigenstates.}  to the gauge bosons, $gV_{lN}$, where
$V_{lN}$ is the mixing angle between active and sterile states. At the
LHC $W$-bosons are produced, which can then decay to $N+l$ via this
non-zero coupling.\footnote{At the LHC $l=e,\mu,\tau$, in $\znbb$
  decay only $l=e$, of course. We will not discuss flavour in detail.}
$N$ will decay itself via an off-shell $W$ to another lepton plus
jets. For a Majorana neutrino, the probability to decay to either
lepton or anti-lepton is the same (at tree-level), thus the final
state for the whole process will contain two leptons and two jets and
the ratio $R = \# events$(same-sign di-lepton plus jets)$/\#
events$(opposite-sign di-lepton plus jets) is equal to one.  Same-sign
di-lepton events are obviously $\Delta(L)=2$ processes, formally the
same as a Majorana neutrino mass. The diagram for this LHC process is
shown in figure~\ref{fig:bblhc} on the left. From this diagram, one
can cut off the quarks and draw the 2-loop Majorana neutrino mass
diagram shown on the right of figure~\ref{fig:bblhc}.

Clearly, if the diagram on the left is present in the model, the
diagram on the right must also exist. Thus, the observation of this
LNV process at the LHC guarantees that neutrinos are Majorana states
{\em for this particular model}. Several comments are in order.

First of all, the connection between the LHC process and Majorana
neutrino masses in this particular example can be trivially understood. The model as discussed is, after all, a type-I seesaw 
and type-I seesaw of course generates Majorana neutrino
masses. However, Majorana neutrino masses in the seesaw are generated
at tree-level, thus the 2-loop diagram in figure~\ref{fig:bblhc} just
represents a ``minimal link'' between the LHC process and the
existence of a Majorana neutrino mass. Numerically it is only a very
minor correction to the total neutrino mass in this model. This is
very similar to the black box theorem for $\znbb$ decay discussed
above.

But, different from the black box theorem for $\znbb$ decay, the
discussion as presented so far is {\em model dependent}, since we have
assumed that the Lagrangian terms of eq.~\eqref{eq:lagSS1} exist. To
make statements as model-independent as possible, we should not a
priori assume that $N$ has a coupling to SM $W$-bosons,\footnote{We
  will come back to this point near the end of this subsection.} nor
that the mass of the $N$ is of Majorana type, but instead consider EFT
operators.

In the mass eigenstate basis, after electro-weak symmetry breaking,
the simplest purely fermionic operator one can write down for the
production of a fermion singlet is either ${\bar d}u\overline{N}e$ or
${\bar d}u\overline{N^c}e$.  Assigning lepton number to $N$ as
$L(N)=1$ ($L(N)=-1$), the former (latter) operator is lepton number
conserving, while the latter (former) represents a LNV object.  If
both types of operators are present, LNV processes will be generated -
independent of the nature of the mass term of $N$.  However, $N$ needs
to be a massive particle, its mass term could be either Dirac,
i.e. lepton number conserving (LNC), or Majorana (LNV), as discussed
above. Thus, overall, one can find $\Delta(L)=2$ processes either with
two LNC operators and a Majorana propagator or with one LNV and one
LNC operator, along with a LNC propagator (no mass flip). For the
discussion to be complete, we need to cover both options.

\begin{figure}[t]
\begin{center}
  \includegraphics[width=0.75\linewidth]{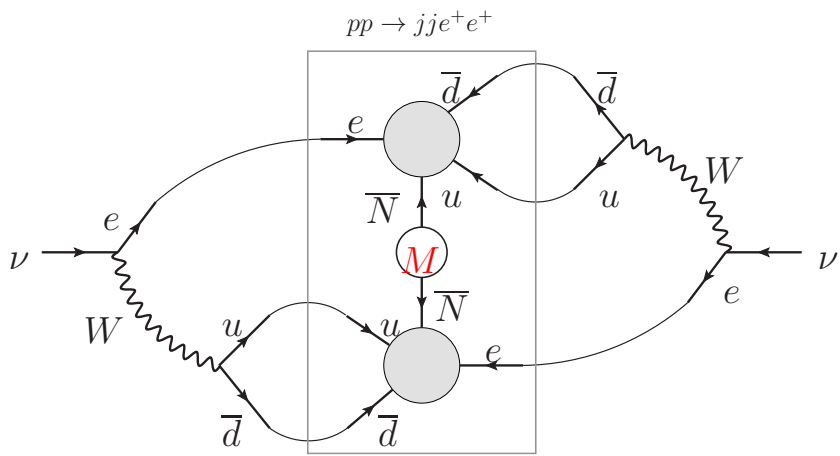}
  \caption{Lepton number violation at the LHC and a 4-loop neutrino
    mass diagram. Different from figure~\ref{fig:bblhc}, this diagram
    does not assume that $N$ has couplings to the $W$-boson.  Instead,
    only electric charge conservation and the existence of some
    operator ${\bar d}u\overline{N}e$ is assumed. Again, the diagram
    is drawn in the mass eigenstate basis and the origin of the LNV is
    assigned to the Majorana mass $M$.}
  \label{fig:bblhc2}
\end{center}
\end{figure}

Figure~\ref{fig:bblhc2} shows, as an example, a 4-loop Majorana
neutrino mass diagram, combining two LNC four-fermion operators with a
LNV Majorana mass insertion in the $N$ propagator. The LHC LNV process
$pp\to l^+l^+ jj$ is contained in the inside of the diagram, just
cutting the diagram at the thin lines.  Thus, the two observables are
always either both present in the theory or none of them is, exactly
as in the originally black box theorem for $\znbb$ decay.

Note that, simple power counting shows that this 4-loop diagram is
divergent. Clearly, a lower order diagram contributing to the neutrino
mass should exist, in order to provide a counter term for this
infinity, and -- apart from some very fine-tuned special cases -- that
lower order diagram will give a larger contribution to the neutrino
mass than the 4-loop diagram. However, given only the effective
operator and not the full UV complete model, it is not possible
to decide at which loop level Majorana masses will appear. We will
come back to discuss this point in section \ref{sect:lnvmdls}.

\begin{figure}[t]
\begin{center}
  \includegraphics[width=0.75\linewidth]{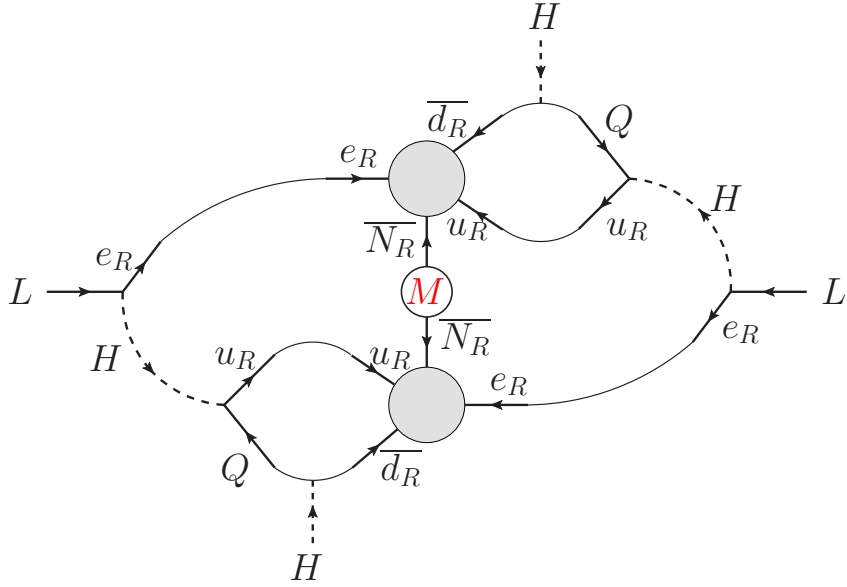}
  \caption{Lepton number violation at the LHC and a 4-loop realisation
    of the Weinberg operator. Different from figure~\ref{fig:bblhc2},
    this diagram is drawn in the electro-weak basis, assuming the
    $d=6$ operator ${\cal O}_{duNe}$ is non-zero.}
  \label{fig:bblhc3}
\end{center}
\end{figure}

Figure~\ref{fig:bblhc2} shows the connection between different LNV
observables in the mass eigenstate basis. The underlying physics,
however, is again more clearly visible in the weak eigenstate
basis. Figure~\ref{fig:bblhc3} shows an example based on the simplest
$d=6$ $N_R$SMEFT operator, ${\cal O}_{duNe}$. Here, the diagram gives
actually a 4-loop realisation of the Weinberg operator, 
${\cal O}_{\rm Wbg}\propto LLHH$ (which will generate the neutrino
mass matrix after symmetry breaking). Similar diagrams can be drawn
for all other single-$N_R$ four-fermion operators in $N_R$SMEFT.

\begin{figure}[t]
\begin{center}
 \includegraphics[width=0.75\linewidth]{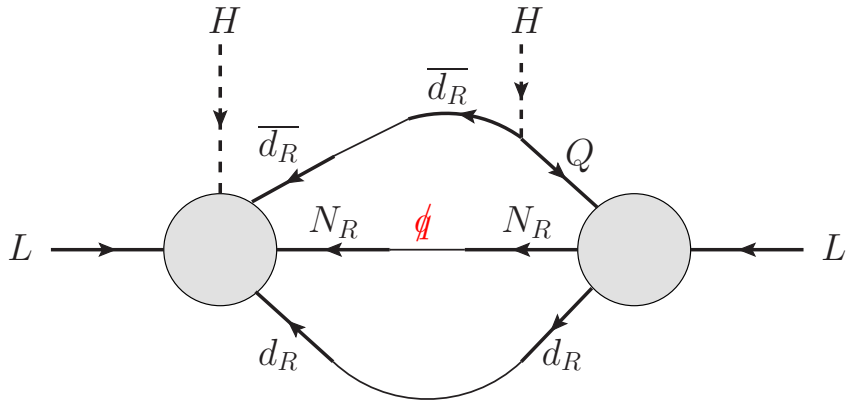}
  \caption{Lepton number violation at the LHC and a 2-loop realisation
    of the Weinberg operator. Different from figure~\ref{fig:bblhc3},
    in this diagram the origin of LNV is the $d=7$ operator
    ${\cal  O}_{dLNH}$.}
  \label{fig:bblhc4}
\end{center}
\end{figure}

In the above discussion it was assumed that the source of the LNV is
the Majorana propagator of the $N_R$. However, the same black box
connection between LNV $@$ LHC and Majorana neutrino mass can be
established also in the case that the necessary LNV is due to a LNV
operator. Let's discuss this in the weak eigenstate basis directly. We
choose as the LNV $d=7$ operator the example of ${\cal O}_{dLNH}$.
(For all other LNV operators the discussion is very similar.) For the
LNC $d=6$ operator we choose ${\cal O}_{dQNL}$.
Figure \ref{fig:bblhc4} shows the resulting 2-loop diagram for 
${\cal O}_{\rm Wbg}$. As before, cutting open the loops defines a LNV
process for the LHC: One could, for example, produce the $N_R$ via
${\cal O}_{dQNL}$, while the final state is produced from the decay of
$N_R$ via ${\cal O}_{dLNH}$. This decay could either contain two jets
plus a Higgs (or $Z^0$) boson and missing momentum, or two jets plus a
$W$ and a charged lepton.\footnote{Only final states {\em without}
  missing energy can be used to determine lepton number
  experimentally, of course.} The additional bosons (relative to the
``standard'' $lljj$ signal) could actually be used to distinguish the
two possibilities (Majorana propagator versus $d=7$ operator) {\em
  experimentally} -- at least in principle.

One can easily check that the chiralities in the diagram in
figure~\ref{fig:bblhc4} are such, that the momentum
$\qslash$ \hspace{-2mm}is picked from the neutrino propagator (no mass
flip).  Again, the resulting integral is divergent, indicating that a
lower order contribution to the neutrino mass should exist in any UV
completion generating this diagram at low energies.  However, whether
the neutrino mass is tree-level or 1-loop can not be decided at the
level of effective operators only.  While we have concentrated here on
a specific combination of operators, the same conclusions can easily
be reached for all other possible combinations: Observation of LNV $@$
LHC guarantees the existence of Majorana neutrino masses for the SM
neutrinos.

Let us return to the comment, stated above, that for generality of our
argument we should not assume that $N$ has a non-zero coupling to
gauge bosons. This statement is motivated by the fact that
experimentally it might not be possible to show that the vertex
$e$-$N$-$W$ exists. However, consider the following: $H$ and $L$ can
always be coupled to a $SU(2)$ singlet, thus, if a $N_R$ is present in
the theory one can always write down a yukawa term
$y_{\nu}\overline{N_R}LH$ -- which is equivalent to a non-zero
$V_{lN}$ in the broken phase.  One might attempt to forbid this
coupling via some extra symmetry beyond those of the SM. An example
could be a $Z_2$ symmetry, under which the $N_R$ is odd, such as in
the famous ``scotogenic'' neutrino mass model
\cite{Ma:2006km}. However, such an extra symmetry is incompatible with
the existence of any of the single-$N_R$ operators in tables
\ref{tab:Opd6} and \ref{tab:Opd7}. This is easy to see: Consider, for
example ${\cal O}_{QuNL}$. We can take this operator and replace
$\overline{Q}u_R$ by $H$ (up type yukawa coupling do exist after all),
thus writing down a term proportional to $\overline{N_R}LH$. Since
similar replacements can be done for any of the single-$N_R$
operators, the observation of any of these will guarantee that some
(although maybe very small) coupling to gauge bosons should be present
in the model as well. We stress, however, that as with the black box,
this argument is purely qualitative. It does not allow to fix the
numerical value of $y_{\nu}$. In particular, both production and decay
of $N_R$ at the LHC could easily be dominated by NRO operators.

We close this subsection with a brief comment about flavour.  While in
double beta decay the charged leptons are always electrons, the LHC
can produce, in principle, any lepton flavour.  Just as $\znbb$ decay
guarantees that the $(m)_{ee}$ entry of the neutrino mass matrix is
non-zero, the observation of different flavour combinations
($\alpha,\beta$) in LNV processes at the LHC would then be related to
the Majorana neutrino mass matrix entry $(m)_{\alpha\beta}$ in the
gauge basis.

\subsection{Neutrino masses in models derived from LNV $d=7$ operators 
\label{sect:lnvmdls}}

In this subsection, we will briefly discuss neutrino mass generation
at the renormalisable level. The aim of this discussion is not to
provide a detailed fit of neutrino masses and mixing angles to
experimental data,\footnote{Once a particular model is specified,
  neutrino fits can be easily done using, for example, the formulas in
  \cite{Cordero-Carrion:2019qtu}.}  but rather to demonstrate that all
models generating LNV operators with $N_R$ also generate active
neutrino masses.\footnote{To simplify the discussion below, we assume
  the $d=7$ operators violate $L$, see previous section.} Given this
connection, one might be tempted to think that $d=7$ operators are not
observable in accelerator experiments, due to the smallness of the
observed neutrino masses. However, as we will discuss now, such a
conclusion holds only for a very specific subset of UV decompositions
and not in general.

First of all, with $N_R$ being a complete SM singlet, all models
giving rise to single-$N_R$ $d=6$ operators necessarily allow to write
down also a neutrino yukawa coupling, as discussed above. If, in
addition, lepton number is violated, also a Majorana mass for $N_R$ is
allowed. In fact, the Majorana mass term is mandatory in this case:
One can easily show that in all $N_R$ models with LNV, $M_M$ is
generated radiatively via diagrams with divergent integrals. Thus, for
consistency, all these models require also the presence of a
tree-level $M_M$ as a counter term. A seesaw type-I contribution to
the neutrino masses is therefore unavoidable in all models with a
$N_R$ and LNV. For the discussion in this subsection, however, this
contribution to the neutrino masses is irrelevant, since it does not
place any restrictions on the Wilson coefficients of the $d=7$
operators.

\begin{figure}[t]
\begin{center}
 \includegraphics[width=0.49\linewidth]{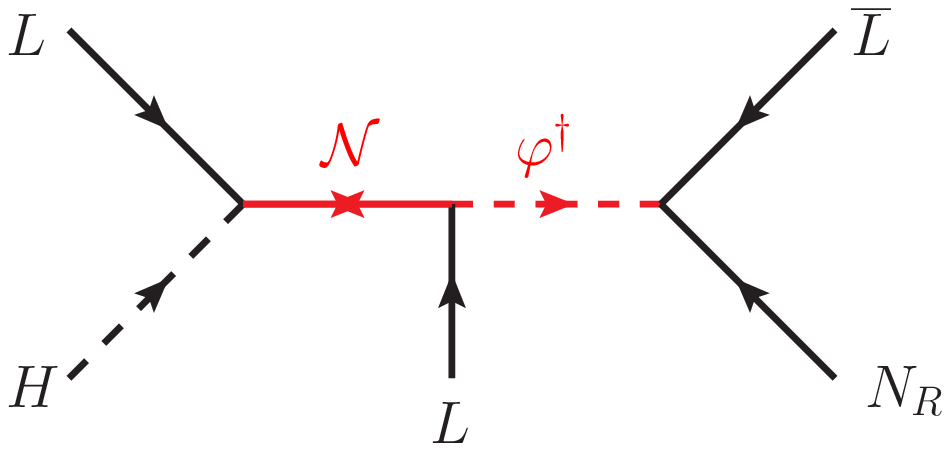}
 \includegraphics[width=0.49\linewidth]{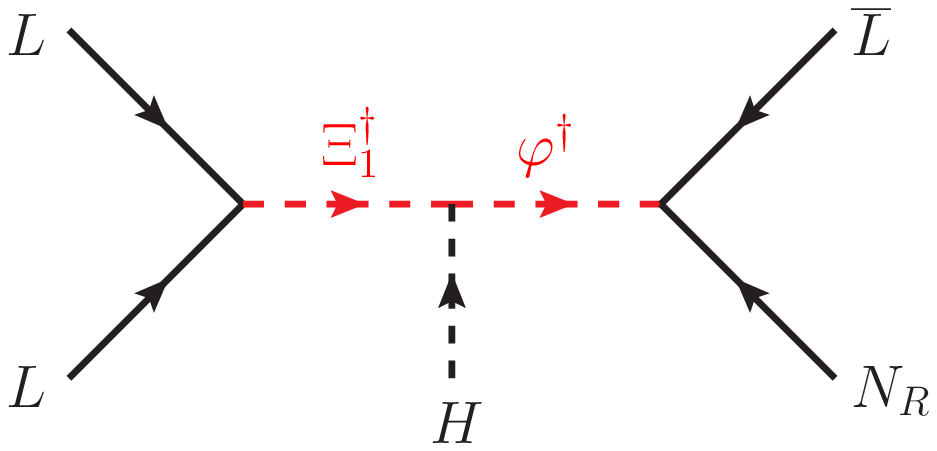}
 \includegraphics[width=0.49\linewidth]{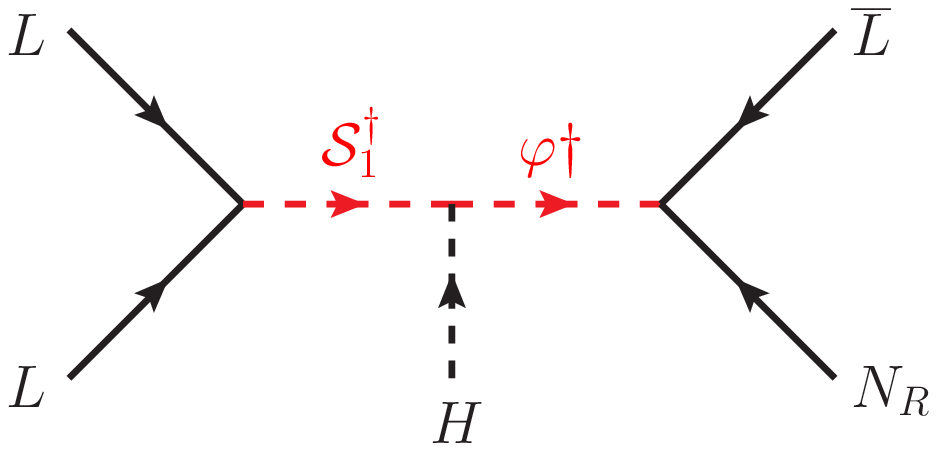}
  \includegraphics[width=0.49\linewidth]{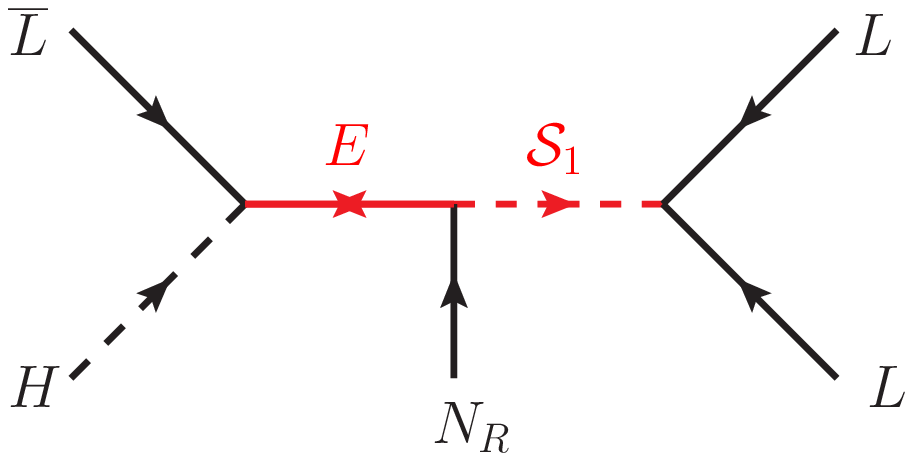}
 \caption{Four example decompositions for ${\cal O}_{LNLH}$. The two
   examples shown in the top row will give tree-level contributions to
   the active neutrino masses via type-I (left) or type-II seesaw
   (right).  The decomposition in the bottom generate radiative
   neutrino masses at 1-loop (left) and 2-loops (right). The quantum
   numbers for the new fields are given in tables~\ref{tab:newS} and
   \ref{tab:newF}.}
  \label{fig:OLNLH}
\end{center}
\end{figure}

As we pointed out in section~\ref{sect:d7}, we found 112 different
models for $d=7$ operators. We will not discuss all these models in
details, but instead focus on just four decompositions for the example
operator ${\cal O}_{LNLH}$, see figure~\ref{fig:OLNLH}.  These four
examples are sufficient to cover essentially all relevant aspects of
neutrino mass generation in the 112 models: The remaining models could
be discussed in much the same way, with adequate replacements for the
corresponding model parameters.

In figure~\ref{fig:OLNLH} we show in the top row two example
decomposition for ${\cal O}_{LNLH}$ containing ${\cal N}$ and
$\Xi_1$. These will give a tree-level seesaw contribution of type-I
and type-II to the neutrino masses. (Note, that one can replace ${\cal
  N}$ by $\Sigma$, and obtain a decomposition with a type-III seesaw
too.) We note in passing that out of the 112 models (14, 8, 8) contain
(${\cal N},\Sigma,\Xi_1$), respectively.  The decompositions in the
bottom row, on the other hand, will generate neutrino masses
radiatively.

Let us discuss the tree-level cases first.  Consider the example
decomposition shown in figure~\ref{fig:OLNLH}, top left.
The Lagrangian contains the terms:
\begin{equation}\label{eq:LagI}
{\cal L} \propto y_{NL}^{\varphi}\left(\overline{N_R}L\right)\varphi
               + y_{{\cal N}L}^{\varphi}\left(\overline{\mathcal{N}}L\right)\varphi
               + y_{{\cal N}L}\left(\overline{\cal N}L\right) H
               + \frac{1}{2}M_{\cal N} \overline{{\cal N}^c}{\cal N}
               + {\rm h.c.}
\end{equation}
where ${\cal N}$ and $\varphi$ are ``heavy'' copies of $N_R$ and
$H$.\footnote{Notice that we have omitted the explicit contraction of
  indices. To simplify the notation we omit them in the rest of the
  manuscript.}  The Wilson coefficient generated from this diagram is
matched via
\begin{equation}\label{eq:mtcM}
c_{LNLH} = -\frac{1}{4}
 \frac{y_{NL}^{\varphi} y_{{\cal N}L}^{\varphi}y_{{\cal N}L}}{M_{\cal N}m_\varphi^2}.
\end{equation}
On the other hand, ${\cal N}$ must be a Majorana field, otherwise 
the diagram can not be closed. Thus, ${\cal N}$ gives a contribution 
to the neutrino mass {\it \'a la} seesaw type-I:
\begin{equation}\label{eq:ssI}
m_{\nu} \propto y_{{\cal N}L}^2\frac{v^2}{M_{\cal N}}.
\end{equation}
We can use this equation to put an upper limit on the coefficient
$c_{LNLH}$:
\begin{equation}\label{eq:limC}
c_{LNLH} \lsim 10^{-6} \frac{y_{NL}^{\varphi} y_{{\cal N}L}^{\varphi}}{\Lambda^3}
           \left(\frac{\Lambda}{v}\right)^{1/2}
              \left(\frac{m_{\nu}}{0.1 {\,\rm eV}}\right),
\end{equation}
where we have assumed $M_{\cal N} \simeq m_\varphi\simeq \Lambda$.
This estimate shows that production of a $N_R$ via this operator in a
lepton collider is completely negligible for this model. We can also
estimate the partial decay width of a $N_R$ via ${\cal O}_{LNLH}$
given this constraint. We find the decay length is roughly
\begin{equation}\label{eq:limLen}
(c\tau) \sim \left(\frac{\Lambda}{{\rm TeV}}\right)^{5} 
          \left(\frac{m_{N_R}}{{\rm 10 \hskip1mm GeV}}\right)^{-5} 
          \left(\frac{0.1 {\,\rm eV}}{m_{\nu}}\right)^2 10^8 {\,\rm m},
\end{equation}
for $y_{NL}^{\varphi}= y_{{\cal N}L}^{\varphi}=1$.  A decay length this
large would render the $N_R$ essentially stable for collider
experiments, unless it is much heavier than indicated in
eq.~\eqref{eq:limLen}. Note that, similar arguments can be presented
for all decompositions of $d=7$ operators containing either ${\cal N}$
or $\Sigma$.

For decompositions containing $\Xi_1$, the situation is slightly more
complicated. If we allow for LNV, the Lagrangian for a model with a
$\Xi_1$ field contains the terms:
\begin{equation}\label{eq:LagII}
{\cal L} \propto y^{\Xi_1}_{L} \left(\overline{L^c}L\right)\Xi_1 + \kappa_{\Xi_1} HH \Xi_1^\dagger + \kappa_{\Xi_1\varphi} \Xi_1^{\dagger}H\varphi + \ldots
\end{equation}
The simultaneous presence of both terms will lead to a seesaw type-II
contribution to the active neutrino mass matrix. We can use this to
rewrite the Wilson coefficient for the decomposition shown in
figure~\ref{fig:OLNLH}, top right as:
\begin{eqnarray}\label{eq:limC2}
|c_{LNLH}| &\lsim & y_{{N}L}^{\varphi}
                 \frac{m_{\nu}}{v_{\Xi_1}}\frac{1}{\Lambda^3}
\\ \label{eq:limC3}
       & \lsim &    10^{-10} y_{{N}L}^{\varphi} 
                   \left(\frac{m_{\nu}}{0.1 \,{\rm eV}}\right)
                   \left(\frac{\rm GeV}{v_{\Xi_1}}\right)\frac{1}{\Lambda^3}.
\end{eqnarray}
where we have assumed
$m_{\Xi_1}=m_{\varphi}=\kappa_{\Xi_1\varphi}=\Lambda$. The SM
$\rho$-parameter puts an upper limit on the induced vacuum expectation
value of the triplet, $v_{\Xi_1}$, of roughly $v_{\Xi_1}\lsim 2$ GeV
\cite{ParticleDataGroup:2022pth}, which motivates the stringent
constraint eq.~\eqref{eq:limC3}.  Neutrino oscillation data, however,
allow $v_{\Xi_1}$ as small as $v_{\Xi_1}\sim 0.1$ eV. Obviously, no
numerically relevant constraint on $|c_{LNLH}|$ can be derived in this
case.

\begin{figure}[t]
\begin{center}
 \includegraphics[width=0.44\linewidth]{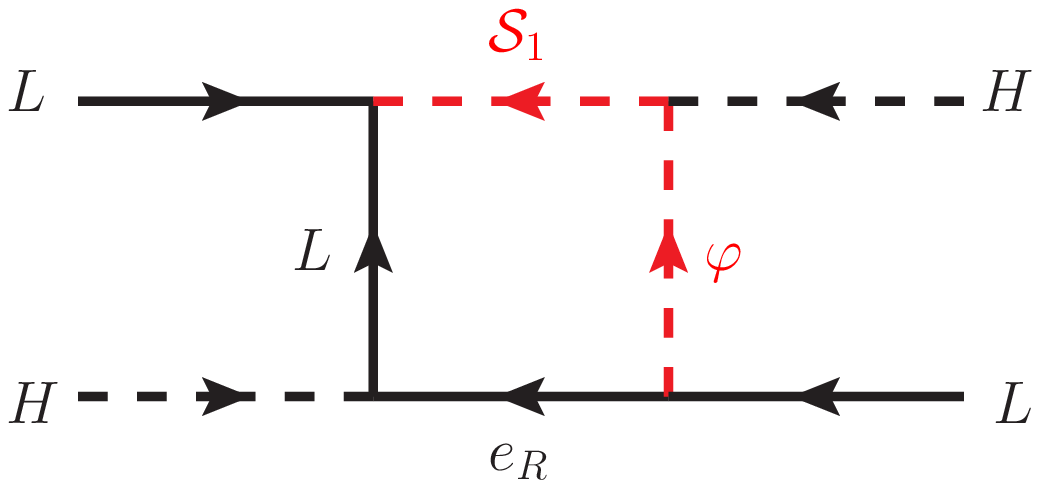}
 \includegraphics[width=0.53\linewidth]{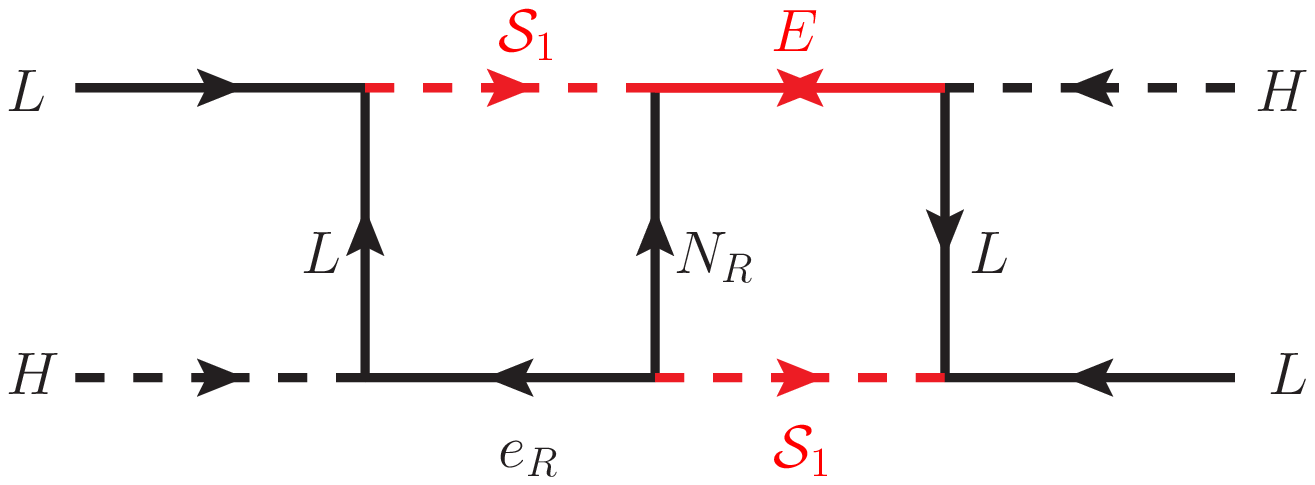}
 \caption{One-loop (left) and two-loop (right) neutrino mass diagrams, 
based on the decomposition of ${\cal O}_{LNLH}$ containing the BSM 
particles (${\cal S}_1,\varphi$) or ($E,{\cal S}_1$).}
  \label{fig:OLNLHMnu}
\end{center}
\end{figure}

Let us turn now to decomposition \#3 in figure~\ref{fig:OLNLH}, 
(${\cal S}_1,\varphi$). The same particle content appears also 
in decompositions of the operators ${\cal O}_{eNLH}$, 
${\cal O}_{LNeH}$,  ${\cal O}_{dQNeH}$ and  ${\cal O}_{QuNeH}$. 
In neutrino physics this combination of BSM particles 
is known as the Zee model \cite{Zee:1980ai}. Neutrino masses are 
generated at 1-loop level, see figure~\ref{fig:OLNLHMnu} to the 
left. The Lagrangian of the Zee model containts the terms:
\begin{equation}\label{eq:LagZee}
{\cal L} \propto y^{\mathcal{S}_1}_L \left(\overline{L^c}L\right) {\cal S}_1 +  y_{Ne}^{\mathcal{S}_1} \left(\overline{N_R^c}e_R\right) {\cal S}_1  + y_{eL}^{\varphi}\left(\overline{e_R}L\right)\varphi^\dagger 
+ \kappa_{\mathcal{S}_1\varphi} {\cal S}_1^\dagger H \varphi + {\rm h.c.} + \ldots
\end{equation}
Disregarding flavour indices for simplicity, the neutrino mass in the
Zee model can be estimated as:
\begin{equation}
  m_{\nu}^{\rm Zee} \simeq - \frac{1}{16\pi^2}  y_L^{\mathcal{S}_1} m_\tau y_{eL}^{\varphi} 
           \frac{\sqrt{2}v\kappa_{\mathcal{S}_1\varphi}}{m_{h_2^+}^2 - m_{h_1^+}^2}
   \log \left( \frac{m_{h_2^+}^2}{m_{h_1^+}^2}\right) \, .
\label{eq:numassZ}
\end{equation}
Here, $h_i$ are the two mass eigenstates formed by ${\cal S}_1$ and
the charged component in $\varphi$.  $m_\tau$ is the mass of the
$\tau$ lepton and we have neglected terms proportional to $m_{\mu,e}$,
which are not relevant for this discussion. Thus, neutrino masses will
put a stringent constraint on the product $y_L^{\mathcal{S}_1}
y_{eL}^{\varphi}(\kappa_{\mathcal{S}_1\varphi}/\Lambda)$, where
$\Lambda \simeq m_{\mathcal{S}_1}\simeq m_{\varphi}$. Logically, this
combination could be small because one, two or all three parameters
are suppressed. The matching of the operators, ${\cal O}_{LNLH}$,
${\cal O}_{eNLH}$, ${\cal O}_{LNeH}$, on the other hand, will be
proportional to:
\begin{align}\label{eq:MatchZee}
c_{LNLH} & \propto y_L^{\mathcal{S}_1} y_{NL}^{\varphi} \frac{\kappa_{\mathcal{S}_1\varphi}}{\Lambda}
\nonumber \\[0.2em] 
c_{eNLH} &\propto y_{Ne}^{\mathcal{S}_1} y_{eL}^{\varphi}\frac{\kappa_{\mathcal{S}_1\varphi}}{\Lambda}
\nonumber \\[0.2em]
c_{LNeH} & \propto y_{Ne}^{\mathcal{S}_1}y_{NL}^{\varphi}\frac{\kappa_{\mathcal{S}_1\varphi}}{\Lambda} \,.
\end{align}
In case all three parameters entering the neutrino mass are small, 
$y_L^{\mathcal{S}_1}\sim y_{eL}^{\varphi}\sim (\kappa_{\mathcal{S}_1\varphi}/\Lambda)\sim \epsilon$,
all of the coefficients in eq.~\eqref{eq:MatchZee} will be suppressed. 
However, in the more optimistic case, where either $y_L^{\mathcal{S}_1}$ or 
$y_{eL}^{\varphi}$ are suppressed by $\epsilon^3$, while the other two 
parameters are order ${\cal O}(1)$, either $c_{LNLH}$ or 
$c_{eNLH}$ and $c_{LNeH}$ can be large. However, given the constraint 
from neutrino masses, it is impossible that all three operators 
are observable at the same time.

Consider now decomposition \#4 in figure~\ref{fig:OLNLH}, ($E, \mathcal{S}_1$). 
This model allows to write the following terms in the Lagrangian:
\begin{equation}\label{eq:LagExa}
{\cal L} \propto y^{\mathcal{S}_1}_L \left(\overline{L^c}L\right) {\cal S}_1 + y_{Ne}^{\mathcal{S}_1} \left(\overline{N_R^c}e_R\right) \mathcal{S}_1 + y_{LE} \left(\overline{L}E\right) H + y_{NE_L}^{\mathcal{S}_1}\left(\overline{N_R}E\right)\mathcal{S}_1 + m_E \overline{E}E +\ldots
\end{equation}
Note that the vertex proportional to $y_{Ne}^{\mathcal{S}_1}$ does not appear in
figure~\ref{fig:OLNLH}, but it is contained in a decomposition for
${\cal O}_{LNeH}$ with the same particle content.  Also,
$y_{Ne}^{\mathcal{S}_1}$ is necessary for the 2-loop diagram in
figure~\ref{fig:OLNLHMnu}. In this diagram, from the $N_R$ propagator
$P_R (\,\qslash \hspace{-2mm}+ M_{N_R})P_L$ the momentum term survives. LNV is due to
the simultaneous presence of $y_{Ne}^{\mathcal{S}_1}$ and
$y_{NE_L}^{\mathcal{S}_1}$. Similar to the discussion for the Zee model, either
$y_{Ne}^{\mathcal{S}_1}$ or $y_{NE_L}^{\mathcal{S}_1}$ (or both) must be small, to fulfill 
the neutrino mass constraint. Thus, either ${\cal O}_{LNLH}$ or 
${\cal O}_{LNeH}$ can have unsupressed Wilson coefficients, but not 
both operators at the same time. 

We close this section summarizing: (1) Models for LNV in $d=7$
$N_R$SMEFT operators will always also lead to active neutrino masses
either at tree-, 1- or 2-loop. (2) For decompositions leading to
seesaw type-I or type-III contributions to the neutrino mass, the
Wilson coefficients will be severely suppressed, rendering the
corresponding operators phenomenologically irrelevant for accelerator
experiments. (3) For decompositions leading to radiative neutrino
masses, the Wilson coefficients for some of the corresponding
operators can be large, but the same UV-decomposition contributes
typically to more than one operator and neutrino mass constraints
exclude the possibility that all corresponding operators are
observable at the same time.

% !TEX root = ../NR_decomp.tex

\section{Conclusions\label{sec:cncl}}

Right-handed neutrinos with electro-weak scale masses have recently
attracted a lot of attention in the literature. From the theoretical
point of view, $N_R$'s represent the simplest extension of the
standard model that can explain the active neutrino masses as observed
in oscillation experiments (via some variant of the seesaw
mechanism). From the experimental side, in the past few years a number
of new experiments have been proposed to search for long-lived
particles with unprecedented sensitivities. $N_R$'s with masses around
(1-100) GeV are prime candidates for long-lived particles, due to the
smallness of the active neutrino masses.

If new physics exists, but at a mass scale outside the reach of the 
LHC, effective field theory is the correct tool to study BSM. 
The relevant EFT involving right-handed neutrinos is $N_R$SMEFT 
and a number of recent papers have studied the phenomenology of 
$N_R$SMEFT.

In this work we discussed a systematic tree-level decomposition of
$N_R$SMEFT operators at $d=6$ and $d=7$, using a diagrammatic
method. The resulting lists of BSM particles provide a complete
dictionary of models, which can be used for studying $N_R$
phenomenology.  We have also briefly compared our lists of particles
to the Granada dictionary for tree-level UV-completions for SMEFT at
$d=6$ \cite{deBlas:2017xtg}.  Our lists of BSM particles are given in
tables~\ref{tab:newS}~-~\ref{tab:newV}. In the appendix we give
the Lagrangian terms for the resulting models for all terms involving
$N_R$. These Lagrangians were calculated with \texttt{Sym2Int}
\cite{Fonseca:2017lem,Fonseca:2019yya}. In the auxiliary file added
to this paper, we give all remaining terms involving the BSM fields 
necessary for the calculation of the matching of the UV models onto 
$N_R$SMEFT. The matching can be done automatically with the help of, 
for example, \texttt{Matchete} \cite{Fuentes-Martin:2022jrf} and we 
added an example notebook for a number of models as an auxiliary 
file to this paper.

We also discussed lepton number violation, that unavoidably appears if
$d=6$ and $d=7$ operators are present in the theory at the same
time. LNV is always linked to Majorana neutrino masses and LNV in
$N_R$SMEFT is no exception, as we discussed in detail.  We also
discussed possible constraints on $d=7$ operators from the observed
neutrino masses. While some of the possible UV-models for $d=7$
operators must have tiny Wilson coefficients, due the neutrino mass
constraint, there exist many models for which $d=7$ $N_R$SMEFT
operators could be observable in future LLP experiments.

% !TEX root = ../NR_decomp.tex

\appendix

\section{Lagrangian \label{sec:app}}

The Lagrangian describing renormalisable interactions among the SM fields, the singlet $N_R$, and the new BSM fields introduced in tables~\ref{tab:newS}~-~\ref{tab:newV}, can be expressed as
\begin{align*}
\mathcal{L}_{\text{UV}} = \mathcal{L}_{light}  + \mathcal{L}_{mixed} + \mathcal{L}_{heavy} \,.
\end{align*}
The first term includes interactions involving only the SM fields and the $N_R$. The second term describes the interactions between these light fields and the heavy BSM fields, while the last term contains the interactions of the heavy fields among themselves. 

In this appendix we write down all renormalisable terms in which the light singlet $N_R$ is involved. Additionally, we provide the Lagrangian terms that include the new vector field $\mathcal{U}_1$ and new interactions for the vector $\mathcal{L}_1$ not considered in Ref.~\cite{deBlas:2017xtg}. Finally, we present the interaction terms belonging to $\mathcal{L}_{heavy}$ that contribute to the model diagrams leading to $N_R$SMEFT operators at $d=7$.

The Lagrangian $\mathcal{L}_{light}$ comprises the renormalisable SM Lagrangian, the mass term for $N_R$,\footnote{Note that we have only written a Majorana mass term for $N_R$. Further discussion regarding this aspect is provided in section~\ref{sec:lnv}.} and the allowed Yukawa interaction for the singlet. It is expressed as\footnote{Explicit $SU(2)$ and $SU(3)$ index contractions have been omitted in the Lagrangians of this appendix.}

\begin{equation}
\mathcal{L}_{light} = \mathcal{L}_{\text{SM}} - \frac{1}{2} M_{\text{M}} \overline{N_R^c} N_R - y_{\nu} \left(\overline{N_R} L \right) H + \text{h.c.}
\end{equation}

The interactions involving $N_R$ in $\mathcal{L}_{mixed}$ can be classified into two categories: fermion-fermion-scalar (\textit{FFS}) and fermion-fermion-vector (\textit{FFV}) interactions. Each category includes two types of terms: A) those with one light field ($N_R$) and two heavy fields, and B) those with two light fields (either one or two $N_R$) and one heavy field. We gather all these interactions in $\mathcal{L}_{mixed}^{N_R}$, which is given by

\begin{equation}
\mathcal{L}_{mixed}^{N_R} = \mathcal{L}_{\scriptscriptstyle FFS,(A)}^{N_R} + \mathcal{L}_{\scriptscriptstyle FFS,(B)}^{N_R} + \mathcal{L}_{\scriptscriptstyle FFV,(A)}^{N_R} + \mathcal{L}_{\scriptscriptstyle FFV,(B)}^{N_R} \,,
\end{equation}
where
{
\allowdisplaybreaks
\begin{align}
\mathcal{L}_{\scriptscriptstyle FFS,(A)}^{N_R}  = & \;   y_{N\mathcal{N}_R}^\mathcal{S} \left( \overline{N_R^c} \mathcal{N}_R \right) \mathcal{S} + y_{N\mathcal{N}_{L}}^\mathcal{S} \left( \overline{N_R} \mathcal{N}_L \right) \mathcal{S} \nonumber \\
& + y_{N E_R}^{\mathcal{S}_1} \left( \overline{N_R^c} E_R \right) \mathcal{S}_1 +  y_{N E_{L}}^{\mathcal{S}_1} \left( \overline{N_R} E_L \right) \mathcal{S}_1  \nonumber \\
& + y_{N \Delta_{1R}}^\varphi \left( \overline{N_R^c} \Delta_{1R} \right) \varphi +  y_{N \Delta_{1L}}^\varphi \left( \overline{N_R} \Delta_{1L} \right) \varphi \nonumber \\
& + y_{N \Sigma_{R}}^{\Xi} \left( \overline{N_R^c} \Sigma_{R} \right) \Xi +  y_{N \Sigma_{L}}^\Xi \left( \overline{N_R} \Sigma_{L} \right) \Xi \nonumber \\
&+  y_{N \Sigma_{1R}}^{\Xi_1} \left( \overline{N_R^c} \Sigma_{1R} \right) \Xi_1 +  y_{N \Sigma_{1L}}^{\Xi_1} \left( \overline{N_R} \Sigma_{1L} \right) \Xi_1 \nonumber \\
& + y_{N D_R}^{\omega_1} \left( \overline{N_R^c} D_R \right) \omega_1^\dagger + y_{N D_L}^{\omega_1} \left( \overline{N_R} D_L \right) \omega_1^\dagger \nonumber \\
& + y_{N U_R}^{\omega_2} \left( \overline{N_R^c} U_R \right) \omega_2^\dagger + y_{N U_L}^{\omega_2} \left( \overline{N_R} U_L \right) \omega_2^\dagger \nonumber\\
& + y_{N Q_{1R}}^{\Pi_1} \left( \overline{N_R^c} Q_{1R} \right) \Pi_1^\dagger + y_{N Q_{1L}}^{\Pi_1} \left( \overline{N_R} Q_{1L} \right) \Pi_1^\dagger \nonumber \\
& + y_{N Q_{7R}}^{\Pi_7} \left( \overline{N_R^c} Q_{7R} \right) \Pi_7^\dagger + y_{N Q_{7L}}^{\Pi_7} \left( \overline{N_R} Q_{7L} \right) \Pi_7^\dagger \nonumber \\
& + y_{N T_{1R}}^{\zeta} \left( \overline{N_R^c} T_{1R} \right) \zeta^{\dagger} + y_{N T_{1L}}^{\zeta} \left( \overline{N_R} T_{1L} \right) \zeta^{\dagger}  + \text{h.c.} \,,	\\[10pt]
\mathcal{L}_{\scriptscriptstyle FFS,(B)}^{N_R}  =  &  \; y_{NN}^\mathcal{S} \left( \overline{N_R^c} N_R \right) \mathcal{S} + y_{Ne}^{\mathcal{S}_1} \left( \overline{N_R^c} e_R \right) \mathcal{S}_1 + y_{NL}^\varphi \left( \overline{N_R} L \right) \varphi \nonumber \\
& + y_{Nd}^{\omega_1} \left( \overline{N_R^c} d_R \right) \omega_1^\dagger + y_{Nu}^{\omega_2} \left( \overline{N_R^c} u_R \right) \omega_2^\dagger + y_{QN}^{\Pi_1} \left( \overline{Q} N_R \right) \Pi_1 \nonumber \\
&  +  y_{N \Delta_{1R}} \left( \overline{N_R^c} \Delta_{1R} \right) H +  y_{N \Delta_{1L}} \left( \overline{N_R} \Delta_{1L} \right) H \text{+ h.c.} \,, \\[10pt]
\mathcal{L}_{\scriptscriptstyle FFV,(A)}^{N_R}  = &  \; g_{N\mathcal{N}_R}^\mathcal{B} \left( \overline{N_R} \gamma_\mu \mathcal{N}_R \right) \mathcal{B}^\mu + g_{N\mathcal{N}_{L}}^\mathcal{B} \left( \overline{N_R^c} \gamma_\mu \mathcal{N}_L \right) \mathcal{B}^\mu \nonumber \\
& + g_{N E_R}^{\mathcal{B}_1} \left( \overline{N_R} \gamma_\mu E_R \right) \mathcal{B}_1^\mu +  g_{N E_{L}}^{\mathcal{B}_1} \left( \overline{N_R^c} \gamma_\mu E_L \right) \mathcal{B}_1^\mu \nonumber \\
& + g_{N \Delta_{1R}}^{\mathcal{L}_1} \left( \overline{N_R} \gamma_\mu \Delta_{1R} \right) \mathcal{L}_1^{\mu} +  g_{N \Delta_{1L}}^{\mathcal{L}_1}\left( \overline{N_R^c} \gamma_\mu \Delta_{1L} \right) \mathcal{L}_1^{\mu } \nonumber \\
& + g_{N \Delta_{3R}}^{\mathcal{L}_3} \left( \overline{N_R} \gamma_\mu \Delta_{3R} \right) \mathcal{L}_3^{\mu\dagger} +  g_{N \Delta_{3L}}^{\mathcal{L}_3}\left( \overline{N_R^c} \gamma_\mu \Delta_{3L} \right) \mathcal{L}_3^{\mu \dagger} \nonumber \\
& + g_{N \Sigma_{R}}^{\mathcal{W}} \left( \overline{N_R} \gamma_\mu \Sigma_{R} \right) \mathcal{W}^{\mu} +  g_{N \Sigma_{L}}^{\mathcal{W}} \left( \overline{N_R^c} \gamma_\mu \Sigma_{L} \right) \mathcal{W}^{\mu} \nonumber \\
& + g_{N \Sigma_{1R}}^{\mathcal{W}_1} \left( \overline{N_R} \gamma_\mu \Sigma_{1R} \right) \mathcal{W}_1^{\mu } +  g_{N \Sigma_{1L}}^{\mathcal{W}_1} \left( \overline{N_R^c} \gamma_\mu \Sigma_{1L} \right) \mathcal{W}_1^{\mu} \nonumber \\
& + g_{N D_R}^{\mathcal{U}_1} \left( \overline{N_R} \gamma_\mu D_R \right) \mathcal{U}_1^{\mu \dagger} + g_{N D_L}^{\mathcal{U}_1} \left( \overline{N_R^c}\gamma_\mu  D_L \right) \mathcal{U}_1^{\mu \dagger} \nonumber\\
& + g_{N U_R}^{\mathcal{U}_2} \left( \overline{N_R} \gamma_\mu U_R \right) \mathcal{U}_2^{\mu \dagger} + g_{N U_L}^{\mathcal{U}_2} \left( \overline{N_R^c} \gamma _\mu U_L \right) \mathcal{U}_2^{\mu \dagger} \nonumber\\
& + g_{N Q_{1R}}^{\mathcal{Q}_1} \left( \overline{N_R} \gamma_\mu Q_{1R} \right) \mathcal{Q}_1^{\mu \dagger} + g_{N Q_{1L}}^{\mathcal{Q}_1} \left( \overline{N_R^c} \gamma_\mu  Q_{1L} \right) \mathcal{Q}_1^{\mu \dagger}\nonumber  \\
& + g_{N Q_{5R}}^{\mathcal{Q}_5} \left( \overline{N_R} \gamma_\mu Q_{5R} \right) \mathcal{Q}_5^{\mu \dagger} + g_{N Q_{5L}}^{\mathcal{Q}_5} \left( \overline{N_R^c} \gamma_\mu  Q_{5L} \right) \mathcal{Q}_5^{\mu \dagger} \nonumber \\
& + g_{N T_{2R}}^{\chi} \left( \overline{N_R} \gamma_\mu T_{2R} \right) \chi^{\mu \dagger} + g_{N T_{2L}}^{\chi} \left( \overline{N_R^c} \gamma_\mu T_{2L} \right) \chi^{\mu \dagger} + \text{h.c.} \,, \\[10pt]
\mathcal{L}_{\scriptscriptstyle FFV,(B)}^{N_R}  = & \;  g_{NN}^\mathcal{B} \left( \overline{N_R} \gamma_\mu N_R \right) \mathcal{B}^\mu + g_{Ne}^{\mathcal{B}_1} \left( \overline{N_R} \gamma_\mu e_R \right) \mathcal{B}_1^\mu + g_{NL}^{\mathcal{L}_1} \left( \overline{N_R^c} \gamma_\mu L \right) \mathcal{L}_1^\mu \nonumber \\
& + g_{Nd}^{\mathcal{U}_1} \left( \overline{N_R} \gamma_\mu d_R \right) \mathcal{U}_1^{\mu \dagger} + g_{Nu}^{\mathcal{U}_2} \left( \overline{N_R} \gamma_\mu u_R \right) \mathcal{U}_2^{\mu \dagger} + g_{QN}^{\mathcal{Q}_1} \left( \overline{N_R^c} \gamma_\mu Q \right) \mathcal{Q}_1^{\mu \dagger} + \text{h.c.}
\end{align}
}
In the previous Lagrangian, we assumed all new heavy fermions are Dirac spinors, although there are two cases, $\mathcal{N}$ and $\Sigma$, which could be Majorana fermions. In the case of those fields being Majorana, one can replace one of the Weyl fermions by its charged conjugated counterpart, since $\mathcal{\psi}_L \equiv \mathcal{\psi}_R^c$ (and $\mathcal{\psi}_R \equiv \mathcal{\psi}_L^c$).  In fact, in one particular model discussed in section~\ref{sec:lnv}, the Majorana nature of $\mathcal{N}$ was required to draw one model diagram. Nevertheless, in this appendix we keep the notation of two distinct chiral components for each fermion.

The Lagrangian terms that involve the new vector $\mathcal{U}_1$ and contain at least one light field are
\begin{align}
\mathcal{L}_{mixed}^{\mathcal{U}_1}  = &  \; g_{N D_L}^{\mathcal{U}_1} \left( \overline{N_R^c} \gamma_\mu D_L \right) \mathcal{U}_1^{\mu \dagger}  + g_{N D_R}^{\mathcal{U}_1} \left( \overline{N_R} \gamma_\mu D_R \right) \mathcal{U}_1^{\mu \dagger} + g_{N d}^{\mathcal{U}_1} \left( \overline{N_R} \gamma_\mu d_R \right) \mathcal{U}_1^{\mu \dagger}  \nonumber \\
& + g_{e U}^{\mathcal{U}_1} \left( \overline{e_R^c} \gamma_\mu U_L \right) \mathcal{U}_1^{\mu \dagger} + g_{L Q_1}^{\mathcal{U}_1} \left( \overline{L^c} \gamma_\mu Q_{1R} \right) \mathcal{U}_1^{\mu \dagger} + g_{L Q_5}^{\mathcal{U}_1} \left( \overline{L} \gamma_\mu Q_{5L} \right) \mathcal{U}_1^{\mu \dagger} \nonumber \\
& + g_{u E}^{\mathcal{U}_1} \left( \overline{u_R^c} \gamma_\mu E_L \right) \mathcal{U}_1^{\mu \dagger} +  g_{Q \Delta_1}^{\mathcal{U}_1} \left( \overline{Q^c} \gamma_\mu \Delta_{1R} \right) \mathcal{U}_1^{\mu \dagger} \nonumber \\
& +  g_{u D}^{\mathcal{U}_1} \left( \overline{u_R^c} \gamma_\mu D_L \right) \mathcal{U}_1^{\mu } +  g_{d U}^{\mathcal{U}_1} \left( \overline{d_R^c} \gamma_\mu U_L \right) \mathcal{U}_1^{\mu } +  g_{Q Q_1}^{\mathcal{U}_1} \left( \overline{Q^c} \gamma_\mu Q_{1R} \right) \mathcal{U}_1^{\mu} \nonumber \\
& + g_{d \mathcal{N}_L}^{\mathcal{U}_1} \left( \overline{d_R^c} \gamma_\mu \mathcal{N}_L \right) \mathcal{U}_1^{\mu \dagger} +  g_{d \mathcal{N}_R}^{\mathcal{U}_1} \left( \overline{d_R} \gamma_\mu \mathcal{N}_R \right) \mathcal{U}_1^{\mu} + \text{h.c.}
\end{align}
For completeness, we have written in the first line the interaction terms containing $N_R$, which were already introduced in $\mathcal{L}_{\scriptscriptstyle FFV}^{N_R}$. 

The renormalisable interactions of the vector $\mathcal{L}_1$ with the light fields are given by 
\begin{align}
\mathcal{L}_{mixed}^{\mathcal{L}_1}  = & \; \ g_{N \Delta_{1R}}^{\mathcal{L}_1} \left( \overline{N_R} \gamma_\mu \Delta_{1R} \right)  \mathcal{L}_1^{\mu} +  g_{N \Delta_{1L}}^{\mathcal{L}_1} \left( \overline{N_R^c} \gamma_\mu \Delta_{1L} \right) \mathcal{L}_1^{\mu} + g_{N L }^{\mathcal{L}_1} \left( \overline{N_R^c} \gamma_\mu L \right) \mathcal{L}_1^{\mu}  \nonumber \\
& + g_{e \Delta_1}^{\mathcal{L}_1} \left( \overline{\Delta_{1R}} \gamma_\mu e_R \right) \mathcal{L}_1^{\mu} + g_{L \mathcal{N}_L}^{\mathcal{L}_1} \left( \overline{\mathcal{N}_{L}} \gamma_\mu L \right) +  g_{L \mathcal{N}_R}^{\mathcal{L}_1} \left( \overline{\mathcal{N}_{R}^c} \gamma_\mu L \right) \mathcal{L}_1^{\mu} \nonumber \\
& + g_{e \Delta_{3}}^{\mathcal{L}_1} \left( \overline{e_R} \gamma_\mu \Delta_{3R} \right)  \mathcal{L}_1^{\mu} +  g_{L \Sigma_{L}}^{\mathcal{L}_1} \left( \overline{\Sigma_L} \gamma_\mu L \right) \mathcal{L}_1^{\mu} +  g_{L \Sigma_R}^{\mathcal{L}_1} \left( \overline{\Sigma_R^c} \gamma_\mu L \right) \mathcal{L}_1^{\mu} \nonumber \\
& + g_{L E }^{\mathcal{L}_1} \left( \overline{L} \gamma_\mu E_L \right) \mathcal{L}_1^{\mu} + g_{L \Sigma_{1} }^{\mathcal{L}_1} \left( \overline{L} \gamma_\mu \Sigma_{1L} \right) \mathcal{L}_1^{\mu} \nonumber \\
& + g_{Q T_2 }^{\mathcal{L}_1} \left( \overline{T_{2L}} \gamma_\mu Q \right) \mathcal{L}_1^{\mu} + g_{u Q_7 }^{\mathcal{L}_1} \left( \overline{Q_{7R}} \gamma_\mu u_R \right) \mathcal{L}_1^{\mu} +  g_{d Q_1 }^{\mathcal{L}_1} \left( \overline{Q_{1R}} \gamma_\mu d_R \right) \mathcal{L}_1^{\mu} \nonumber \\
& +  g_{Q U }^{\mathcal{L}_1} \left( \overline{U_{L}} \gamma_\mu Q \right) \mathcal{L}_1^{\mu} +  g_{Q D }^{\mathcal{L}_1} \left( \overline{Q} \gamma_\mu D_L \right) \mathcal{L}_1^{\mu} +  g_{Q T_1 }^{\mathcal{L}_1} \left( \overline{Q} \gamma_\mu T_{1L} \right) \mathcal{L}_1^{\mu} \nonumber \\
& +  g_{d Q_5 }^{\mathcal{L}_1} \left( \overline{d_R} \gamma_\mu Q_{5R} \right) \mathcal{L}_1^{\mu} +  g_{u Q_1 }^{\mathcal{L}_1} \left( \overline{u_R} \gamma_\mu Q_{1R} \right) \mathcal{L}_1^{\mu} + \text{h.c.}
\end{align}
Again, we have included in the first line the three interactions with $N_R$, which we already presented in $\mathcal{L}_{\scriptscriptstyle FFV}^{N_R}$. The remaining terms involve at least one SM field.

Finally, we write down the 3-point interaction terms among heavy BSM fields that are needed in different opening diagrams of the operator $\mathcal{O}_{NH^4}$. These are
\begin{align}
\mathcal{L}_{heavy}^{\text{diag}} = & \; \bigg\lbrace y_{\Delta_1}^{\mathcal{S}} \left( \overline{\Delta_{1L}} \Delta_{1R} \right) \mathcal{S} + y_{\Delta_1}^{\Xi} \left( \overline{\Delta_{1L}} \Delta_{1R} \right) \Xi  \nonumber \\
& + y_{\Delta_{1L}}^{\Xi_1} \left( \overline{\Delta_{1L}^c} \Delta_{1L} \right) \Xi_1  + y_{\Delta_{1R}}^{\Xi_1} \left( \overline{\Delta_{1R}^c} \Delta_{1R} \right) \Xi_1  \bigg\rbrace + \text{ h.c.}  \nonumber \\
& + \mu_{\mathcal{S} \Xi_1} \left( \mathcal{S} \Xi_1^\dagger \Xi_1 \right) + \mu_{\mathcal{S} \Xi} \left(\mathcal{S} \Xi \Xi\right) + \mu_{\mathcal{S}} \left( \mathcal{S}  \mathcal{S} \mathcal{S} \right) \,.
\end{align}
The complete set of interactions in $\mathcal{L}_{\text{UV}}$ can be found in the ancillary file added to this paper.

\section*{Acknowledgements}

We would like to thank Jos\'e Santiago for help with the tool
MatchMakerEFT~\cite{Carmona:2021xtq} and Javier Fuentes-Mart\'in for
help with \texttt{Matchete} \cite{Fuentes-Martin:2022jrf}.  M.H. and
R.B. acknowledge support by grants PID2020-113775GB-I00 (AEI/10.13039/501100011033) and CIPROM/2021/054 (Generalitat Valenciana).  R.B. also
acknowledges financial support from the Generalitat Valenciana (grant
ACIF/2021/052). R.C. is supported by the Alexander von Humboldt
Foundation Fellowship.

\bibliographystyle{JHEP}
\bibliography{NR_decomp}

\end{document}